\documentclass[a4paper,11pt]{article}
\pdfoutput=1 % if your are submitting a pdflatex (i.e. if you have
\usepackage{jcappub} % for details on the use of the package, please
\usepackage[T1]{fontenc} % if needed
\usepackage{graphicx}
\usepackage{subfig}
\usepackage[export]{adjustbox}
\usepackage{svg}
\usepackage{hyperref}

\title{Fast Lightcones for Combined Cosmological Probes}
\author[,a]{Rapha\"{e}l Sgier,\footnote{Corresponding author, Email: raphael.sgier@phys.ethz.ch}}
\author[a]{Janis Fluri,}
\author[a]{J\"org Herbel,}
\author[a]{Alexandre R\'efr\'egier,}
\author[a,b]{Adam Amara,}
\author[a]{Tomasz Kacprzak,}
\author[c]{Andrina Nicola}
\affiliation[a]{Institute for Particle Physics and Astrophysics, Department of Physics, ETH Zurich,\\Wolfgang-Pauli-Strasse 27, 8093 Zurich, Switzerland}
\affiliation[b]{Institute of Cosmology \& Gravitation, University of Portsmouth, Dennis Sciama Building, Burnaby Road, Portsmouth PO1 3FX, UK}
\affiliation[c]{Department of Astrophysical Sciences, Princeton University, Peyton Hall, Princeton NJ08544-0010, USA}

\abstract{The combination of different cosmological probes offers stringent tests of the $\Lambda$CDM model and enhanced control of systematics. For this purpose, we present an extension of the lightcone generator \textsc{UFalcon} first introduced in Sgier \textit{et al.} \cite{Sgier2019}, enabling the simulation of a self-consistent set of maps for different cosmological probes. Each realization is generated from the same underlying simulated density field, and contains full-sky maps of different probes, namely weak lensing shear, galaxy overdensity including RSD, CMB lensing, and CMB temperature anisotropies from the ISW effect. The lightcone generation performed by \textsc{UFalcon} is parallelized and based on the replication of a large periodic volume simulated with the GPU-accelerated $N$-Body code \textsc{PkdGrav3}. The post-processing to construct the lightcones requires only a runtime of about 1 walltime-hour corresponding to about 100 CPU-hours. We use a randomization procedure to increase the number of quasi-independent full-sky \textsc{UFalcon} map-realizations, which enables us to compute an accurate multi-probe covariance matrix. Using this framework, we forecast cosmological parameter constraints by performing a multi-probe likelihood analysis for a combination of simulated future stage-IV-like surveys. We find that the inclusion of the cross-correlations between the probes significantly increases the information gain in the parameter constraints. We also find that the use of a non-Gaussian covariance matrix is increasingly important, as more probes and cross-correlation power spectra are included.
A version of the \textsc{UFalcon} package currently including weak gravitational lensing is publicly available.\footnote{\textsc{UFalcon}: \href{https://cosmology.ethz.ch/research/software-lab/UFalcon.html}{https://cosmology.ethz.ch/research/software-lab/UFalcon.html}}}
\begin{document}
\maketitle
\flushbottom

\section{Introduction}

The beginning of the era of precision cosmology has lead to the establishment of the $\Lambda$CDM model for cosmology. In spite of this remarkable success, the nature of dark energy and dark matter (DM) is still poorly understood and remains one of the greatest challenges in physics nowadays. A promising approach to increase the knowledge about our Universe is based on the combination of different cosmological probes tracing the Large-Scale Structure (LSS) of the Universe, such as the cosmic microwave background (CMB) temperature anisotropies, galaxy clustering, weak gravitational lensing, galaxy-galaxy lensing and galaxy clusters. 
\smallbreak
Current surveys such as the Dark Energy Survey (DES\footnote{\href{http://www.darkenergysurvey.org}{http://www.darkenergysurvey.org}}), the Canada France Hawaii Telescope Lensing Survey (CFHTLenS\footnote{\href{https://www.cfhtlens.org}{https://www.cfhtlens.org}}), the Kilo-Degree Survey (KiDS\footnote{\href{http://kids.strw.leidenuniv.nl}{http://kids.strw.leidenuniv.nl}}) and the Dark Energy Spectroscopic Instrument (DESI\footnote{\href{http://desi.lbl.gov}{http://desi.lbl.gov}}) continue to set tighter constraints on the cosmological model and its components. Future surveys such as the Legacy Survey of Space and Time (LSST\footnote{\href{http://www.lsst.org}{http://www.lsst.org}}), Euclid\footnote{\href{http://sci.esa.int/euclid/}{http://sci.esa.int/euclid/}} and the Wide Field Infrared Telescope (WFIRST\footnote{\href{http://wfirst.gsfc.nasa.gov}{http://wfirst.gsfc.nasa.gov}}) are expected the provide more and deeper data and therefore significantly advance our understanding of cosmology. Analyses based on individual cosmological probes are straightforward if they are assumed to be statistically independent \cite{Kilbinger2013},\cite{Abbott2016}. But since these surveys will cover large and overlapping regions of the sky, the retrieved high-quality imaging data is statistically not independent. Combining various probes in the analysis therefore allows us to consider non-negligible correlations between the probes, which have to be taken into account. Considering the cross-correlations between the statistics of the various cosmological probes not only significantly tightens the constraints on the cosmological model, but also allows us to have a better control of systematic effects and test for inconsistencies among the different probes \cite{Weinberg2013}. Moreover, the increase in sensitivity of current and future CMB experiments such as Planck\footnote{\href{https://www.cosmos.esa.int/web/planck}{https://www.cosmos.esa.int/web/planck}}, ACT\footnote{\href{https://act.princeton.edu/}{https://act.princeton.edu/}} and SPT\footnote{\href{https://pole.uchicago.edu/}{https://pole.uchicago.edu/}}, allowing us to resolve CMB temperature fluctuations down to arcmin scale, could potentially lead to a correlation of the different sources of secondary CMB anisotropies with other LSS probes, such as weak gravitational lensing and the Sunyaev-Zel'dovic (SZ) effect \cite{SunyaevZeldovich1972} (see e.g \cite{Bielby2010},\cite{Cooray2002},\cite{Fosalba2003}).
\smallbreak
Various earlier studies have conducted joint analyses of different cosmological probes. For example, parameter constraints from a joint analysis of galaxy-galaxy lensing and galaxy clustering have been derived in Mandelbaum \textit{et al.} \cite{Mandelbaum2013} and Cacciato \textit{et al.} \cite{Cacciato2013}. An extensive integrated analysis of CMB temperature anisotropies, galaxy clustering and cosmic shear has been performed in Nicola \textit{et al.} \cite{Nicola2016} and additionally included CMB lensing maps, supernovae data and local Hubble parameter measurement data from the Hubble Space Telescope (HST) in their joint analysis in Nicola \textit{et al.} \cite{Nicola2017}. 
\smallbreak
Cosmological parameter inference not only requires the observed data vector and an accurate theoretical prediction, but also a realistic estimation of the covariance matrix $\Sigma$. The simplest approach for the latter is to use a Gaussian approximation, which gives the minimum contribution to the covariance. A Gaussian covariance matrix $\Sigma_G$ would be the only contribution if the density field is a homogeneous and isotropic Gaussian random field, i.e. all information is contained in the power spectrum. The Gaussian term can then easily be computed given the survey area and the nonlinear matter power spectrum. Such an approximation for the covariance estimation is valid on sufficiently large scales (for multipoles $\ell \lesssim 100 - 200$ for galaxy sources at redshift $z_s \approx 1$ concerning cosmic shear), but becomes insufficient on smaller scales where nonlinear structure formation induces non-Gaussianities in the density field \cite{Barreira2018},\cite{Krause2017}. The non-Gaussian terms in the covariance $\Sigma_{NG}$ that arise from the connected higher-order moments of the density field have been found to have a significant impact on the error bars when considering the convergence power spectrum \cite{Hilbert2011}, \cite{Sato2009}. Likewise, invoking a non-Gaussian covariance matrix also significantly increases the size of the parameter constraints obtained from a Fisher matrix analysis based on the convergence power spectrum and bispectrum \cite{Sato2013}. 
\smallbreak
In contrast to using a Gaussian approximation, the evaluation of the non-Gaussian term is nontrivial. The first approach for the latter consists in directly evaluating the nonlinear matter trispectrum, which represents the analytical form of the non-Gaussian contribution to the covariance matrix \cite{Scoccimarro1999}. Numerous analytical methods such as Standard Perturbation Theory (SPT) \cite{Scoccimarro1999}, Effective Field Theory of Large-Scale Structure (EFTofLSS) \cite{Bertolini2016a},\cite{Bertolini2016b},\cite{Bertolini2016c} and approaches based on the halo model \cite{Cooray2002_halomodel} have been used to calculate the trispectrum. An analytical calculation of a non-Gaussian multi-probe covariance matrix based on the halo model has been calculated in Krause \& Eifler \cite{Krause2017}, which incorporates cosmic shear, galaxy-galaxy lensing, galaxy clustering, cluster number counts and cluster weak lensing to perform a joint analysis using the \textsc{CosmoLike}\footnote{\href{https://github.com/CosmoLike}{https://github.com/CosmoLike}} software.
\smallbreak
The second common approach to evaluate the non-Gaussian term consists in generating a sufficiently large ensemble of statistically independent realizations of the matter density field, from which the power spectrum sample covariance can be measured (see e.g. \cite{Sato2009},\cite{Takahashi2009},\cite{Sato2011},\cite{Deraps2012},\cite{Blot2015},\cite{Takahashi2017},\cite{Petri2016},\cite{Klypin2018}). Following such an approach, recent work done in Harnois-D\'eraps \textit{et al.} \cite{Deraps2018} introduced a multi-probe covariance matrix for CMB lensing, cosmic shear, galaxy-galaxy lensing and galaxy lensing from 844 independent realizations.

Combining these two approaches has led to the development of hybrid methods, which analytically compute the well-understood parts of the covariance matrix and use simulations to estimate the remaining parts \cite{Friedrich2017}. A fourth approach measures simulation covariances from different subsamples of the underlying dataset, such that their combination is less affected by noise when estimating the diagonal elements of the covariance matrix \cite{Joachimi2016}.

The drawback of the numerical ensemble-approach is its computational cost: For each statistically independent realization of the non-Gaussian matter density field, a $N$-Body simulation has to be performed. This conflicts with the need of a large ensemble of realizations required to estimate a statistically well-converged covariance matrix. Future surveys are expected to measure in hundreds of data bins, which will require an order of $10^4$ simulation realizations to prevent 5-10\% degradation in the parameter constraints \cite{Dodelson2013}. To this end, various alternative methods to fully realized $N$-Body codes have been developed, such as the Comoving Lagrangian Acceleration method (COLA; \cite{Tassev2013},\cite{Koda2016}) and applications thereof (e.g. L-PICOLA \cite{Howlett2015},\cite{Winther2017},\cite{Wright2017},\cite{Sgier2019} and ICE-COLA \cite{Izard2016},\cite{Izard2017}). Although these methods are several orders of magnitude faster than full $N$-Body codes, they lack accuracy on small scales. Another interesting alternative method consists in reusing the same $N$-Body simulation output by applying random transformations to generate multiple quasi-independent maps. Concerning the cosmic shear field, it has been shown by Petri \textit{et al.} \cite{Petri2016} that one single $N$-Body output is sufficient to produce $\sim 10^4$ realizations. 
\bigbreak
This work presents a extension of the \textsc{UFalcon} package, which was first introduced in our paper \cite{Sgier2019} and initially designed to generate full-sky convergence maps with a minimal runtime. \textsc{UFalcon} has recently been applied in Kapcprzak \textit{et al.} \cite{Kacprzak2019} for the covariance estimation based on the approximate $N$-Body code \textsc{L-PICOLA}. Furthermore, it was used to generate convergence maps used as training data for Convolutional Neural Networks (CNN) for cosmological parameter inference in Fluri \textit{et al.} \cite{Fluri2018a},\cite{Fluri2018b},\cite{Fluri2019} and used to study non-Gaussian statistics such as peak counts, minimum counts and Minkowski functionals in Z\"urcher \textit{et al.} \cite{Zuercher2020}. The presented extension adds the functionality to the code to generate a complete \textit{set} of full-sky maps for different cosmological probes, such as weak lensing shear, galaxy overdensity including the effects of redshift-space distortions (RSD), CMB lensing and CMB temperature anisotropies from the ISW effect. Furthermore, we invoke a randomization procedure based on random transformation applied to the 3-dimensional matter density field in order to enhance the number of realizations at least by a factor of $\sim 20$. %Our randomization procedure operates on large periodic simulation volumes, such that we expect to capture all super-sample covariance effects (SSC) and 
Our pipeline is designed to post-process output of the DM-only $N$-Body simulation code \textsc{PkdGrav3} (Stadel \textit{et al.} \cite{Stadel2001}), but is in general able to use output of different $N$-Body codes. \textsc{UFalcon} applied to \textsc{PkdGrav3} output combines accuracy and minimal computational runtime: The simulation of the density field guarantees to satisfy a certain force accuracy and is therefore not an approximate $N$-Body code. Furthermore, the \textsc{PkdGrav3} code is highly efficient and can be run with graphics processing units (GPU) support. The subsequent post-processing with \textsc{UFalcon} can be parallelized on a computer cluster and has a runtime of $\sim 1$ hour walltime for one set of full-sky maps.
\smallbreak
Furthermore, \textsc{UFalcon} offers the possibility to generate continuous full-sky CMB lensing potential and deflection angle maps directly from the gravitational potential of the density field. Our approach of using large periodic simulation volumes together with an interpolation routine enables the generation of continuous lensing potential maps which include nonlinear and non-Gaussian effects. These maps can be of use to help separating the different contributions to the CMB anisotropies and improve on an accurate and complete interpretation of the latest Planck data \cite{Seljak2004}.
\smallbreak
In this paper, we first describe the extension of the \textsc{UFalcon} package, namely the possibility to compute full-sky maps of different cosmological probes from the same underlying density field. We describe the application of \textsc{UFalcon} to the DM-only $N$-Body simulation output generated using the \textsc{PkdGrav3} code and the statistical analysis of the generated full-sky maps by considering the 4 auto- and the 6 cross- spherical harmonic power spectra between the probes. Furthermore, we estimate a multi-probe covariance matrix based on 630 quasi-independent realizations and perform a forecast analysis for a stage-IV-like survey geometry based on spherical harmonic power spectra for multipoles between $\ell = 10^2$ and $10^3$. Hereby we focus on quantifying the impact of including different probe-combinations, the effect of including the cross-correlations between the probes and the use of a fully non-Gaussian simulation-based covariance matrix on cosmological parameter constraints. The obtained results are compared to the case where we use a Gaussian approximation for the covariance matrix.
\bigbreak
This paper is organized as follows. In section \ref{theopred}, we review the analytical framework required for our spherical harmonic power spectrum predictions. The numerical framework is discussed in section \ref{nummet}, which includes a description of our past-lightcone construction for the different probes and an overview of various systematic effects present. In section \ref{statanal}, we show various statistical quantities including a multi-probe covariance matrix we compute based on our full-sky maps. Based on our findings, we perform a forecast analysis for a stage-IV-like survey geometry in section \ref{forecast} and quantify the impact of the non-Gaussian contribution to the covariance matrix on the cosmological parameter constraints.

\section{Analytical Predictions}
\label{theopred}
The analytical predictions for the different cosmological probes considered are calculated in the same way as in Nicola \textit{et al.} \cite{Nicola2016} using \textsc{PyCosmo} (see R\'efr\'egier \textit{et al.} \cite{pycosmo2018}, Tarsitano \textit{et al.} \cite{Tarsitano2020}), which is a Python-based framework to solve the Einstein-Boltzmann equations governing the evolution of the linear perturbations. We consider the statistical properties of two probes $i$ and $j$ by computing the spherical harmonic power spectra $C_{\ell}^{ij}$ using the \textit{Limber approximation} (\cite{Limber1953},\cite{Kaiser1992},\cite{Kaiser1998}) at multipole $\ell$ given by
\begin{equation}
C_{\ell}^{ij} = \int \mathrm{d}z \, \frac{c}{H(z)} \, \frac{W^{i} (\chi(z)) \, W^{j} (\chi(z))}{\chi^{2} (z)} \,\, P^{\mathrm{nl}}_{\delta \delta} \left(k = \frac{\ell + 1/2}{\chi(z)}, z \right)\, ,
\label{cl}
\end{equation}
where $H(z)$ is the Hubble parameter, $\chi(z)$ the comoving distance and $c$ is the speed of light. Note that equation (\ref{cl}) is not only used to make analytical predictions for the auto-correlations, but also for the cross-correlations between different probes. The Limber approximation has been used in order to speed up the calculation and is valid for multipoles $\ell \gtrsim 10$ (i.e. small angular scales) and broad redshift bins \cite{Nicola2016},\cite{Peacock1999}. The nonlinear matter power spectrum $P^{\mathrm{nl}}_{\delta \delta} \left(k, z \right)$ is calculated using the fitting function from Mead \textit{et al.} \cite{Mead2015},\cite{Mead2016}, whereas the linear matter power spectrum is obtained from the transfer function derived by Eisenstein \& Hu \cite{EisensteinHu}. In the following we describe the window functions implemented in \textsc{PyCosmo} for the cosmological probes $i, j \in \{\gamma, \delta_g, \kappa_\mathrm{CMB}, \Delta T_\mathrm{ISW}\}$, described below for a flat cosmological model. Note that \textsc{UFalcon} does not depend on the code used to calculate analytical predictions. The $N$-Body simulation-based results from our pipeline can therefore be compared to analytical predictions from other codes, such as \textsc{Class} \cite{Lesgourgues2011}.\\
\\
\textbf{Weak lensing shear ($\gamma$).} The window function for cosmic shear is given by \cite{Nicola2016}
\begin{equation}
W^{\gamma} (\chi(z)) = \frac{3}{2} \frac{\Omega_\mathrm{m} H_0^2}{c^2} \frac{\chi(z)}{a} \int^{\chi_\mathrm{h}}_{\chi(z)} \mathrm{d}z' n(z') \frac{\chi(z') - \chi(z)}{\chi(z')} \, ,
\label{wshear}
\end{equation}
where $\Omega_\mathrm{m}$ and $H_0$ denote the present day values of the matter density and the Hubble parameter respectively. Furthermore, $\chi_\mathrm{h}$ is the comoving distance to the horizon, $n(z)$ is the normalised redshift selection function of the lensed galaxies and $a$ denotes the scale factor. Note that a simple $\ell$-dependent pre-factor of the form $(\ell + 2)(\ell + 1)\ell(\ell -1)(\ell + 0.5)^{-4}$ can be multiplied to the Limber-approximated cosmic shear power spectrum formula to alleviate some of the inaccuracies of the Limber approximation on multipoles $\ell \lesssim 10$ \cite{Kitching2017}. Since we are interested in computing the Limber approximated power spectra of different probes in a consistent way (i.e. based on equation (\ref{cl})) and focusing our analysis on multipoles between $\ell = 100$ and $1000$, we choose not to apply the pre-factor to our power spectrum calculations.\\
\\
\textbf{Galaxy clustering ($\delta_g$).} Regarding galaxy clustering, the window function can be written as \cite{Nicola2016}
\begin{equation}
W^{\delta_g} (\chi(z)) = \frac{H(z)}{c} b(z) n(z) \, .
\label{wdelta}
\end{equation}
In our analysis, we use a simple approach based on a constant, linear, scale- and redshift-independent galaxy bias $b(z) \equiv b$. This is a valid assumption on large scales, which are well-described by linear theory. More complicated galaxy bias models could also be implemented. The observed galaxy redshifts are typically converted to radial distances by using Hubble's law, which neglects the peculiar velocities of the galaxies. This leads to redshift-space distortions (RSD) between the clustering of galaxies along the line-of-sight and perpendicular to it. The true comoving position of a galaxy $\vec{r}$ is distorted along the line-of-sight due to its peculiar velocity $\vec{v}$ according to \cite{Torre2012}
\begin{equation}
\vec{s} = \vec{r} + \frac{v_{||}(\vec{r})\, \hat{e}_{||}}{aH(a)} \, ,
\end{equation}
where $\hat{e}_{||}$ denotes the unit vector along the line-of-sight. We refer to \cite{Percival2011} and \cite{Torre2012} for a more detailed description of galaxy clustering and RSD. The above description of the power spectrum for galaxy clustering ignores the effect of the peculiar velocities of the galaxies on the nonlinear power spectrum $P^{\mathrm{nl}}_{\delta \delta} \left(k, z \right)$. In the presence of RSD, the window function for galaxy clustering has an additional component given by \cite{Padmanabhan2008}
\begin{equation}
\begin{split}
W^{\delta_g}_\mathrm{RSD} &=  \frac{H(z)}{c} b(z) n(z) \left[ \beta \frac{(2 \ell^2 + 2 \ell - 1)}{(2 \ell + 3)(2 \ell - 1)} \right. \\
 & \left. - \beta \frac{\ell (\ell - 1)}{(2 \ell - 1)(2 \ell + 1)}  - \beta \frac{(\ell + 1)(\ell + 2)}{(2 \ell + 1)(2 \ell + 3)} \right] \quad ,
\end{split}
\end{equation}
where $\beta \approx \Omega_m^{0.6} / b(z)$ is the approximated redshift distortion parameter. Note that the additional term vanishes in the limit of $\ell \gg 1$, i.e. $W^{\delta_g}_\mathrm{RSD} \xrightarrow{\ell \gg 1} 0$, and mostly contribute to the largest scales ($\ell  \lesssim 30$). We refer to \cite{Padmanabhan2008} for a detailed derivation of the window function for galaxy clustering in presence of RSD.\\
\\
\textbf{CMB temperature anisotropies ($\Delta T$)}. The spherical harmonic power spectrum of the temperature anisotropies $T$ of the CMB can be related to the primordial power spectrum generated during inflation $P^{\mathrm{lin}}_{\delta \delta} (k)$ and is given by \cite{Dodelson2003}
\begin{equation}
C_{\ell}^{TT} = \frac{2}{\pi} \int \mathrm{d} k \,  k^2 P^{\mathrm{lin}}_{\delta \delta} (k) \left| \frac{\Delta T_\ell (k)}{\delta (k)} \right| \, ,
\end{equation}
where $\Delta T_\ell (k)$ represents the CMB temperature anisotropies,  $\delta \equiv (\rho  - \bar{\rho}) / \bar{\rho}$ is the density contrast and $\bar{\rho}$ the mean density of the universe. Secondary temperature perturbations in the CMB radiation can be generated by the linear (integrated Sachs-Wolfe or ISW effect \cite{SachsWolfe1967}) and nonlinear (Rees-Schiama or RS effect \cite{ReesSchiama1968}) decay of large-scale gravitational potential fluctuations. These temperature fluctuations arising from the ISW + RS effect can be written as the integral over conformal time $\eta$ of the time derivative of the gravitational potential (\cite{Padmanabhan2005},\cite{SachsWolfe1967},\cite{Gonzalez1990})
\begin{equation}
\Delta T_\mathrm{ISW}(\hat{n}) = T_\mathrm{CMB} \, \delta T_\mathrm{ISW} = 2\, T_\mathrm{CMB} \int^{\eta_0}_{\eta_r} \mathrm{d} \eta \frac{\partial \Phi}{\partial \eta}\, ,
\label{iswdef}
\end{equation}
where $\eta_0$ and $\eta_r$ represents the conformal time today and at recombination respectively. CMB temperature anisotropies are correlated through the ISW effect to tracers of the LSS such as weak lensing shear or galaxy clustering \cite{SachsWolfe1967}. In the present work, we focus on the temperature anisotropies due to the ISW effect $\Delta T_\mathrm{ISW}$.
\smallbreak
This source of secondary CMB anisotropies is especially challenging to measure, since the ISW signal suffers from cosmic variance at large scales and is entangled with the nonlinear RS signal at smaller scales. A disentanglement between the two signals at smaller scales is thus crucial to correctly retrieve cosmological information \cite{Cai2010}. On larger scales, where linear theory holds, the spherical harmonic power spectrum between the CMB temperature anisotropies and a tracer $i \in \{ \gamma, \delta_g, \kappa_\mathrm{CMB} \}$ of the LSS can be written as \cite{Crittenden1996}
\begin{equation}
\begin{split}
C_{\ell}^{iT} &= T_\mathrm{CMB} \left( \frac{3 \Omega_m H_0^2}{c^2} \right) \frac{1}{(\ell + 1/2)^2} \int \mathrm{d}z \frac{\mathrm{d}}{\mathrm{d} z} \left[ D(z) (1 + z) \right] \\
 & \times D(z) W^{i} (\chi (z)) \, P^{\mathrm{lin}}_{\delta \delta} \left(k = \frac{\ell + \frac{1}{2}}{\chi(z)}, 0 \right) \quad ,
\end{split}
\label{iswcross}
\end{equation}
where the window function $W^{i}  (\chi (z))$ is given by equation (\ref{wshear}), (\ref{wdelta}) or (\ref{wkappa}). The linear matter power spectrum $P^{\mathrm{lin}}_{\delta \delta} (k, z)$ has been split up into a time-dependent growth factor $D(z)$ and a scale-dependent part $P^{\mathrm{lin}}_{\delta \delta} (k, 0)$ \cite{Nicola2016}. Furthermore, the mean temperature of the CMB today is denoted by $T_\mathrm{CMB}$. A detailed derivation of equation (\ref{iswcross}) for galaxy clustering and weak lensing shear can be found in Padmanabhan \textit{et al.} \cite{Padmanabhan2005} and Appendix F in Nicola \textit{et al.} \cite{Nicola2016} respectively.
\smallbreak
We additionally implemented the auto-correlation of the temperature anisotropies from the ISW effect $\Delta T_\mathrm{ISW}$ in order to verify more directly our results obtained from the numerical $N$-Body simulation, which can be written as
\begin{equation}
\begin{split}
C_{\ell}^\mathrm{ISW} &= T_\mathrm{CMB}^2 \left( \frac{3 \Omega_m H_0^2}{c^2} \right)^2 \frac{1}{(\ell + 1/2)^4} \int \mathrm{d}z \frac{\mathrm{d}}{\mathrm{d}z} \left[ D(z) (1+z) \right]^2 \chi(z)^2\\
& \times P^{\mathrm{lin}}_{\delta \delta} \left(k = \frac{\ell + 1/2}{\chi(z)}, 0 \right) \, .
\end{split}
\label{iswauto}
\end{equation}
A derivation of equation (\ref{iswauto}) is given in Appendix \ref{appendixA}. In the present work we consider the $\Delta T_\mathrm{ISW}$ auto-correlation and all 3 cross-correlations between $\Delta T_\mathrm{ISW}$ and the tracers $\{ \gamma, \delta_g, \kappa_\mathrm{CMB} \}$ of LSS. The cross-correlations between the CMB temperature anisotropies and the galaxy overdensity and weak lensing shear are mainly due to the ISW effect \cite{SachsWolfe1967}. As pointed out in \cite{Nicola2017}, the cross-correlation between the CMB temperature anisotropies and the CMB lensing convergence is also mostly dominated by the ISW effect but additionally obtains contributions from the Sunyaev-Zel'dovic effect (SZ) \cite{Sunyaev1980} and the Doppler effect due to the bulk velocities of electrons scattering the CMB photons (see e.g. \cite{Goldberg1999} and \cite{Cooray2000} for more details). We do not take into account these further contributions in the present analysis.\\
\\
\textbf{CMB lensing ($\kappa_\mathrm{CMB}$).} The CMB can be described by primary anisotropies imprinted on the last scattering surface and by secondary anisotropies, which originate from the scattering of the CMB photons by matter inhomogeneities and on electrons along the way to us. Amongst the most important effects leading to secondary anisotropies is the weak gravitational lensing of the CMB, which arises from the deflection of CMB photons through potential gradients along our line of sight \cite{Lewis2006}.
\smallbreak
In the subsequent discussion, we approximate recombination as an instantaneous process at redshift $z_\ast \approx 1100$ located at a single source plane at a comoving distance $\chi_\ast \approx 1.5 \times 10^4 \, \mathrm{Mpc}$. Working with conformal time $\eta$ in a flat universe, we define the lensing potential (induced by scalar perturbations with no anisotropic stress) as a function of the line-of-sight pointing in direction $\hat{n} = (\theta, \phi)$ as \cite{Ma1995},\cite{Carbone2008}
\begin{equation}
\psi(\hat{n}) = -2 \int_0^{\chi_\ast} \frac{\chi_\ast - \chi}{\chi_\ast \chi} \frac{\Phi (\chi \hat{n}; \eta_0 - \chi )}{c^2} \mathrm{d}\chi \, ,
\label{psi}
\end{equation}
where $\eta_0 - \chi$ corresponds to the conformal time at which the photon was at comoving coordinates $\chi \hat{n}$ and $\Phi$ is the physical peculiar gravitational potential generated by matter inhomogeneities. The deflection angle of a source located at $\chi_\ast$ is given in terms of the lensing potential as
\begin{equation}
\alpha(\hat{n}) = -2 \int_0^{\chi_\ast} \frac{\chi_\ast - \chi}{\chi_\ast \chi} \nabla_{\hat{n}} \frac{\Phi (\chi \hat{n}; \eta_0 - \chi )}{c^2} \mathrm{d}\chi \, , 
\end{equation}
with $\nabla_{\hat{n}}$ representing the divergence operator on the sky, i.e. the two-dimensional derivative transverse to the direction $\hat{n}$ (see e.g. \cite{Lewis2006},\cite{Hu2000},\cite{BartelmannSchneider2001},\cite{Refregier2003}). The scalar spherical harmonic coefficients of the CMB lensing potential $\psi_{\mathrm{CMB}, \ell m}$ are related to the CMB lensing convergence $\kappa_{\mathrm{CMB}, \ell m}$ through \cite{Wallis2017}
\begin{equation}
\kappa_{\mathrm{CMB}, \ell m} = \frac{\ell (\ell + 1)}{2} \psi_{\mathrm{CMB}, \ell m}\, .
\end{equation}
As described in \cite{Nicola2017}, the window function for the CMB lensing convergence can be expressed as the single-plane limit of the weak lensing shear window function (equation (\ref{wshear})) and be written as
\begin{equation}
W^{\kappa_\mathrm{CMB}} (\chi(z)) = \frac{3}{2} \frac{\Omega_\mathrm{m} H_0^2}{c^2} \frac{\chi(z)}{a} \frac{\chi(z_\ast) - \chi(z)}{\chi(z_\ast)} \, .
\label{wkappa}
\end{equation}

\section{Numerical Methods}
\label{nummet}

The \textsc{UFalcon} pipeline was first introduced in \cite{Sgier2019} and designed to generate full-sky convergence maps for weak lensing shear. In this section, we describe the novel extension to \textsc{UFalcon}, which includes the functionality to generate a complete set of full-sky maps for the cosmological probes $\gamma$, $\delta_g$, $\kappa_\mathrm{CMB}$ and $\Delta T_\mathrm{ISW}$ from the same $N$-Body simulation output. We describe the underlying $N$-Body simulation, the lightcone construction and the map-making procedure for the different probes. We further highlight the construction of CMB lensing potential maps $\psi_\mathrm{CMB}$ and the deflection angle $\alpha$ form first principles. Moreover, we give a short description of the publicly available \textsc{UFalcon} package, which contains the weak gravitational lensing part of the pipeline.

\subsection{N-Body Simulation}
\label{nbody}
As of now, $N$-Body simulations provide the most accurate method to predict a wide dynamical range of structure formation. In the context of current and future LSS surveys, smaller angular scales indeed need to be under control in order to correctly infer cosmological parameters.
We use the DM-only $N$-Body code \textsc{PkdGrav3} (Stadel \textit{et al.} \cite{Stadel2001}) to simulate the matter density field. \textsc{PkdGrav3} is based on the Fast Multipole Method (FMM) to accurately compute the forces between the particles and scales as $\mathcal{O}(N)$. Furthermore, it is highly optimised for usage on the Piz Daint supercomputer (CSCS\footnote{\href{https://www.cscs.ch/}{https://www.cscs.ch/}}, Switzerland) and can be further accelerated when run with graphics processing unit (GPU) support. \textsc{PkdGrav3} has been successfully run with more than a trillion particles in Potter \textit{et al.} \cite{Potter2017} and used as training data for CNN's used for cosmological inference \cite{Fluri2019}.\\
\\
\textbf{Simulation configuration.} In this study, we assume a flat $\Lambda$CDM cosmological model with fiducial parameter values given by 
\begin{equation}
\boldsymbol{\theta}_\mathrm{fid} = \{ h, \Omega_m, \Omega_b, n_s, \sigma_8, T_\mathrm{CMB}\} = \{ 0.7, 0.3, 0.05, 0.961, 0.8, 2.275\, \mathrm{K} \} \, .
\label{fid_params}
\end{equation}
We aim at accurately resolving angular scales corresponding to multipoles between $\ell = 10^2$ and $10^3$, which imposes the need to have a sufficiently high resolution in the simulation to include the effects of nonlinear structure formation. At the same time, we need a sufficiently large simulation volume in order to incorporate large scale perturbation modes (see section \ref{SSC} for a discussion of super-sample covariance effects) and to cover an appropriate redshift range. We therefore ran a total of 30 simulations with $N_p = 1024^3$ particles in a volume of $V_\mathrm{sim} = (1600\, \mathrm{Mpc})^3$ (with a mass per DM-particle of $3.43 \times 10^{11}\, M_\odot$). Each simulation was generated with a different random seed for the initial conditions at redshift $z_\mathrm{init} = 50$ using the initial conditions generator \textsc{Music} \cite{Hahn2011}. Note that the redshift-interval between the outputs of the simulation is not constant and is defined by the choice of $z_\mathrm{init}$, $z_\mathrm{final}$, the number of timesteps and the cosmological parameters. The simulations were subsequently run with 150 timesteps between $z_\mathrm{init} = 50$ and $z_\mathrm{final} = 0.0$, with more outputs for lower redshifts. The simulation outputs a snapshot at every timestep. We store and post-process 54 snapshots output between redshift $z=0.0$ to $1.75$ using \textsc{UFalcon}. Although the redshift-interval between the outputs is not constant, we have a mean redshift-spacing of about $\sim 0.03$ in the interval of our interest. We set the accuracy parameters given by the tree opening angles for different redshifts to the default values $\theta = 0.4$ for $z >20$, $\theta_{20} = 0.55$ for $20>z>2$ and $\theta_2 = 0.7$ for $2>z$. The force calculation in \textsc{PkdGrav3} is realized through particle-particle and particle-cell interactions. Choosing a smaller value of $\theta$ results in a larger tree opening radius and more particle-particle interactions are calculated, while a high value of $\theta$ results in less particle-particle interactions. At very early times ($z>20$), when the Universe is very homogeneous, a smaller value of $\theta$ is necessary to obtain a certain accuracy in the force calculation. Otherwise, small errors in the initial nonlinear growth at early times could be amplified during the evolution and potentially lead to errors greater than $1\%$ in the power spectrum at the end of the simulation \cite{Stadel2001},\cite{Potter2017}.\\
\\
\textbf{Runtime.} The above simulation configuration leads to a walltime-runtime of about $\sim 2$ hours for lightcone-output (i.e. one output-file containing the particles at different redshifts, concentrically arranged) and $\sim 4$ hours ($\sim 32$ node-hours) for snapshot output per simulation with GPU-support on our computer cluster.

\subsection{Lightcone Construction}
\label{lightcone}

\begin{figure}[htbp!]
\centering
\includegraphics[width=7cm]{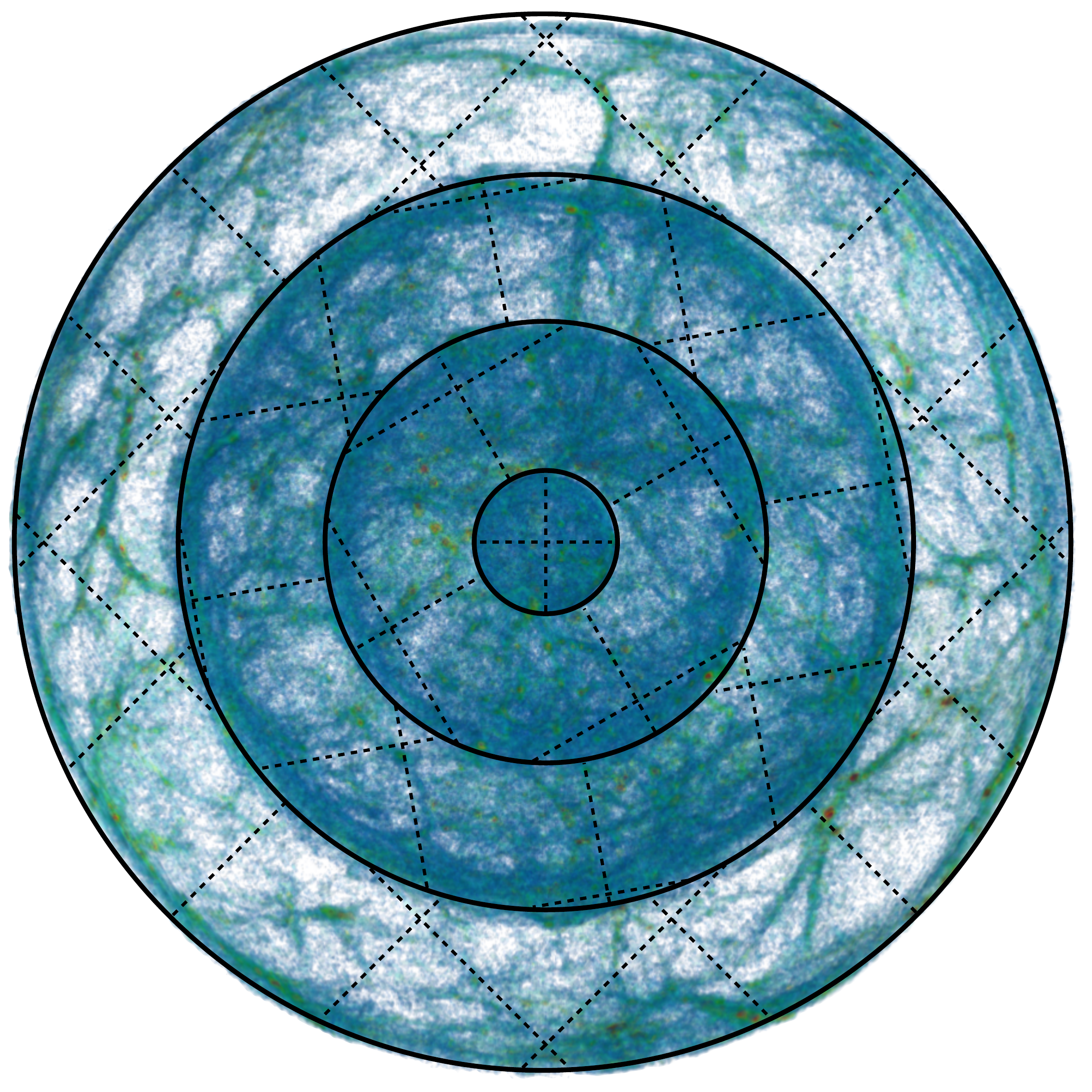}
\caption{Sketch of the full-sky lightcone construction obtained by replicating the density field. The simulation output at a discrete number of redshifts are used to generate concentrically arranged shells around the observer located at the center. The shell randomisation procedure, here depicted by the different orientation of the simulation volume (dashed lines) in each shell, is described in section \ref{randomisation}.} \label{sketch}
\end{figure}

Our aim is to generate full-sky maps covering a redshift range from $z=0.0$ to $1.75$, i.e. we have to construct a lightcone spanning a large survey volume. To this end, we decided to adopt a replication scheme in order to guarantee a sufficiently high resolution of the simulated density field across the full survey volume. Note that in our first paper \cite{Sgier2019}, we adopted an alternative lightcone-construction scheme by nesting two simulation boxes with different volumes and resolutions.
\smallbreak
We replicate the simulation volume 6 times in $x$, $y$ and $z$ direction, leading to a total of 216 replicas for the whole volume. The lightcone is constructed by concentrically stacking shells at different redshift of the replicated density field around the observer located at $z=0$. A sketch of the lightcone construction is shown in Figure \ref{sketch}. To construct our maps we make use of the Hierarchical Equal Area Iso-Latitude Pixelization tool\footnote{\href{http://healpix.sourceforge.net}{http://healpix.sourceforge.net}} (\textsc{Healpix}) \cite{Gorski2005} to pixelize the sphere with a resolution of nside = 1024.
In order to construct full-sky maps for weak lensing shear, CMB lensing convergence and galaxy overdensity we project the integrand in radial direction. On the other hand, the temperature anisotropies maps from the ISW effect and the CMB lensing potential maps are constructed by interpolating the density field and integrating the quantity along each ray. Importantly, the different maps are constructed from the same underlying density field, such that when computing the spherical harmonic power spectra, one can consider the auto- \textit{and} the cross-correlations between the different probes.\\
\\
\textbf{Runtime.} Our pipeline works by parallelising the processing of the simulation output (i.e. each snapshot-output is processed simulatenously) and takes $\sim 1$ hour walltime ($\sim 100$ CPU-hours) to generate one set of full-sky maps when run on a CPU-based computer cluster.

\subsubsection{Projected Lightcone}

\begin{figure}[htbp!]
\centering
  \begin{minipage}[b]{\textwidth}
    \centering
    \includegraphics[width=0.455\paperwidth]{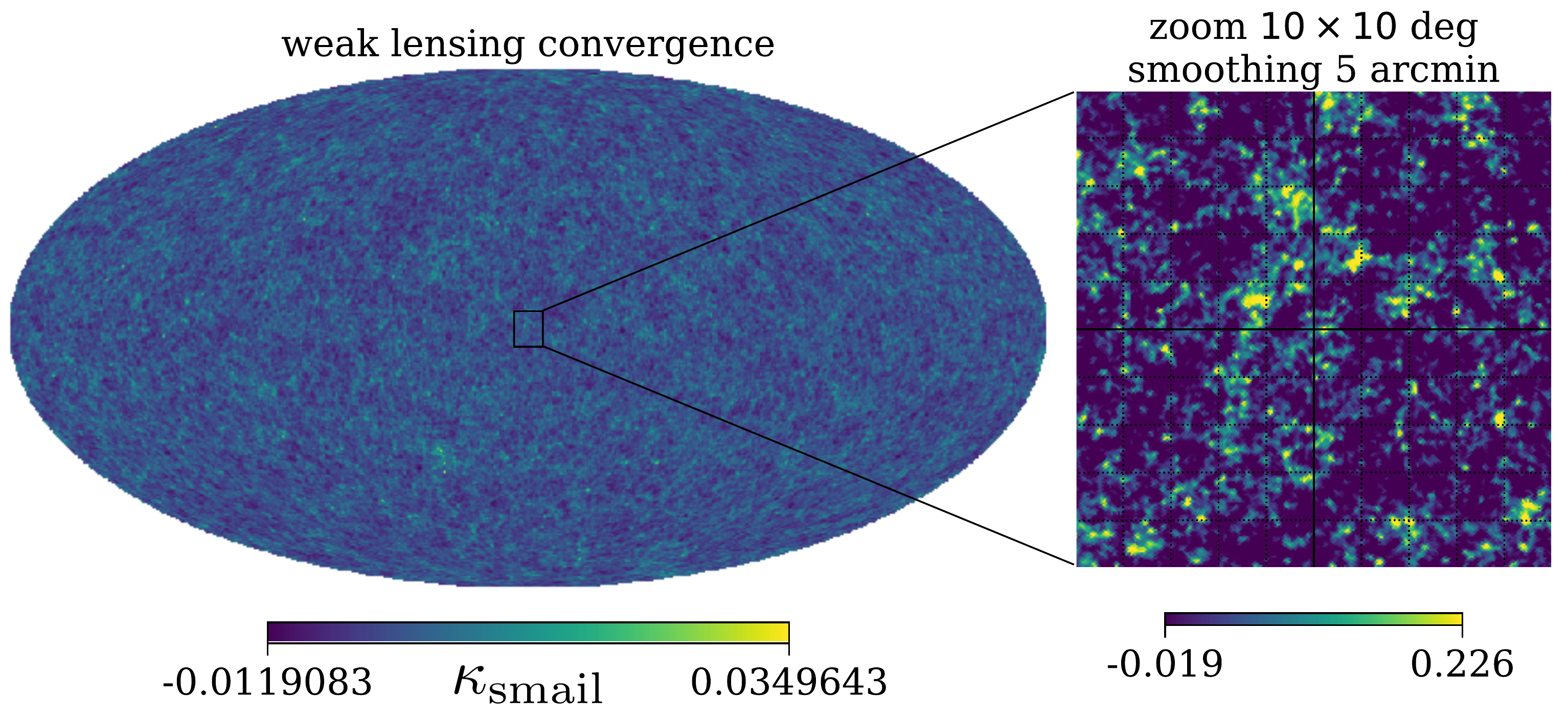} 
  \end{minipage}%%
  \vspace{6 mm}
 
  \begin{minipage}[b]{\textwidth}
    \centering
    \includegraphics[width=0.455\paperwidth]{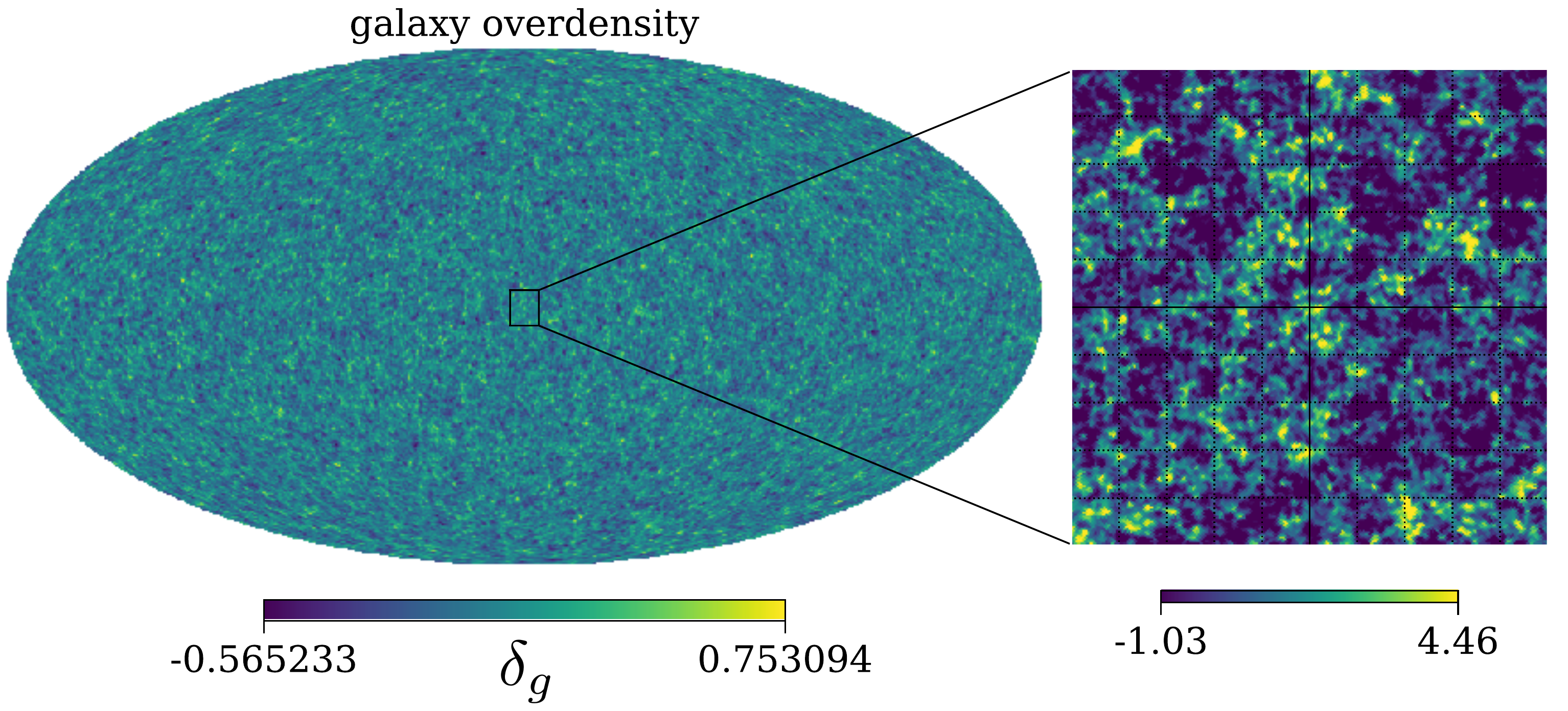} 
  \end{minipage} %%
  \vspace{0.005 mm}
  
  \begin{minipage}[b]{\textwidth}
    \centering
    \includegraphics[width=0.455\paperwidth]{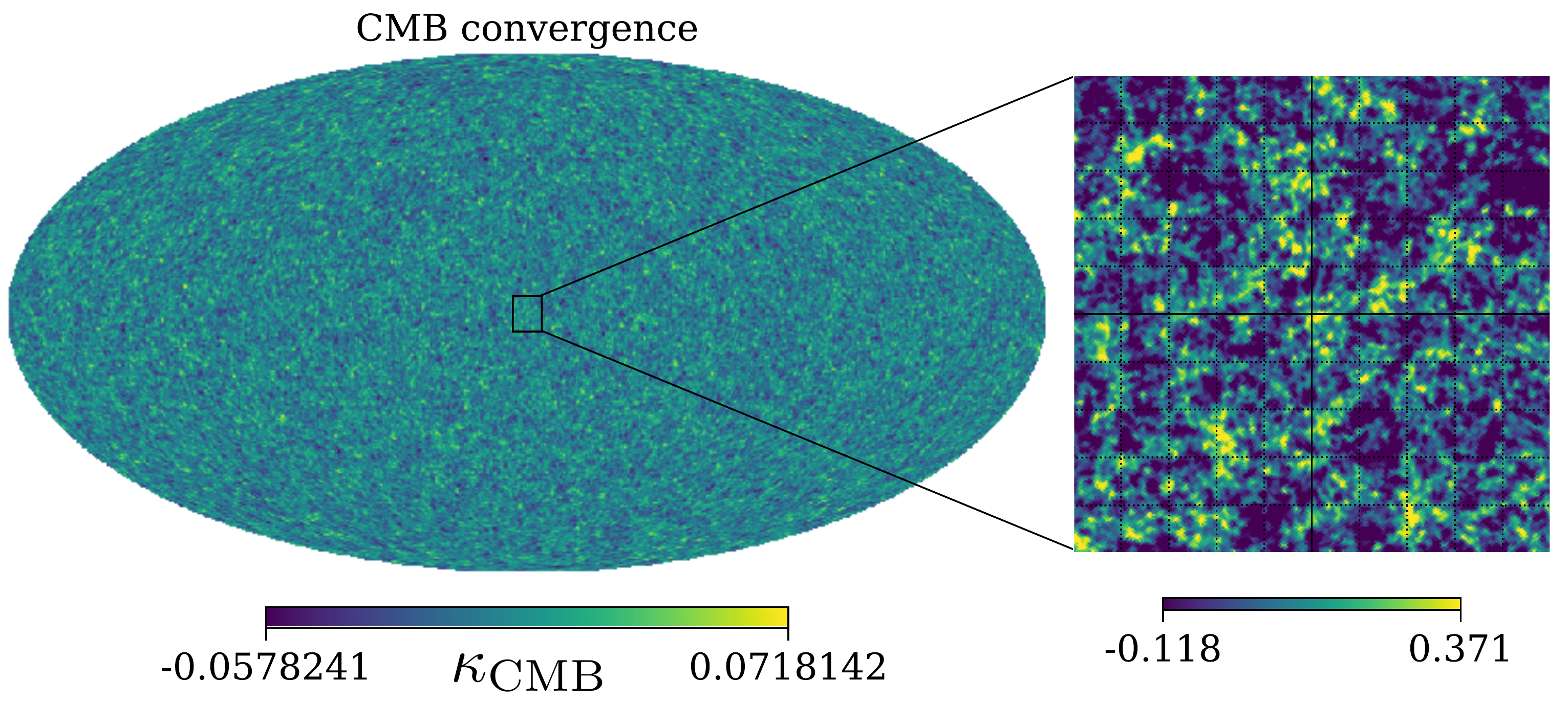} 
  \end{minipage}%% 
  \vspace{4 mm}
  
  \begin{minipage}[b]{\textwidth}
    \centering
    \includegraphics[width=0.455\paperwidth]{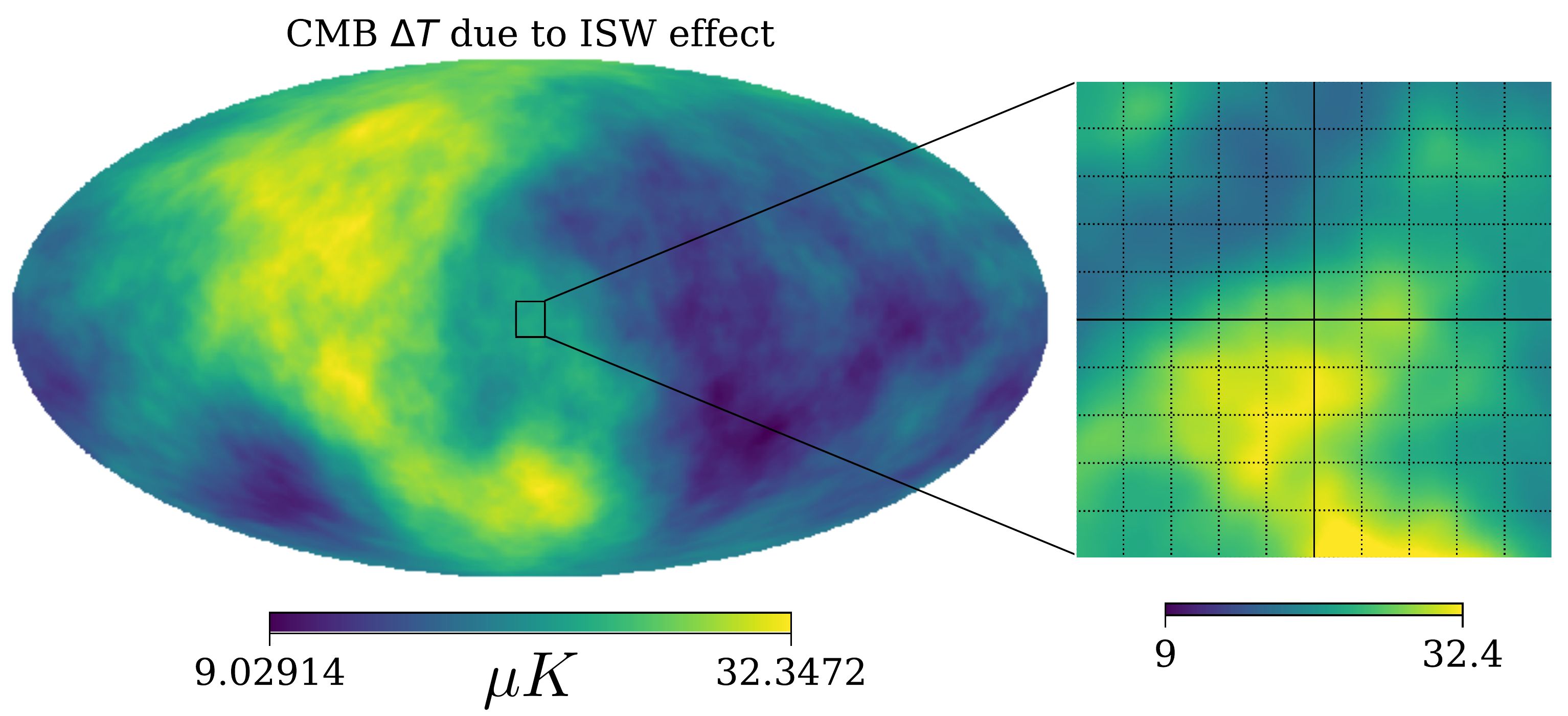} 
  \end{minipage} %% 
\caption{Full-sky maps generated from one $N$-Body realization for $z = 0.0 - 1.75$. From top to bottom: (a)  Weak lensing convergence map based on a Smail \textit{et al.} redshift distribution. (b) Galaxy clustering map including RSD based on the DM density field. (c) CMB convergence map. (d) $\Delta T_\mathrm{ISW}$ map obtained by interpolating the density field and evaluating for each pixel on the sphere. The zoom-in regions span $\sim 100\, \, \mathrm{deg}^2$, have been smoothed using a Gaussian kernel with a FWHM of 5 arcmin and increased in contrast.}
\label{projmaps}
\end{figure}

\textbf{Convergence.} The following treatment is based on the \textit{Born approximation}, which is valid when the change in the comoving separation between the light rays, being deflected through gravitational lensing, is small compared to the comoving separation between undeflected light rays (i.e. the \textit{small-angle scattering limit}). As detailed in \cite{Sgier2019}, we follow the procedure described in the Appendix A of \cite{teyssier2009}. The convergence value of a given pixel $\theta_\mathrm{pix}$ is thus given by
\begin{equation}
\kappa (\theta_\mathrm{pix}) \approx \frac{3}{2} \Omega_m \sum_b W_b \frac{H_0}{c} \int_{\Delta z_b} \frac{c \mathrm{d} z}{H_0 E(z)} \delta \left( \frac{c}{H_0} \mathcal{D}(z) \hat{n}_\mathrm{pix}, z \right) \, ,
\label{kappanum}
\end{equation}
where $\hat{n}_\mathrm{pix}$ is the unit vector pointing to the center of the pixel and the dimensionless comoving distance is given by $\mathcal{D} (z) = (H_0 / c) \chi(z)$. The sum in equation (\ref{kappanum}) runs over all the timesteps with simulation output (i.e. redshift-shell with index $b$) between $z=0.0$ to $1.75$. The integral over the density contrast can be recast as a function of the number of particles $n_p$ in shell $b$ with thickness $\Delta \chi_b$ located in pixel $\theta_\mathrm{pix}$, giving a practical expression for the convergence \cite{Sgier2019}
\begin{equation}
\kappa (\theta_\mathrm{pix}) \approx \frac{3}{2} \Omega_m \sum_b W_b \left( \frac{H_0}{c} \right)^3 \frac{N_\mathrm{pix}}{4 \pi} \frac{V_\mathrm{sim}}{N_\mathrm{p}} \frac{n_p (\theta_\mathrm{pix}, \Delta \chi_b)}{\mathcal{D}^2(z_b)} \, ,
\label{kappaproj}
\end{equation}
where $N_\mathrm{pix}$ is the total number of pixels on the sky. Each shell $\Delta \chi_b$ is multiplied with a weight $W_b$, which depends on the redshift distribution of the source galaxies. For an arbitrary redshift distribution $n(z)$, the weights can be written as
\begin{equation}
W_b^{n(z)} = \left( \int_{\Delta z_b} \frac{\mathrm{d} z}{E(z)} \int_z^{z_s} \mathrm{d} z' n(z') \frac{\mathcal{D}(z) \mathcal{D}(z, z')}{\mathcal{D}(z')} \frac{1}{a(z)} \right) / \left( \int_{\Delta z_b} \frac{\mathrm{d} z}{E(z)} \int_{z_0}^{z_s} \mathrm{d} z' n(z')\right) \, ,
\label{wnz}
\end{equation}
where $\mathcal{D}(z, z') = \mathcal{D}(z') - \mathcal{D}(z)$. We used equation (\ref{wnz}) together with a source galaxy distribution based on Smail \textit{et al.} \cite{Smail1994} given by
\begin{equation}
n(z) = z^\alpha e^{- \left(z / z_s\right)^\beta} \, ,
\label{nz}
\end{equation}
where $\alpha = \beta = 2.0$, $z_s = 0.7$ and normalized it to unity, i.e. $\int \mathrm{d}z\, n(z) = 1$. The full-sky convergence maps we obtained have then been converted to weak lensing shear maps using \cite{Wallis2017},
\begin{equation}
 _{_2} \gamma_{\ell m} = \frac{-1}{\ell (\ell + 1)} \sqrt{\frac{(\ell + 2)!}{(\ell - 2)!}} \kappa_{\ell m} \, ,
 \label{kappa2gamma}
\end{equation}
where $ _{_2} \gamma_{\ell m}$ are the spin-2 spherical harmonic coefficients of the weak lensing shear. In order to compute the CMB lensing convergence field from equation (\ref{kappaproj}), we use the weights
\begin{equation}
W_b^{z_s} = \left( \int_{\Delta z_b} \frac{\mathrm{d} z}{ E(z)} \frac{\mathcal{D}(z) \mathcal{D}(z, z_s)}{\mathcal{D}(z_s)} \frac{1}{a(z)} \right) / \left( \int_{\Delta z_b} \frac{\mathrm{d} z}{ E(z)} \right) \, ,
\end{equation}
with a single source located at $z_s = z_\ast$.\\
\\
\textbf{Galaxy clustering.} Analogously to the expression for the convergence given by equation (\ref{kappaproj}) and the approximations described in section \ref{theopred} , we can approximate the galaxy clustering by
\begin{equation}
\delta_g ( \theta_\mathrm{pix}) \approx \sum_b W_b^{\delta_g} \left( \frac{H_0}{c} \right)^2 \frac{N_\mathrm{pix}}{4 \pi} \frac{V_\mathrm{sim}}{N_p} \frac{n_p (\theta_\mathrm{pix}, \Delta \chi_b)}{\mathcal{D}^2(z_b)} \, ,
\end{equation}
where the weights are given by
\begin{equation}
W_b^{\delta_g} = \left( \int_{\Delta z_b} \frac{\mathrm{d}z}{E(z)} H(z) b(z) n(z) \right) / \left( \int_{\Delta z_b} \frac{\mathrm{d}z}{E(z)} \right) \, .
\end{equation}
Full-sky maps for galaxy clustering are constructed by projecting the particles within each redshift-shell (no. of particles in shell $b$ given by $n_p (\Delta \chi_b)$) onto the sky, before weighting and summing up the shells over the full lightcone. Taking into account the peculiar velocities of the individual particles $v_{||}$, we can alter the positions of the particles in radial direction due to RSD (no. of particles in shell $b$ due to RSD given by $n^\mathrm{RSD}_p (\Delta \chi_b)$). Therefore, in our lightcone construction pipeline, the magnitude of the RSD effect depends on the thickness of the redshift shells. The thinner the shell, the more likely it is that a particle changes the shell and is associated to a different redshift. Note that shells closer to the observer are thinner and provide therefore a better resolution for the lightcone construction (see section \ref{sim_res}). In Figure \ref{projmaps} we show one realisation of the full-sky maps $\kappa_\mathrm{smail}$, $\kappa_\mathrm{CMB}$, $\delta_g$ and $\Delta T_\mathrm{ISW}$, where $\kappa_\mathrm{smail}$ represents the convergence field based on a source galaxy distribution given by equation (\ref{nz}).
\clearpage
\subsubsection{Interpolated Lightcone}
\begin{figure}[htbp!]
\centering
\includegraphics[width=1\textwidth]{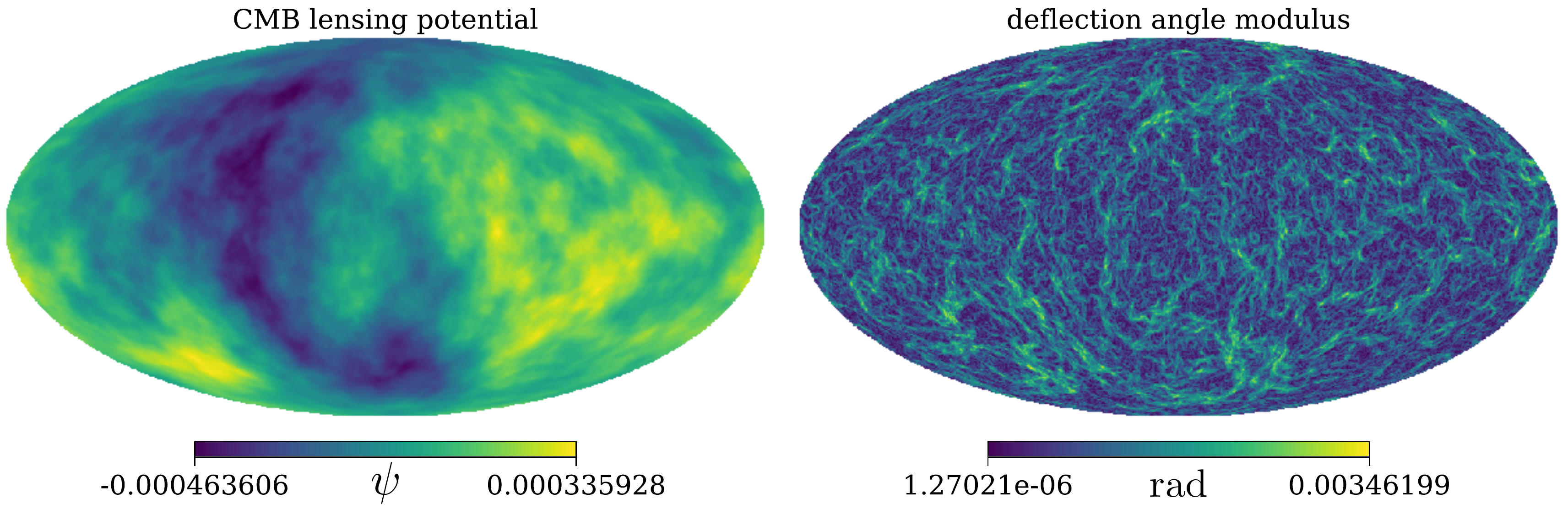}
\caption{Left panel: Full-sky CMB lensing potential map obtained by interpolating the density field and evaluating for each pixel on the sphere for $z = 0.0 - 1.75$. Right panel: Corresponding map of the deflection angle modulus $|\vec{\alpha}| = \sqrt{(\Delta \theta)^2 + (\Delta \phi)^2}$, given by the angular derivative of the lensing potential map.}
\label{intmaps}
\end{figure}

In this section, we present our procedure based on Cai \textit{et al.} \cite{Cai2010} and Smith \textit{et al.} \cite{Smith2009} to construct full-sky maps of the temperature fluctuations from the ISW effect and the CMB lensing potential from our simulations which are \textit{continuous} on the sky. We first construct an estimate of $\Phi$ and $\dot{\Phi}$ on a cubic grid, interpolate the quantity and integrate $\psi_\mathrm{CMB}$ and $\Delta T_\mathrm{ISW}$ respectively along each ray of the past-lightcone.\\
\\
\textbf{ISW effect.} The integral in equation (\ref{iswdef}) can be rewritten as a function of comoving distance as
 \begin{equation}
\Delta T_\mathrm{ISW}(\hat{n}) = \frac{2}{c^3} \int_0^{\chi_\ast} \dot{\Phi}(\chi \hat{n}; \eta_0 - \chi )\, a\, \mathrm{d}\chi \, .
\label{iswcom}
\end{equation}
We use Poisson's equation in comoving coordinates $\nabla^2 \Phi (\vec{x}, t) = 4 \pi G \bar{\rho}(t) a^2 \delta (\vec{x}, t)$ to write the gravitational potential in Fourier space given by
\begin{equation}
\Phi (\vec{k}, t) = - \frac{3}{2} \left( \frac{H_0}{k} \right)^2 \Omega_m \frac{\delta (\vec{k}, t)}{a} \, .
\end{equation}
The time derivative of $\Phi (\vec{k}, t)$ together with the Fourier space form of the continuity equation $\dot{\delta}(\vec{k}, t) + i\, \vec{k} \cdot \vec{p}(\vec{k}, t) = 0$ can be written as
\begin{equation}
\dot{\Phi} (\vec{k}, t) = \frac{3}{2} \left(\frac{H_0}{k}\right)^2 \Omega_m \left[\frac{H(a)}{a} \delta(\vec{k}, t) + \frac{i\, \vec{k} \cdot \vec{p}(\vec{k}, t)}{a} \right] \, ,
\label{ISWRS}
\end{equation}
where $\vec{p}(\vec{k}, t) = [1 + \delta(\vec{k}, t)]\,\vec{v}(\vec{k}, t)$ is the momentum density. Equation (\ref{ISWRS}) contains the contributions from the linear ISW effect as well as the non-linear RS effect and relates the evolution of the gravitational potential to the time-evolution of the matter fluctuations. In order to isolate the linear ISW effect, one can work in the linear regime and use $\dot{\delta}(\vec{k}, t) = \dot{D}(t) \delta(\vec{k}, z = 0)$, where $D(t)$ is the linear growth factor. In this case, equation (\ref{ISWRS}) can be rewritten as
\begin{equation}
\dot{\Phi} (\vec{k}, t) = \frac{3}{2} \left(\frac{H_0}{k}\right)^2 \Omega_m \frac{\dot{a}}{a^2} \delta(\vec{k}, t) [1 - \beta(t)] \, ,
\label{psinum}
\end{equation}
where the linear growth rate is given by $\beta(t) \equiv d\, \mathrm{ln}\, D(t) / d\, \mathrm{ln}\, a$. Here the overdensity field directly determines the potential field and its time derivative.
\smallbreak
The quantity $\dot{\Phi} (\vec{k}, t)$ given by equation (\ref{psinum}) is then constructed as follows: First, we use a cloud-in-cell mass assignment scheme (CIC) \cite{Hockney1981} to obtain the density field $\delta (\vec{x})$ on a 3D cubic grid with $1024^3$ cells. The density field is then Fast Fourier transformed to compute $\dot{\Phi}$ in Fourier space. Second, we perform an inverse Fourier transform to obtain $\dot{\Phi} (\vec{x})$ and interpolate linearly. Lastly, we approximate the integral given by equation (\ref{iswcom}) as a sum over discrete set of steps in comoving radial distance
\begin{equation}
\Delta T_\mathrm{ISW} (\theta_\mathrm{pix}) = \frac{2}{c^2} \sum_{b,\delta z} \dot{\Phi}(\chi \hat{n}_\mathrm{pix}, z)\, a\, \Delta \chi \, ,
\label{iswsum}
\end{equation}
where the steps are given by $\Delta \chi = \chi (z_b + \delta z) -  \chi (z_b)$ and we use a finer redshift-spacing $\delta z = 0.01$ within each shell $\Delta z_b$ related to the timesteps of the simulation. We then evaluate $\Delta T_\mathrm{ISW}$ for all the rays pointing to the center of the $12 \times \mathrm{nside}^2$ pixels of our map. In the bottom panel of Figure \ref{projmaps} we show a full-sky map of $\Delta T_\mathrm{ISW}$.\\
\\
\textbf{CMB lensing potential.} Full-sky maps of the lensing potential of the CMB can be computed using equation (\ref{psi}). Therefore, we can directly use the gravitational potential field $\Phi$ output from the \textsc{PkdGrav3} simulation or compute it from the overdensity field $\delta$ by solving Poisson's equation. The following discussion is based on using $\Phi$ directly from the simulation. Analogously to the construction of the $\Delta T_\mathrm{ISW}$ maps described above, we use a CIC mass assignment scheme to get the gravitational potential on a 3D cubic grid with $1024^3$ cells. The $\Phi$-field is then interpolated linearly and the expression given by equation (\ref{psi}) is approximated by a sum over discrete steps corresponding to a redshift-spacing of $\delta z = 0.01$ (analogous to equation (\ref{iswsum})). In order to integrate over all the redshifts relevant for CMB lensing, one needs to cover a large enough volume. In the present work, we replicate the simulation volume 6 times along each axis, reaching a redshift of $z \sim 1.75$. According to Carbone \textit{et al.} \cite{Carbone2008}, a contribution to the lensing power from redshifts higher than $z \sim 11.22$ is negligible for the lensing of CMB photons. For our setup this would suggest to double the number of replications along each axis, which is computationally more expensive. Figure \ref{intmaps} shows our results for full-sky maps of the CMB lensing potential and the corresponding deflection angle modulus $|\vec{\alpha}| = \sqrt{(\Delta \theta)^2 + (\Delta \phi)^2}$ (see Appendix \ref{cmb_lensing_appendix} for a discussion of CMB lensing).
\subsection{Systematic Effects}
\subsubsection{Simulation Resolution}
\label{sim_res}
\textsc{PkdGrav3} is a mesh-less tree-code, such that one can not directly infer the resolution of the simulation by considering the mean particle separation. In this case the minimally resolved scale depends on the number of particles and can be inferred by looking at the spherical harmonic power spectrum after the subtraction of the shot-noise contribution (see section \ref{shotnoise} for a discussion of shot noise). For our simulation setup we are able to resolve angular scales within 2\% compared to the analytical prediction up to $\ell \sim 1500$. Increasing the number of particles by keeping the same simulation volume would enable the resolution of even smaller scales. Eventually, this would require using a higher nside-value in order to decrease the pixel size and therefore obtain a better map resolution. A comparison of our results for the spherical harmonic power spectra using our simulation setup with analytical predictions are presented in section \ref{PS}.
\smallbreak
These considerations are altered when constructing maps of the temperature fluctuations from the ISW effect. In this case we first assign the overdensity to a cubic grid with $1024^3$ grid cells before integrating $\Delta T_\mathrm{ISW}$ given by equation (\ref{iswcom}). For our simulation volume with a side length of 1.6 Gpc, we are able to resolve structures down to $\sim$ 1.6 Mpc. The resolution of our grid and the redshift-steps for the integration can have an impact on the resolution of our $\Delta T_\mathrm{ISW}$ maps. Our \textsc{Healpix} maps with nside = 1024 have a pixel size of 3.43$'$ and correspond to the linear size of our grid cell at a radial comoving distance of 1120 Mpc, i.e. at a redshift of $z \sim 0.28$. We therefore expect not to compromise the resolution of the maps beyond this redshift with our choice of number of grid cells. 

We have further investigated the impact of the number of timesteps used in the simulation, which determines the thickness of the redshift shells $\Delta z_b$ in the lightcone construction. The $N$-Body simulation code \textsc{PkdGrav3} relies on adaptive time stepping, which ensures a certain time-resolution of the simulation by automatically increasing the number of substeps when choosing a lower number of timesteps (see subsection 3.2 in Potter \textit{et al.} 2017 \cite{Potter2017}). Therefore, as long as we use a minimum of about $\sim 100$ timesteps, we obtain a accurate simulation output. The number of timesteps is therefore mostly used for the number of snapshot output files, which can be postprocessed using \textsc{UFalcon}.

Concerning the lightcone construction using \textsc{UFalcon}, we observed only marginal improvements in our power spectrum results when choosing redshift shells finer than $0.05$. Increasing the number of timesteps from 150 ($\Delta z_b \sim 0.03$) to 200 ($\Delta z_b \sim 0.02$) between $z_\mathrm{init}$ and $z_\mathrm{final}$ only marginally changed our power spectrum results by $\lesssim 1\%$. Choosing an even higher number of timesteps (e.g. 500 or 1500), we observe nearly zero improvements ($\ll 1\%$) in the agreement between our simulation results and the analytical predictions.

\subsubsection{Super-Sample Covariance Effects}
\label{SSC}
The finite volume of our simulation box can introduce systematic errors on the power spectrum, caused by missing modes larger than the simulation volume and their coupling to small-scale modes. This effect introduces an additional super-sample covariance term $\Sigma_{SSC}$ to the covariance matrix \cite{Li2014}. According to Schneider \textit{et al.} \cite{Schneider2016}, the error on the matter power spectrum when using a box with a length of $L = 512\, h^{-1} \mathrm{Mpc}$ compared to $L = 1024\, h^{-1} \mathrm{Mpc}$ is within 1\% for the wavenumbers between $k = 10^{-2}\, h \, \mathrm{Mpc}^{-1}$ and $10^{1}\, h \, \mathrm{Mpc}^{-1}$. Considering our choice of box-length given by $L = 1120\, h^{-1} \mathrm{Mpc}$, we expect the errors to lie well within 1\% for the same scales. We examined the finite volume effects on the spherical harmonic power spectrum of our full-sky maps constructed by replicating the simulation box (described in section \ref{nummet}) and observed a significant drop in power for very large scales $\ell \lesssim 10$ for all the probes considered. Note that such a lack of power on large scales could also stem from the underlying \textsc{PkdGrav3} code, which is a mesh-less tree-code and interpolates very large scales. Since the present forecast analysis is focused on scales between $\ell = 10^2$ and $10^3$, we leave investigations of this effect to future work.

\subsubsection{Shot Noise Estimation}
\label{shotnoise}
Particle simulations in general contain a certain amount of shot noise due to the finite number of particles involved, which can be statistically described by a Poisson distribution. Concerning the matter power spectrum, one can write the shot noise contribution as $P_\mathrm{sn} = V_\mathrm{sim} / N_p$ \cite{Schneider2016}. In this section we investigate the shot noise contribution to the spherical harmonic power spectrum. The most direct approach is to use the shot noise contribution to the matter power spectrum $P_\mathrm{sn}$ and perform the weighted integrals given by equations (\ref{cl}), (\ref{iswcross}) and (\ref{iswauto}), all based on the Limber approximation \cite{Calabrese2015}. The expression for the shot noise on the weak lensing convergence power spectrum is then given by
\begin{equation}
C_{\ell}^{\mathrm{sn},\, \mathrm{Limber}} = \int \mathrm{d}z \, \frac{c}{H(z)} \, \frac{W^{\kappa} (\chi(z)) \, W^{\kappa} (\chi(z))}{\chi^{2} (z)} \,\, \frac{V_\mathrm{sim}}{N_p} \, ,
\label{snlimber}
\end{equation}
where the integration is done using \textsc{PyCosmo}. In order to relate the shot noise contribution to our numerical results more accurately, we estimate the shot noise based on the Born approximation, i.e. in a way analogous to the lightcone construction used in \textsc{UFalcon}. The shot noise for the weak lensing convergence can then be written as a weighted sum over the number of particles in redshift-shell $b$ as
\begin{equation}
C_{\ell}^{\mathrm{sn}, \, \mathrm{Born}} = \left[\frac{3}{2} \Omega_m \left( \frac{H_0}{c} \right)^3 \frac{1}{4 \pi} \right]^2 \sum_b \left[ W_b \frac{V_b}{\mathcal{D}^2(z_b)} \right]^2 \frac{4 \pi}{N_b} \, ,
\label{snborn}
\end{equation}
where $N_b$ is the expected number of particles in the shell $b$ and can be expressed as
\begin{equation}
N_b = \frac{N_p^\mathrm{sim}}{V_\mathrm{sim}} V_b \, .
\end{equation}
A derivation of equation (\ref{snborn}) is given in Appendix \ref{snappendix}, where we compare the shot noise contribution to the weak lensing convergence power spectrum based on the Limber and Born approximations. The shot noise estimates based on both approximations become equal in the limit of infinitesimal shell-thickness. We observe that the shot noise contributions using the Limber approximation and using the \textsc{PkdGrav3} redshift-spacing have the same order of magnitude and both are 2-3 orders of magnitude smaller in amplitude than our \textsc{UFalcon} results (see Figure \ref{shotnoise_label} in Appendix \ref{snappendix}). So the choice of method does not have an impact on our final results. We therefore use the Limber approximation for the estimation of the shot noise contribution.
\subsubsection{Shell Randomization Procedure}
\label{randomisation}
\begin{figure}[htbp!]
  \begin{minipage}[b]{0.37\paperwidth}
    \centering
    \includegraphics[width=.38\paperwidth]{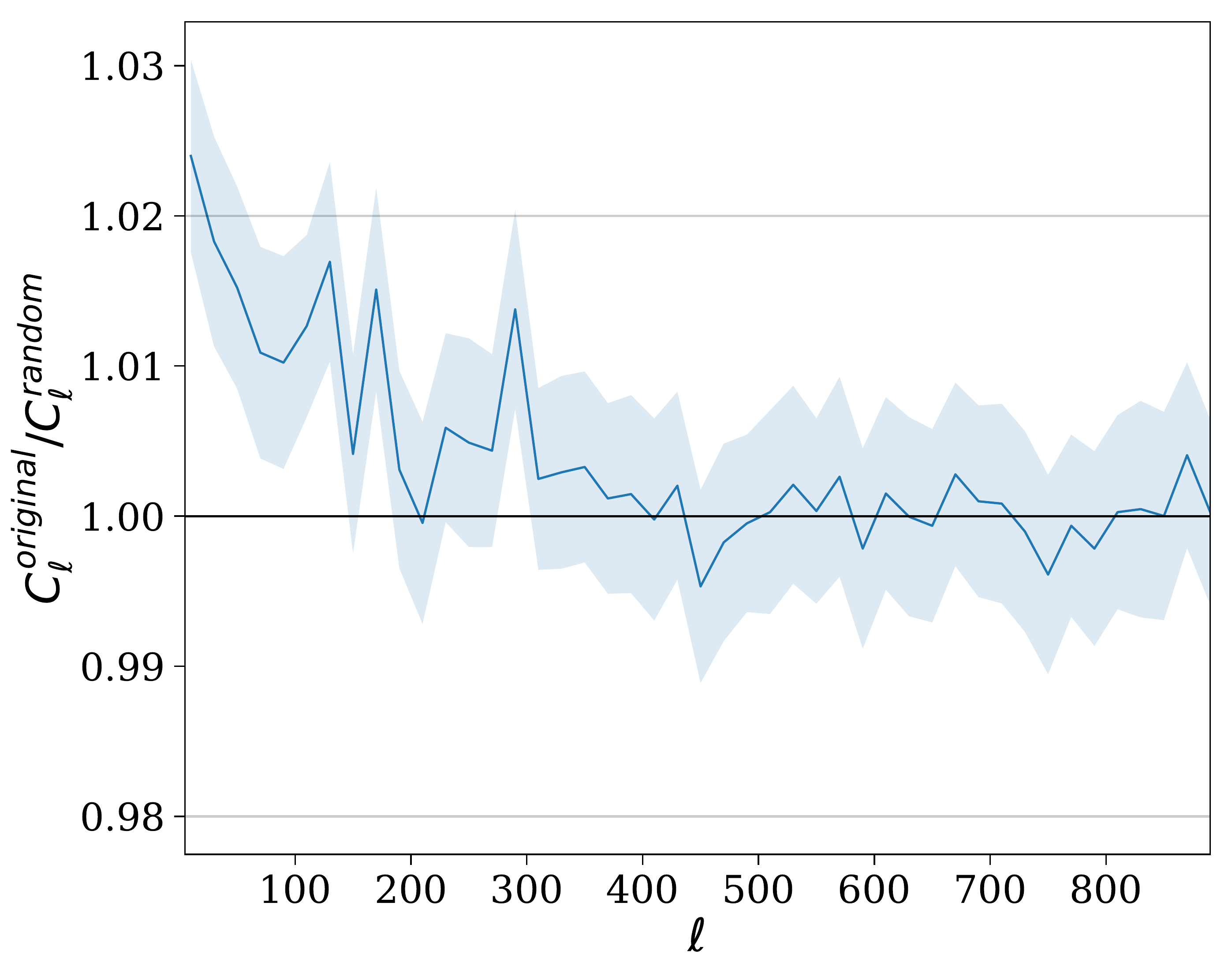} 
  \end{minipage}%%
  \begin{minipage}[b]{0.37\paperwidth}
    \centering
    \includegraphics[width=0.35\paperwidth]{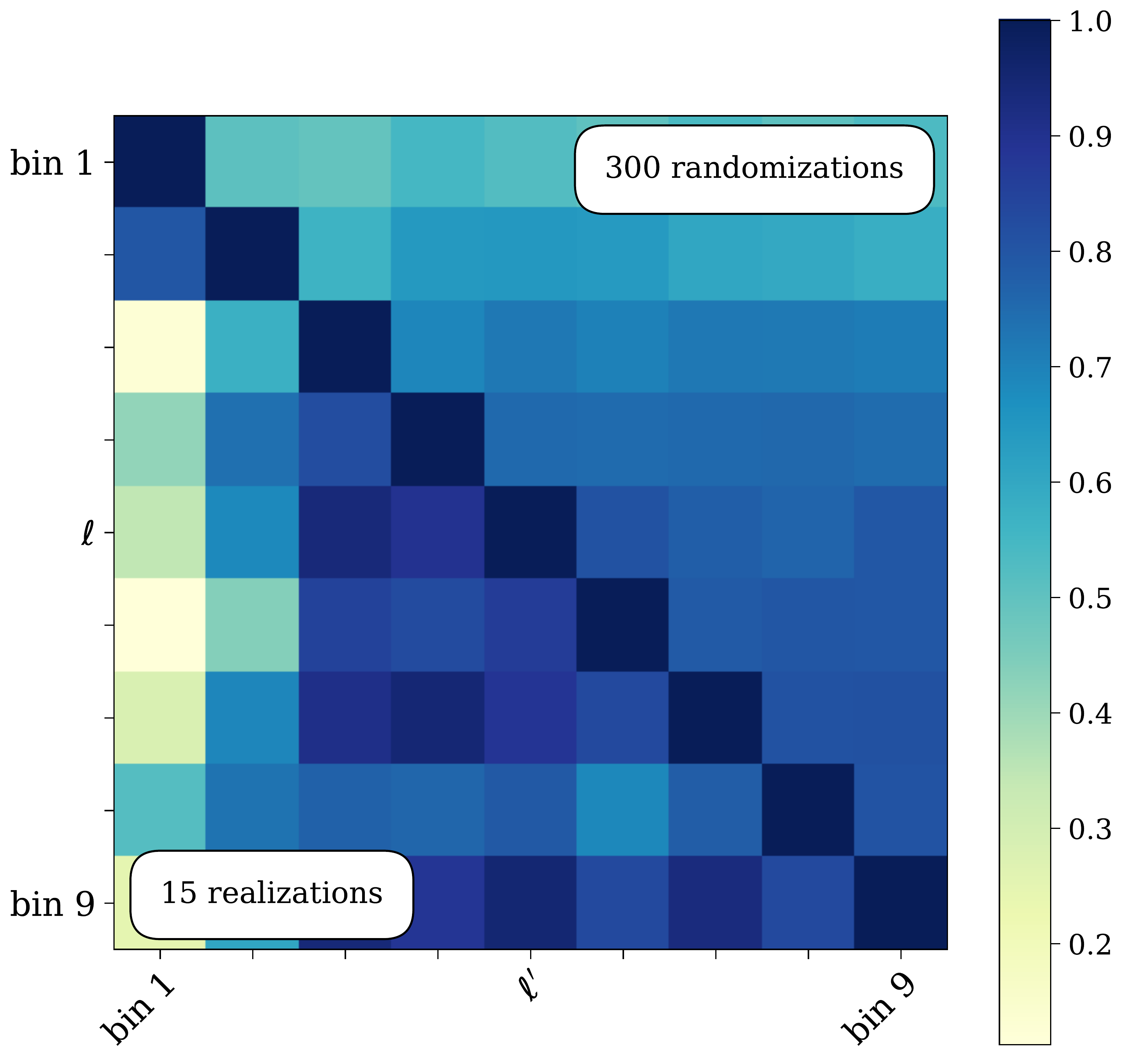} 
  \end{minipage} 
  \caption{Left panel: Ratio between the weak lensing shear power spectrum of 15 individual simulation runs and 300 new randomized realizations between $\ell = 10^2$ and $10^3$. The choice of our shell-bundle thickness leads to a loss of power in the new realizations up to 2\% at $\ell \sim 100$. Right panel: Lower triangle: Unconverged correlation matrix of weak lensing power spectrum binned between $\ell = 10^2$ and $10^3$ with $\delta  \ell = 100$ from 15 simulation runs only. Upper triangle: More converged correlation matrix by using 300 randomized realizations.}
\label{Figrandom}
\end{figure}

As sketched in Figure \ref{sketch}, we adopt a randomization scheme to increase the number of realizations from one underlying simulation run. Similarly to Carbone \textit{et al.} \cite{Carbone2008}, we apply random operations to the particle positions or to the $\delta$ and $\Phi$ fields on the 3D grids consisting of rotations by $90^\circ$ (interchanging the axes), translations and parity flips \cite{Springel2001}. Such a randomization of the simulation boxes avoids the repetition of the same structures present in the density field along the line of sight. We bundle neighbouring shells corresponding to adjacent output-redshifts together to have a thickness of about $\mathrm{d}z \sim 0.1$. In this way we avoid breaking correlations within the density field on scales smaller than the comoving distance $\chi (z + \mathrm{d}z) - \chi (z)$. Note that different shell-bundles are randomized using different random operations and are therefore uncorrelated. With our choice of shell-bundle we observe a loss of power of about 2\% at $\ell \sim 100$ when considering the spherical harmonic power spectrum of weak lensing shear after construction of the full lightcone (left hand side of Figure \ref{Figrandom}). Importantly, all replicated boxes within one shell-bundle are randomized in a coherent way leading to a 3D tessellation (see Figure \ref{sketch}). This procedure conserves the continuity of the density field across the boundaries of the replicated boxes, which is particularly important to obtain continuous maps of the CMB temperature fluctuations and lensing potential across the sky.
\smallbreak
In the present work, we randomize each individual simulation run 20 times in order to increase the number of realizations for the covariance matrix estimation. On the right hand side of Figure \ref{Figrandom}, we show two different triangles of the correlation matrices for the weak lensing shear power spectrum binned between $\ell = 10^2$ and $10^3$ with a bin-width of $\delta  \ell = 100$: The lower triangle shows the unconverged correlation matrix for only 15 simulations runs. On the upper triangle, we show a more converged correlation matrix computed using the 20 new randomized realizations from the 15 individual runs.
\subsubsection{Further Effects}

\textbf{Born approximation.} The lightcone construction in our pipeline is based on the Born approximation, i.e. only the lowest-order expansion of the gravitational potential is used to compute the convergence. In this way the convergence field is integrated and weighted along the unperturbed light-rays. This stands in contrast to the fully ray-traced computation of the convergence, where the exact solution to Poisson's equation is needed and higher order terms in $\Phi$ are included. Depending on the considered statistics and angular scales, using the Born approximation can have a non-negligible impact on cosmological parameter constraints \cite{Petri2017}.

Depending on the shape of the lensing kernel, the galaxy and CMB lensing convergence power spectra can be affected differently by the post-Born corrections. For CMB lensing the contributions could be relatively much more important than for galaxy lensing, since the CMB lensing kernel is broadly peaked at high redshift and the lensing potentials are nearly linear and Gaussian \cite{Pratten2016},\cite{Barthelemy2020}. Results reported in Hagstotz \textit{et al.} 2015 \cite{Hagstotz2015} show that there are large effects on the CMB lensing power spectrum on scales well below $\ell \approx 1000$ (with contributions comparable to the power generated by nonlinear structure formation on small scales). The majority of research in fact report that the effect on the CMB lensing convergence is indeed small, at least for current and near-future observations (see e.g. \cite{Pratten2016}, \cite{Barthelemy2020}, \cite{Cooray2002b}, \cite{Krause2010}). More specifically, second- and higher-order corrections in the gravitational potential to the galaxy and CMB convergence are expected to give small corrections to the spherical harmonic power spectrum, which are at least two orders of magnitude lower than the leading order expansion on scales $\ell = 10^2$ to $10^4$ (see e.g. \cite{Petri2017},\cite{Cooray2002b},\cite{Hirata2003},\cite{Shapiro2006},\cite{Hilbert2009}). For example, Pratten \& Lewis 2016 \cite{Pratten2016} show that post-Born corrections make $\lesssim 0.2 \%$ contribution to the CMB convergence power spectrum up to $\ell \lesssim 3000$, being well below cosmic variance.

However, recent work showed that post-Born corrections might become relevant when considering cross-correlations, for example between galaxy counts and CMB lensing \cite{Boehm2020},\cite{Fabbian2019}.

We conclude that invoking a fully ray-traced lightcone involving higher-order corrections becomes relevant when considering higher-order statistics of the convergence, smaller angular scales or considering cross-correlations between CMB lensing and other probes. Concerning the auto-power spectra, we expect the Born-induced errors on the power spectra to be well below 1\% for the scales $\ell = 10^2$ to $10^4$ \cite{Hilbert2019}.\\
\\
\textbf{Baryonic effects.} Feedback processes from baryons are expected to have a significant impact on the weak lensing power spectrum \cite{Weiss2019},\cite{Schneider2019},\cite{Osato2015}. Depending on the baryonic model considered, the effects on the weak lensing power spectrum can be up to 10\% at $\ell \sim 10^3$. The present work is based on DM-only simulations, although the incorporation of baryonic effects, e.g. by using the \textit{baryonic correction model} introduced in Schneider \textit{et al.} \cite{Schneider2019} to mimic baryonic effects on the DM-only density field, remains an important extension of our pipeline and left for future work.

\subsection{Codebase}
\label{codebase}

We publish the \textsc{UFalcon}\footnote{\textsc{UFalcon}: \href{https://cosmology.ethz.ch/research/software-lab/UFalcon.html}{https://cosmology.ethz.ch/research/software-lab/UFalcon.html}} code, which contains the weak gravitational lensing part of the pipeline. The package is written in \texttt{Python 3} and is publicly available on the Python Package index \textsc{PyPi}. The package documentation and some example-functions showing the user how to implement \textsc{UFalcon} are given in the repository.
\smallbreak
The features of the released code include the fast computation of full-sky maps containing particle counts (particle-shells) from lightcone output and the subsequent fast construction of convergence maps for user-specific source galaxy redshift distributions and single-source redshifts. The released version of \textsc{UFalcon} currently supports the post-processing of $N$-Body simulation output in lightcone mode generated using the codes \textsc{PkdGrav3}\footnote{\href{https://bitbucket.org/dpotter/pkdgrav3/}{https://bitbucket.org/dpotter/pkdgrav3/}} (Stadel \textit{et al.} \cite{Stadel2001}) and \textsc{L-PICOLA}\footnote{\href{https://cullanhowlett.github.io/l-picola/}{https://cullanhowlett.github.io/l-picola/}} (Howlett \textit{et al.} \cite{Howlett2015}).

%%%%%%%%%%%%%%%%%%%%%%%%%%%%%%%%%%%%%%%%%%%%%%%%%%%%%%%%%%%

\section{Statistical Analysis}
\label{statanal}
In this section, we perform several quantitative analyses on the full-sky maps of the different cosmological probes we obtained by applying our pipeline \textsc{UFalcon} on \textsc{PkdGrav3} simulation output, as described in section \ref{nummet}. All the maps are constructed for a redshift range between $z=0.0$ and $1.75$ and every set of $\gamma_1$, $\delta_g$, $\kappa_\mathrm{CMB}$ and $\Delta T_\mathrm{ISW}$ maps are based on the same simulation output. We test the consistency of our simulation results by comparing to analytical predictions.%{\color{red} (Effects of RSD not visible for our shell thickness due to same range as SSC)}

\subsection{1-Point Distribution}
\begin{figure}[htbp!]
  \begin{minipage}[b]{0.37\paperwidth}
    \centering
    \includegraphics[width=.37\paperwidth]{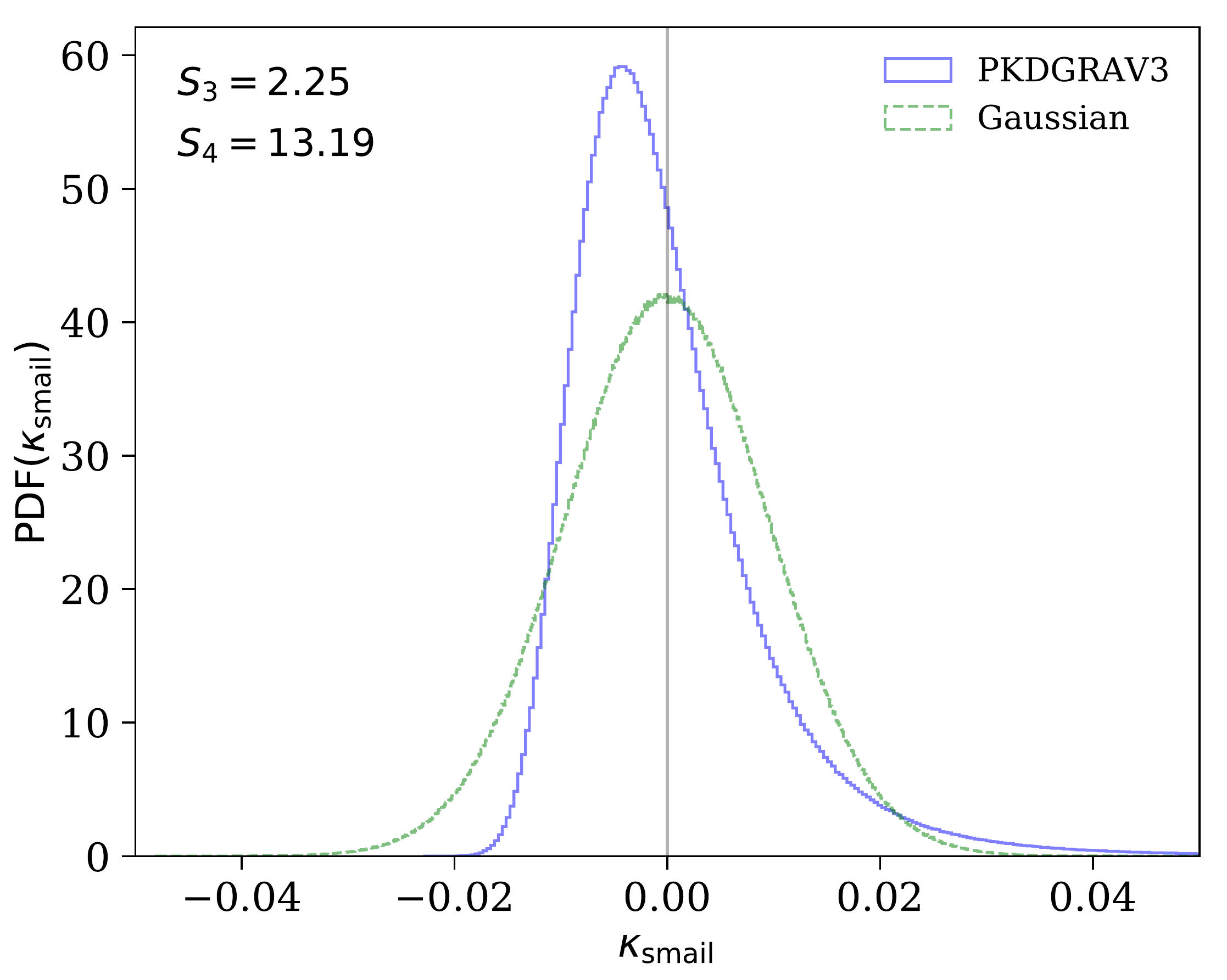} 
  \end{minipage}%%
  \begin{minipage}[b]{0.37\paperwidth}
    \centering
    \includegraphics[width=.38\paperwidth]{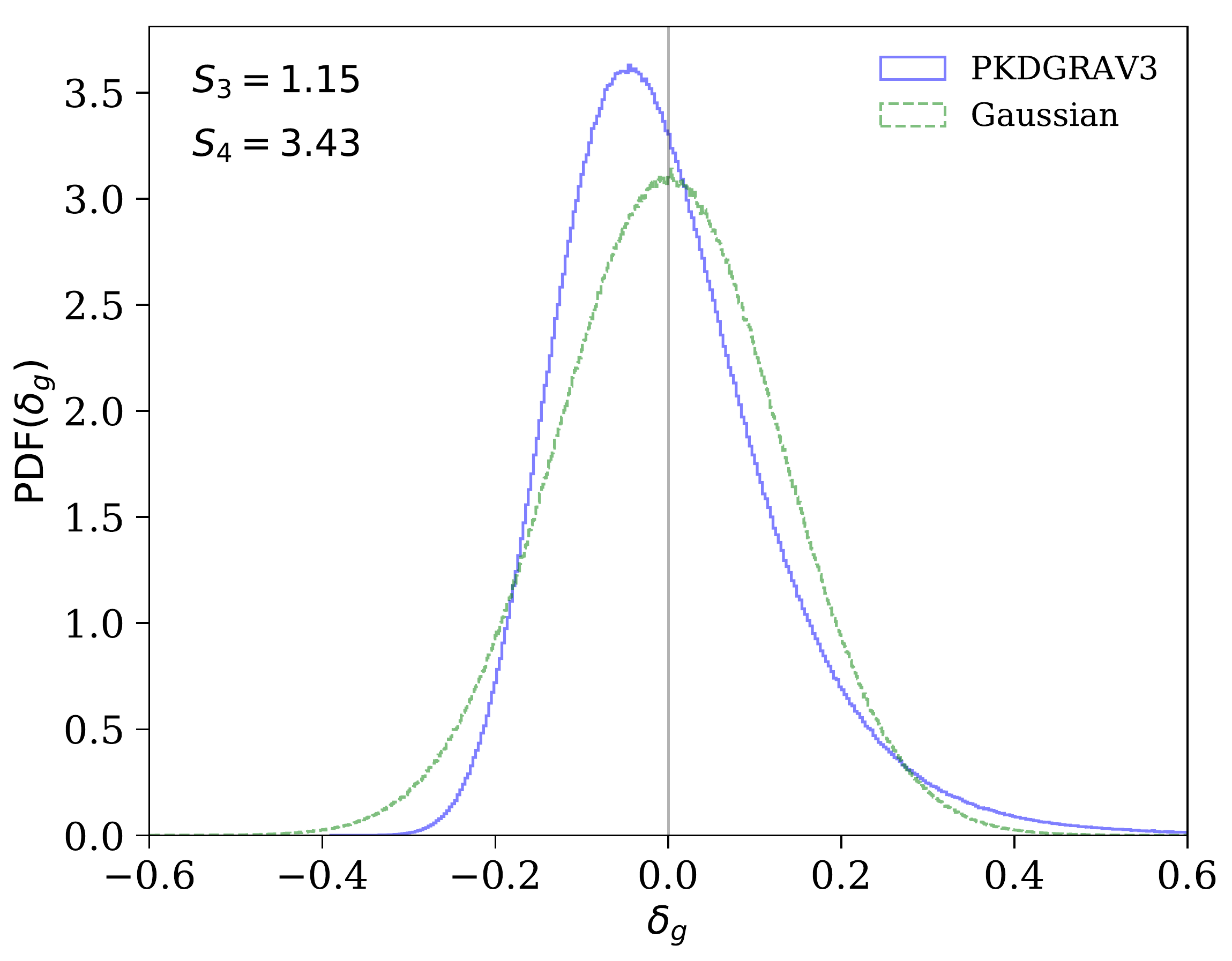} 
  \end{minipage} 
  \begin{minipage}[b]{0.37\paperwidth}
    \centering
    \includegraphics[width=.37\paperwidth]{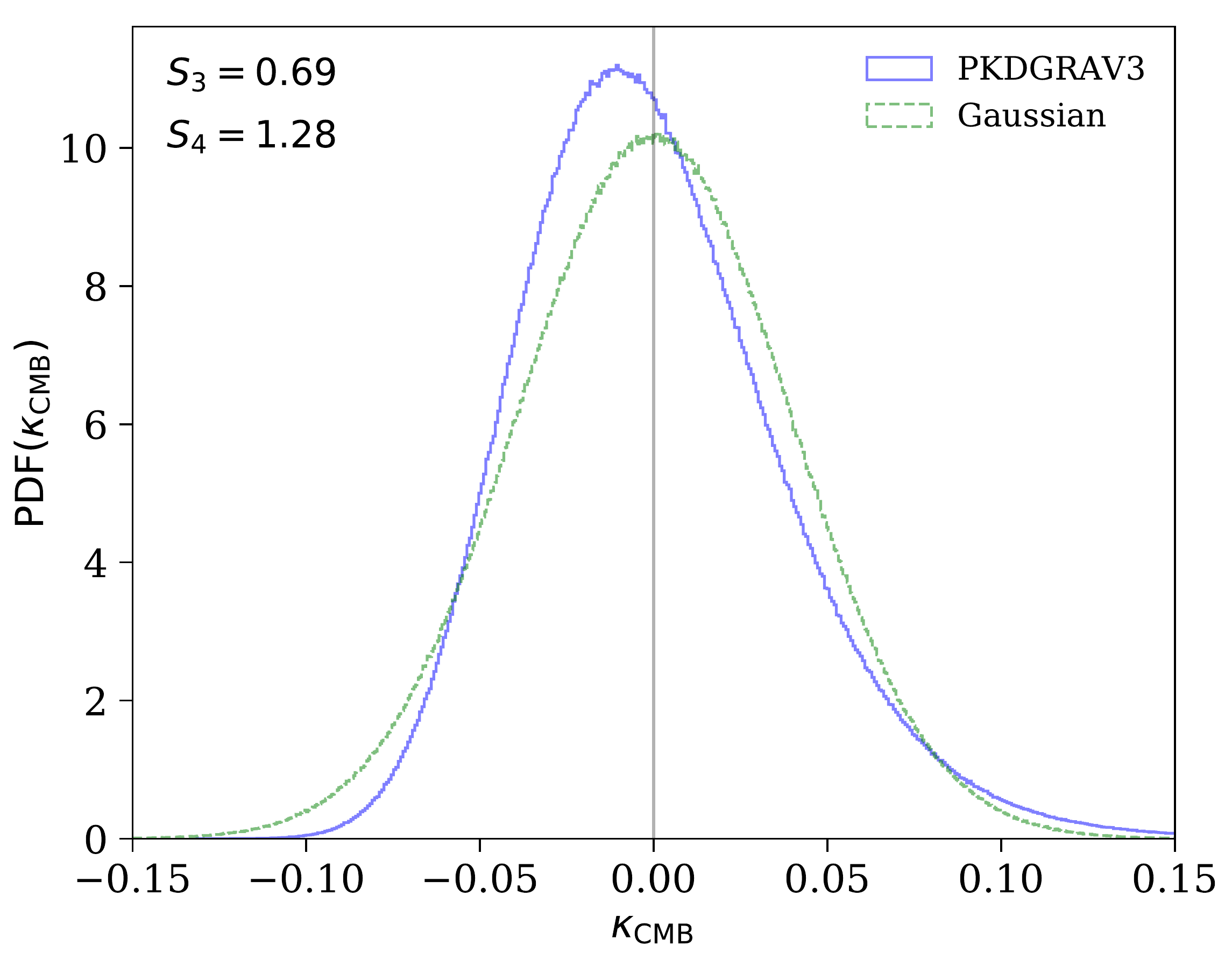} 
  \end{minipage}%% 
  \begin{minipage}[b]{0.37\paperwidth}
    \centering
    \includegraphics[width=.38\paperwidth]{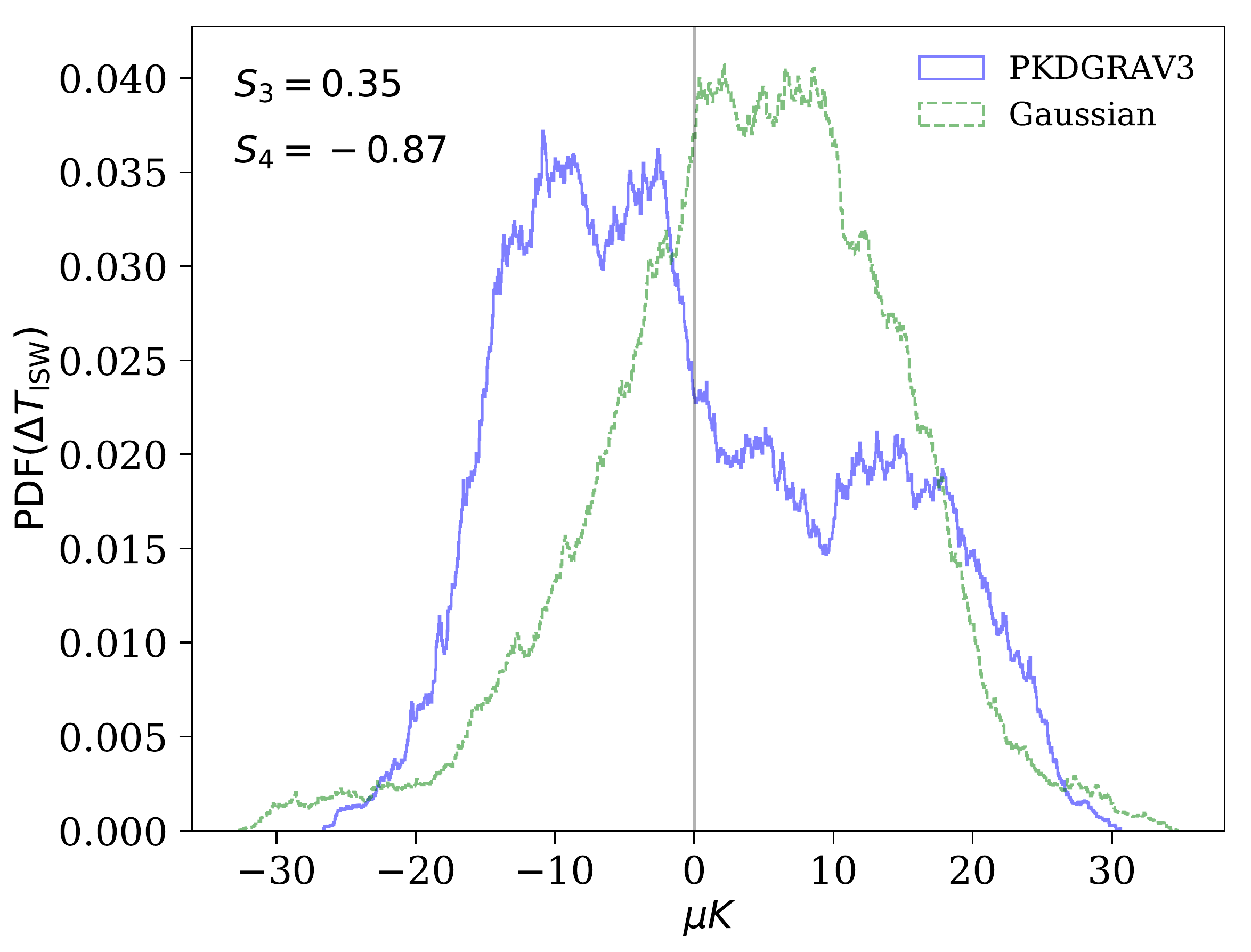} 
  \end{minipage} 
\caption{PDFs of the four different cosmological probes considered with a map resolution of nside = 1024. The blue solid lines show the non-Gaussian results by applying our pipeline to \textsc{UFalcon} / \textsc{PkdGrav3} output and the green dashed lines show the distributions of Gaussian fields. The values for the skewness $S_3$ and excess kurtosis $S_4$ for the non-Gaussian distributions are shown in each panel.}
\label{pdf_multi}
\end{figure}

We investigate the probability density function (PDF) of each map with nside = 1024 by calculating the histogram as a function of pixel values. Figure \ref{pdf_multi} shows a comparison of the PDF's of the different probes obtained with \textsc{UFalcon} from \textsc{PkdGrav3} simulations with the PDF's generated from synthetic Gaussian maps by applying the \textsc{Healpix} subroutine \texttt{synfast} on analytical spherical harmonic power spectra computed with \textsc{PyCosmo}. The distributions based on $N$-Body simulation output clearly deviates from a purely Gaussian field, which is also quantified by the skewness $S_3 = \mu_3 / \sigma^3$ and the excess kurtosis $S_4 = \mu_4 / \sigma^4 - 3$ of the non-Gaussian maps shown in each panel. The third and fourth central moment of the distributions are given by $\mu_3$ and $\mu_4$ respectively. Note that a Gaussian distribution has a skewness and an excess kurtosis of zero.

\subsection{Spherical Harmonic Power Spectrum}
\label{PS}
\begin{figure}[htbp!]
\centering
\makebox[0pt]{%
\includegraphics[width=0.75\paperwidth]{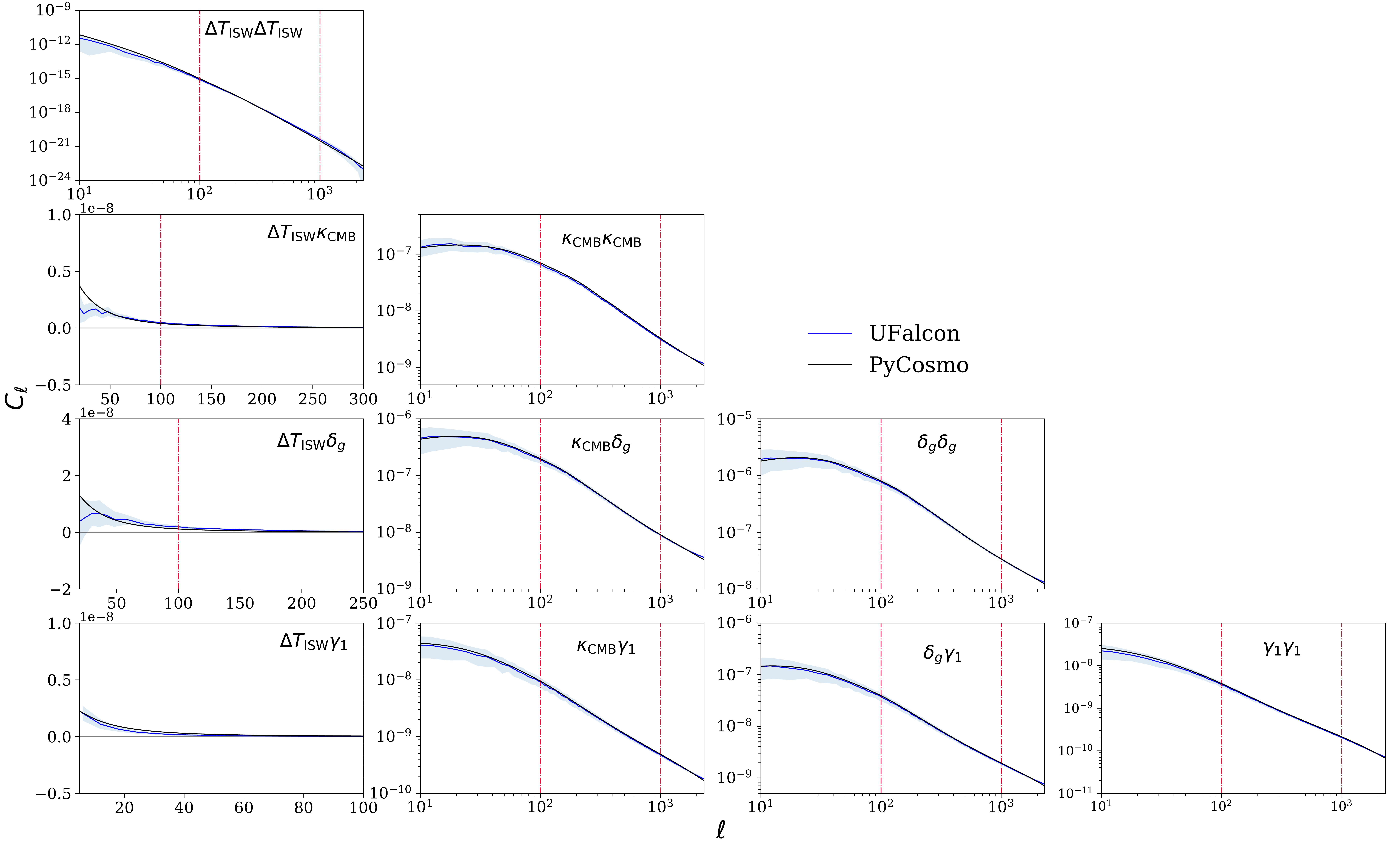}}
\caption{Blue solid lines: Mean auto- and cross- power spectra from 630 sets of full-sky maps for $\Delta T_\mathrm{ISW}$, $\kappa_\mathrm{CMB}$, $\delta_g$ and $\gamma_1$ based on \textsc{UFalcon} / \textsc{PkdGrav3} output. The shaded blue area represent the standard deviation from 630 realizations. The vertical dot-dashed red lines indicate the angular scales of our interest $\ell = 10^2$ and $\ell = 10^3$. Black solid lines: Analytical predictions calculated with \textsc{PyCosmo} based on the Limber approximation and using the fitting function from Mead \textit{et al.} \cite{Mead2015},\cite{Mead2016}.}
\label{multi_cls}
\end{figure}

\begin{figure}[htbp!]
\centering
\makebox[0pt]{%
\includegraphics[width=0.75\paperwidth]{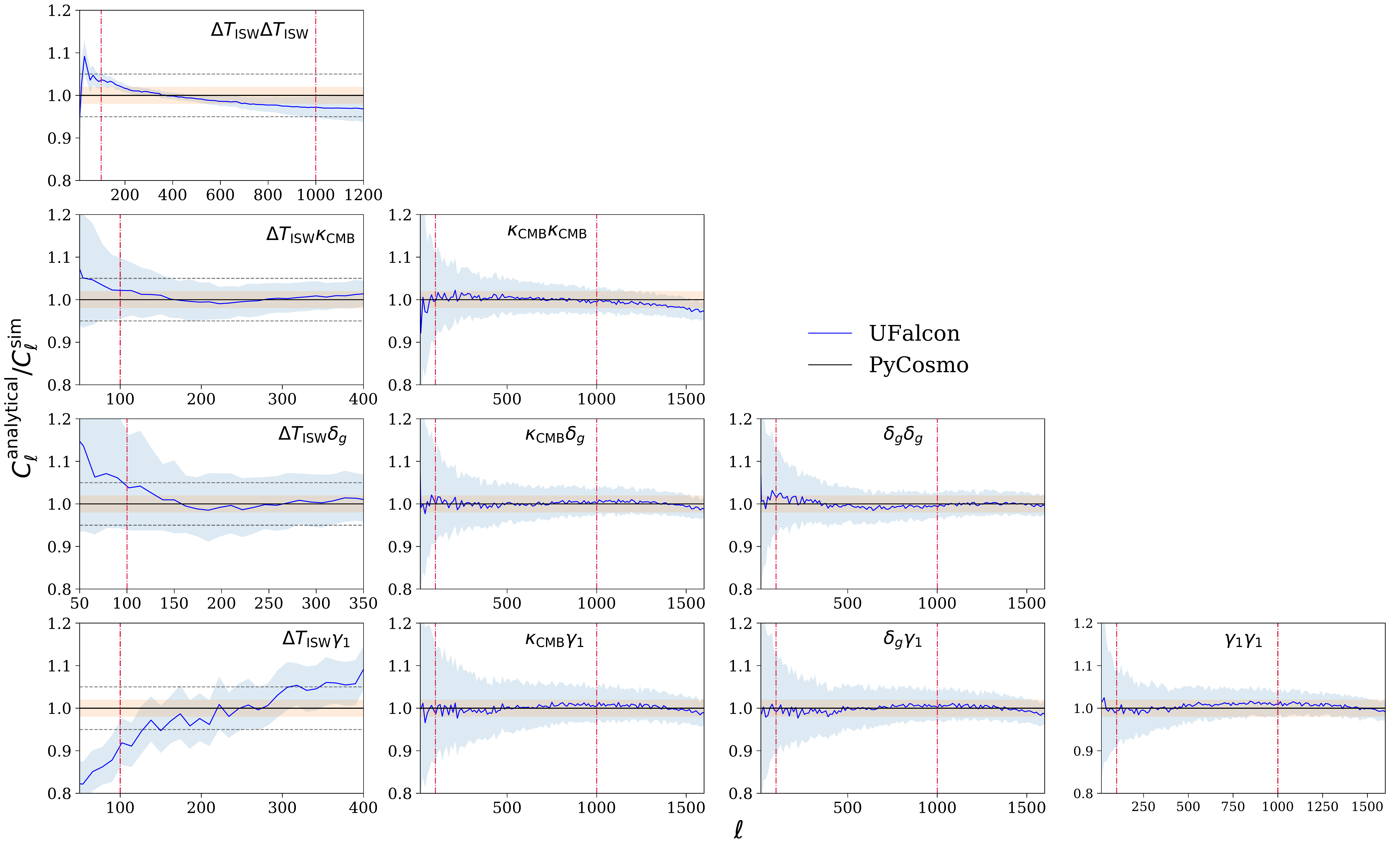}}
\caption{Ratio between analytical predictions based on \textsc{PyCosmo} and the mean auto- and cross- power spectra from 630 sets of full-sky maps for $\Delta T_\mathrm{ISW}$, $\kappa_\mathrm{CMB}$, $\delta_g$ and $\gamma_1$ based on \textsc{UFalcon} / \textsc{PkdGrav3} output. The shaded blue area represent the standard deviation from 630 realizations. The shaded orange area and the dashed grey lines represent 2\% and 5\% deviation from the analytical predictions respectively. The vertical dot-dashed red lines indicate the angular scales of our interest $\ell = 10^2$ and $\ell = 10^3$.}
\label{multi_ratio}
\end{figure}

In this section, we present a comparison between our simulation results and analytical predictions for the spherical harmonic power spectra. Figure \ref{multi_cls} shows the average power spectra of 630 sets of full-sky maps containing weak lensing shear, galaxy clustering, CMB lensing and temperature anisotropies from the ISW effect. The full-sky maps have been computed as described in section \ref{nummet} and have a resolution of nside = 1024. We first used the \textsc{Healpix} subroutine \texttt{anafast} to compute the auto- and cross- power spectra between the maps and then subtracted the corresponding shot noise contribution based on the Limber approximation given by equation (\ref{snlimber}). Furthermore, we adopt the correct deconvolution rules for the \textsc{Healpix} pixel window function and use a linear binning of $\delta \ell = 12$ to plot the power spectra. The blue solid lines show the simulation results and the black solid lines represent the analytical predictions computed with \textsc{PyCosmo} based on a Limber approximation as described in section \ref{theopred}. The auto-power spectra of the different probes are shown on the diagonal and the cross-power spectra on the off-diagonal panels of the triangle plot. In Figure \ref{multi_ratio}, we show the ratio between the analytical prediction and our simulation results for the spherical harmonic power spectrum. All the auto- and cross power spectra between the probes $\gamma$, $\delta_g$ and $\kappa_\mathrm{CMB}$ agree within 2\% to the \textsc{PyCosmo} prediction up to a multipole of $\ell \sim 1500$. The auto- and cross-power spectra including the temperature anisotropies from the ISW effect $\Delta T_\mathrm{ISW}$ agree within 5\% to the \textsc{PyCosmo} prediction within the range form $\ell = 10^2$ to $10^3$ with the exception of the cross-power spectrum $\left< \Delta T_\mathrm{ISW}\,\gamma_1 \right>$, which has a lower agreement. In general, we attribute the lower agreement of the cross-power spectrum between $\Delta T_\mathrm{ISW}$ and the other probes to the potentially insufficient resolution of our $\Delta T_\mathrm{ISW}$ maps for redshifts $z \lesssim 0.28$. As discussed in section \ref{sim_res}, redshift $z \sim 0.28$ corresponds to the radial comoving distance where the linear size of the grid cells (for our choice of $1024^3$ grid cells in the cubic grid used for the mass assignment scheme) corresponds to the pixel size of our \textsc{Healpix} maps when using nside = 1024. The weak lensing shear window function $W^{\gamma}$ peaks at around $z \sim 0.3$ for a Smail \textit{et al.} distribution (shown in the right panel of Figure \ref{survey_geometry}), whereas the weights of the other probes we consider peak at higher redshifts. The cross-correlation between $\Delta T_\mathrm{ISW}$ and $\gamma_1$ is thus weighting the integrand more strongly for lower redshifts (which is also where the ISW-maps are potentially lacking resolution) and therefore has lower agreement to the theoretical predictions than the other cross-power spectra.

Our simulation results are lacking power on very large scales, i.e. on multipoles $\ell \lesssim 10$, which we attribute to missing large scale modes larger than our simulation volume (super sample covariance effects, see section \ref{SSC} for a discussion thereof) and the inaccuracy of the Limber approximation for multipoles $\ell \lesssim 10$. This effect is not visible on the plots due to our choice of binning. Applying our randomization procedure to increase the number of realizations additionally enhances the power loss by 1-2 \% on very large scales (see section \ref{randomisation}). Note that since the present analysis is only considering scales between multipoles $\ell = 10^2$ and $10^3$, the observed missing power on multipoles $\ell \lesssim 10$ does not negatively affect the power spectrum covariance matrix.

\subsection{Multi-Probe Covariance Matrix}
\label{covariance}

\begin{figure}[htbp!]
\centering
\includegraphics[width=0.5\textwidth]{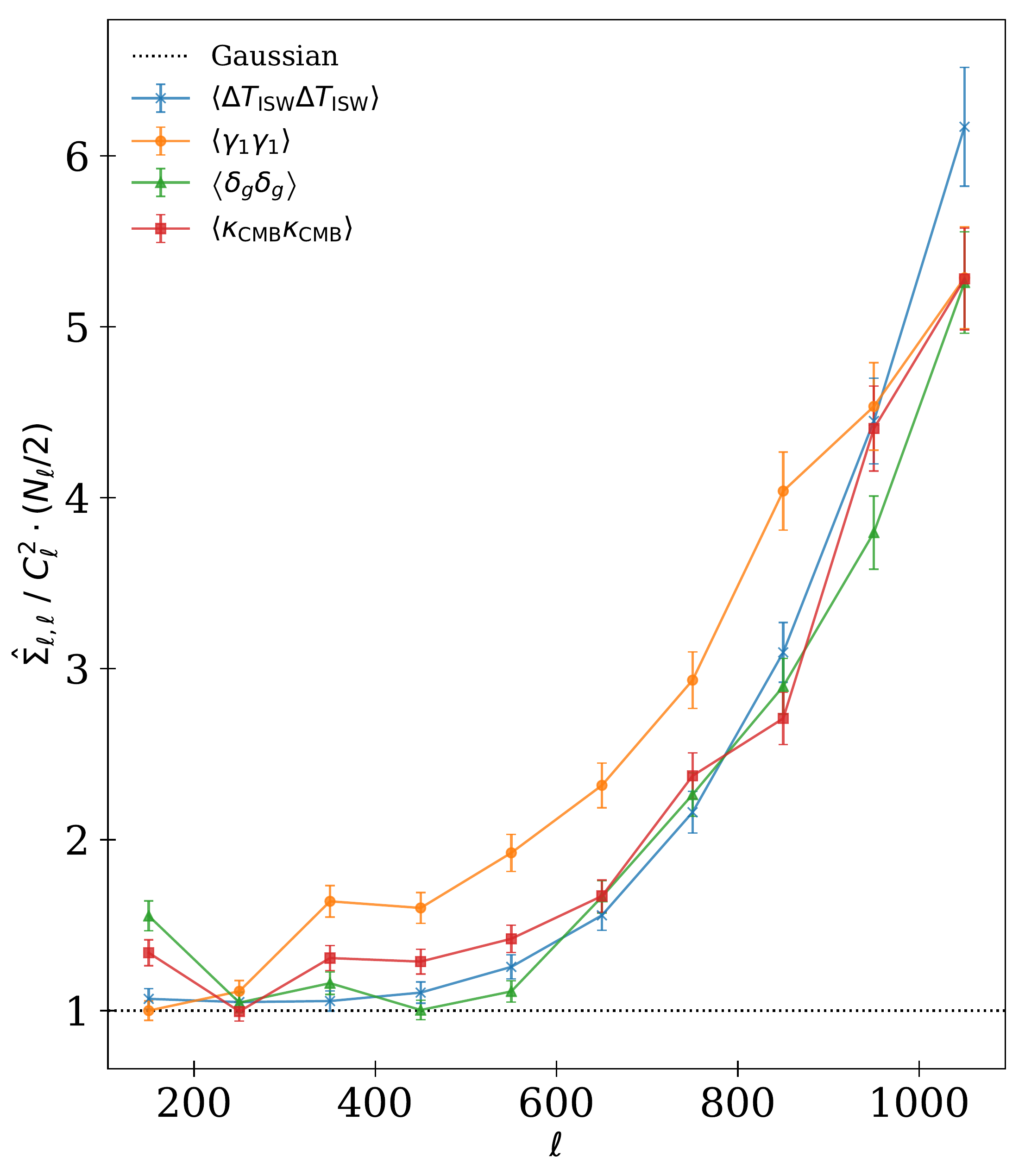}
\caption{Diagonal components of the auto-power spectrum covariances for the simulated fields $\Delta T_\mathrm{ISW}$, $\gamma_1$, $\delta_g$ and $\kappa_\mathrm{CMB}$ divided by the Gaussian covariances given by $\hat{\Sigma}_{\ell, \ell'} / C_{\ell}^2 \cdot (N_{\ell} / 2)$. The errorbars correspond to the standard deviation obtained by using equation (\ref{coverr}) and assuming that our covariance matrix estimate follow a Wishart distribution.}
\label{covdiag}
\end{figure}

\begin{figure}[htbp!]
\centering
\includegraphics[width=0.7\textwidth]{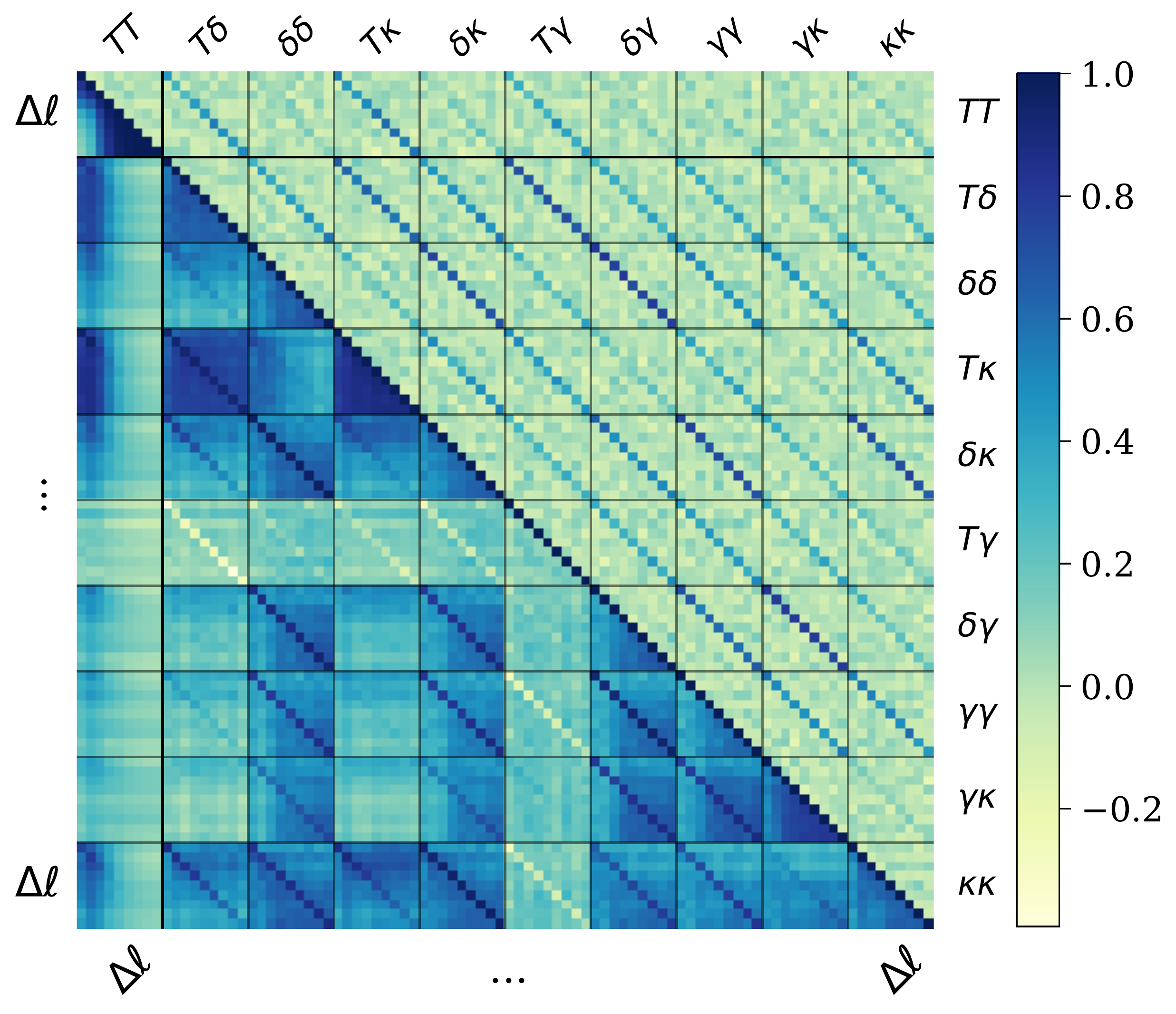}
\caption{Multi-probe covariance correlation matrix for the spherical harmonic power spectra derived using 630 full-sky realizations. Each panel covers the multipole range $\Delta \ell = [10^2, 10^3]$ with a linear binning of $\delta \ell = 100$. Lower triangle: Non-Gaussian correlation matrix obtained by applying \textsc{UFalcon} to \textsc{PkdGrav3} simulations. Upper triangle: Correlation matrix based on synthetic Gaussian realizations.}
\label{corr_allprobes}
\end{figure}

%The covariance matrix of spherical harmonic power spectra can be written as \cite{Scoccimarro1999}, \cite{Meiksin1999}
%\begin{equation}
% \left< (C_\ell - \left< C_\ell \right>)(C_\ell' - \left< C_\ell' \right>) \right>
%\Sigma (\ell, \ell ') = \frac{2}{N_{\ell}} (C_\ell)^2 \delta_{\ell, \ell'} + T_{\ell, \ell'} \, ,
%\label{cov1}
%\end{equation}
%where $N_\ell = 2 \pi \ell \Delta \ell$ is the number of modes between $\ell$ and $\ell + \Delta \ell$, $\delta_{\ell, \ell'}$ is the Kronecker delta and %$T_{\ell, \ell'}$ is the lensing trispectrum. The first term in equation (\ref{cov1}) represents the Gaussian variance and the second term is the %non-Gaussian contribution originating from non-linear structure formation. 
We compute the joint covariance matrix from the $N_s = 630$ \textsc{UFalcon} full-sky realizations of the spherical harmonic power spectra $\hat{C}^{ij}(\ell)$ discussed in section \ref{PS} using the sample covariance estimator
\begin{equation}
\hat{\Sigma}_{\ell, \ell '} = \frac{1}{N_s - 1} \sum_{k = 1}^{N_s} \left[\hat{C}_{k}^{ij}(\ell) - \bar{C}_{k}^{ij}(\ell) \right] \left[\hat{C}_{k}^{i'j'}(\ell') - \bar{C}_{k}^{i'j'}(\ell') \right] \, ,
\label{cov2}
\end{equation}
where $i,j,i',j' \in \{ T_\mathrm{ISW}, \kappa_\mathrm{CMB}, \delta_g, \gamma_1 \}$ denote the different cosmological probes and $\bar{C}_{k}^{ij}(\ell)$ is the mean over all realizations. For the case of Gaussian distributed data, the probability distribution of the sample covariance matrix is described by a Wishart distribution \cite{Wishart1928}. The variance of the sample covariance is then given by \cite{Taylor2013},\cite{Blot2016}
\begin{equation}
\sigma^2 (\hat{\Sigma}_{\ell, \ell '}) = \frac{1}{N_s - 1} \left( \Sigma^2_{\ell, \ell '} + \Sigma_{\ell, \ell} \Sigma_{\ell', \ell '}\right)\quad .
\label{coverr}
\end{equation}
In Figure \ref{covdiag} we show the diagonal components of the full-sky auto-power spectrum covariances divided by their corresponding Gaussian variances, denoted by $\hat{\Sigma}_{\ell, \ell'} / C_{\ell}^2 \cdot (N_{\ell} / 2)$. The number of modes between $\ell$ and $\ell + \Delta \ell$ is given by $N_\ell = A_s \Omega_s / (2 \pi)^2$ with $A_s = 2 \pi \ell \Delta \ell$ being the area of the 2D shell around the bin labelled with $\ell$. The deviations from unity arise from the non-Gaussian error contribution present in our simulation results. Note that we obtain different results for the binned non-Gaussian covariance matrix, depending whether we first calculate the covariance with unit bin ($\delta \ell = 1$) and then manually re-bin it or when we directly calculate it at the bin-centers using broader $\ell$-bins. Although for the case of a strongly off-diagonal covariance matrix we expect the difference to be small.
%Note that the Gaussian term in the covariance matrix depends on the width of the multipole-bins chosen, whereas the non-Gaussian term %does not. Therefore, choosing smaller values for $\Delta \ell$ increases the Gaussian errors relative to the non-Gaussian contributions.
\smallbreak
In the case of working with full-sky maps, only the presence of non-Gaussian contributions introduce off-diagonal components to the power spectrum covariance matrix. The relative strength of the off-diagonal terms to the diagonal terms in the covariance matrix can be quantified through the correlation coefficient as
\begin{equation}
\mathrm{Corr} (\ell, \ell ') = \frac{\hat{\Sigma}_{\ell, \ell '}}{\sqrt{\hat{\Sigma}_{\ell, \ell} \hat{\Sigma}_{\ell', \ell '}}} \, ,
\end{equation}
which is unity for $\ell = \ell '$. The correlation coefficient implies strong correlation for $\mathrm{Corr} \rightarrow 1$, no correlation for $\mathrm{Corr} \rightarrow 0$ or strong anti-correlation for $\mathrm{Corr} \rightarrow -1$ between the spectra at different mulipoles. In Figure \ref{corr_allprobes}, we show the joint covariance correlation matrix by using 630 sets of auto- and cross- spherical harmonic power spectra based on full-sky \textsc{PkdGrav3} simulations, whereas each spectrum spans a multipole range $\Delta \ell = [10^2, 10^3]$ with a linear binning of $\delta \ell = 100$. 
We additionally compute an analogous covariance matrix using equation (\ref{cov2}) based on 630 sets of synthetic Gaussian maps. To this end, we follow the approach outlined in Giannantonio \textit{et al.} \cite{Giannantonio2008} and applied in Nicola \textit{et al.} \cite{Nicola2016}, \cite{Nicola2017} in order to generate a set of correlated Gaussian maps. This approach consists of first generating correlated realizations of the spin-0 fields $\Delta T_\mathrm{ISW}$, $\kappa_\mathrm{CMB}$, $\delta_g$ and $\kappa_\mathrm{smail}$. The obtained convergence map $\kappa_\mathrm{smail}$ based on the redshift distribution given by equation (\ref{nz}) is then transformed to the spin-2 weak lensing shear fields $\gamma_1$ and $\gamma_2$ using equation (\ref{kappa2gamma}). Further details about our implementation of this algorithm are given in Appendix \ref{GiannantonioRoutine}.

\section{Forecast}
\label{forecast}
\begin{figure}[htbp!]
  \begin{minipage}[b]{0.5\textwidth}
    \centering
    \includegraphics[width=1\textwidth]{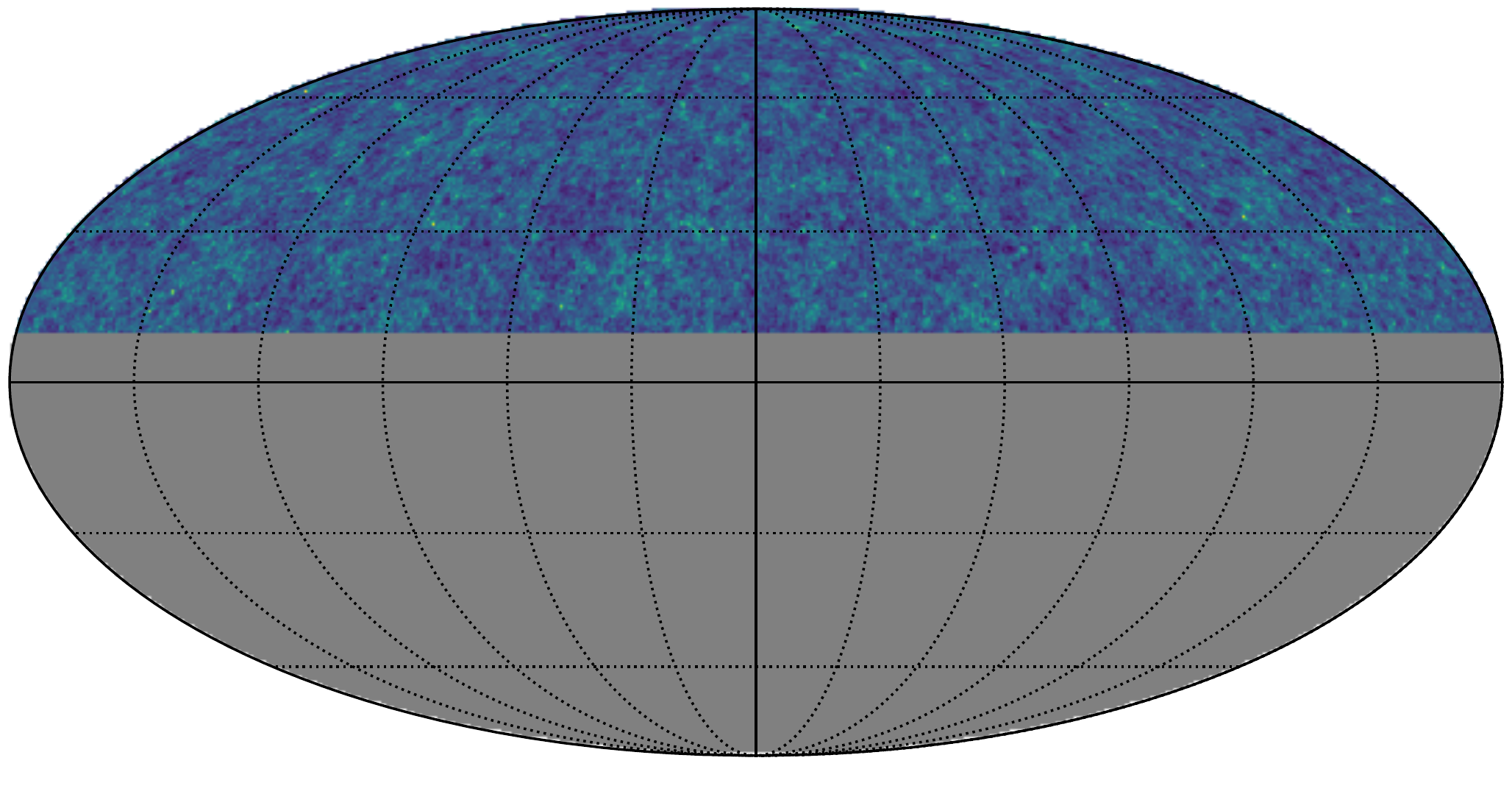} 
  \end{minipage}%%
  \begin{minipage}[b]{0.5\textwidth}
    \centering
    \includegraphics[width=1\textwidth]{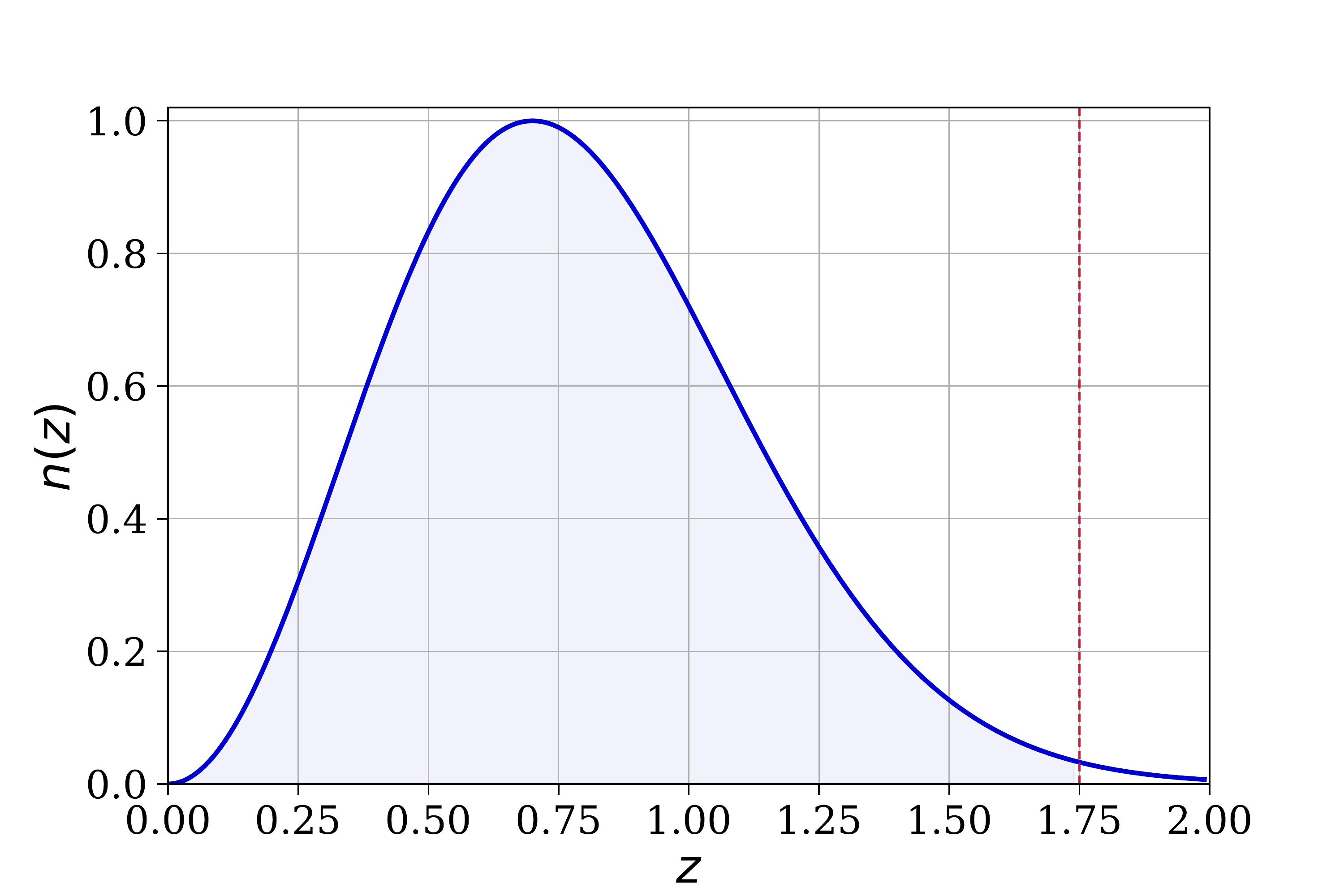} 
  \end{minipage} 
  \caption{Left panel: Stage-IV-like survey area covering $\sim 17000\, \, \mathrm{deg}^2$. Right panel: Smail \textit{et al.} \cite{Smail1994} source galaxy redshift distribution given by equation (\ref{nz}) used to generate the $\gamma$ and $\delta_g$ maps. The vertical dashed red line at $z_\mathrm{max} = 1.75$ indicates the upper redshift bound, up to which our analysis has been performed.}
\label{survey_geometry}
\end{figure}

\begin{figure}[htbp!]
\centering
\includegraphics[width=10cm]{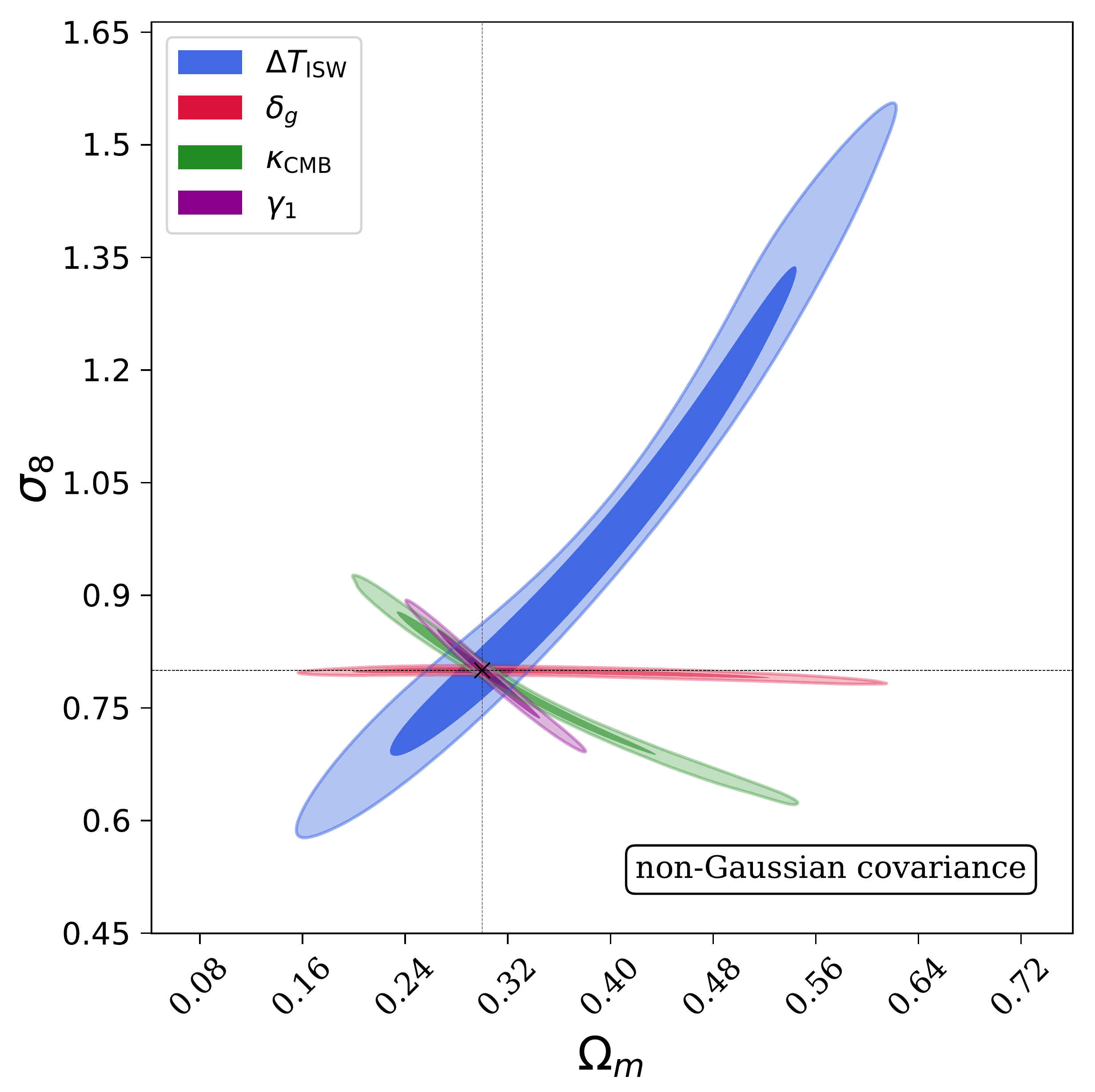}
\caption{Parameter constraints for $\Omega_m$ and $\sigma_8$ derived from separate single-probe analyses based on the auto-power spectrum using a simulation-based non-Gaussian covariance matrix. The constraints are marginalized over the nuisance parameters. The inner (outer) contours depict the 68\% (95\%) confidence levels. Note that our forecast analysis is only limited by cosmic variance, i.e. no survey-specific noise has been added to our simulated maps.}
\label{single_probes_sim}
\end{figure}

\begin{figure}[htbp!]
  \begin{minipage}[b]{0.37\paperwidth}
    \centering
    \includegraphics[width=.3\paperwidth]{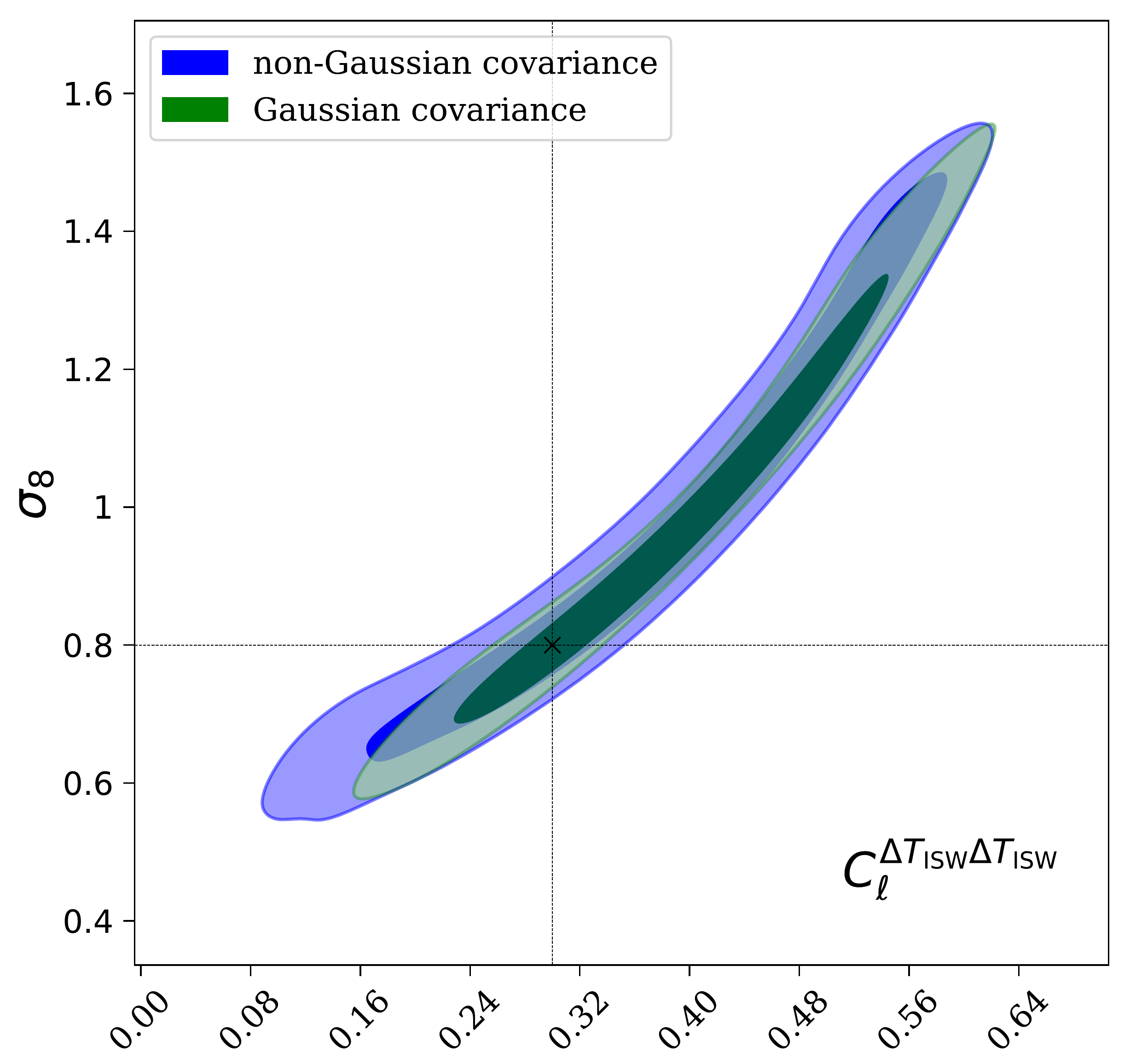} 
  \end{minipage}%%
  \begin{minipage}[b]{0.28\paperwidth}
    \centering
    \includegraphics[width=.28\paperwidth]{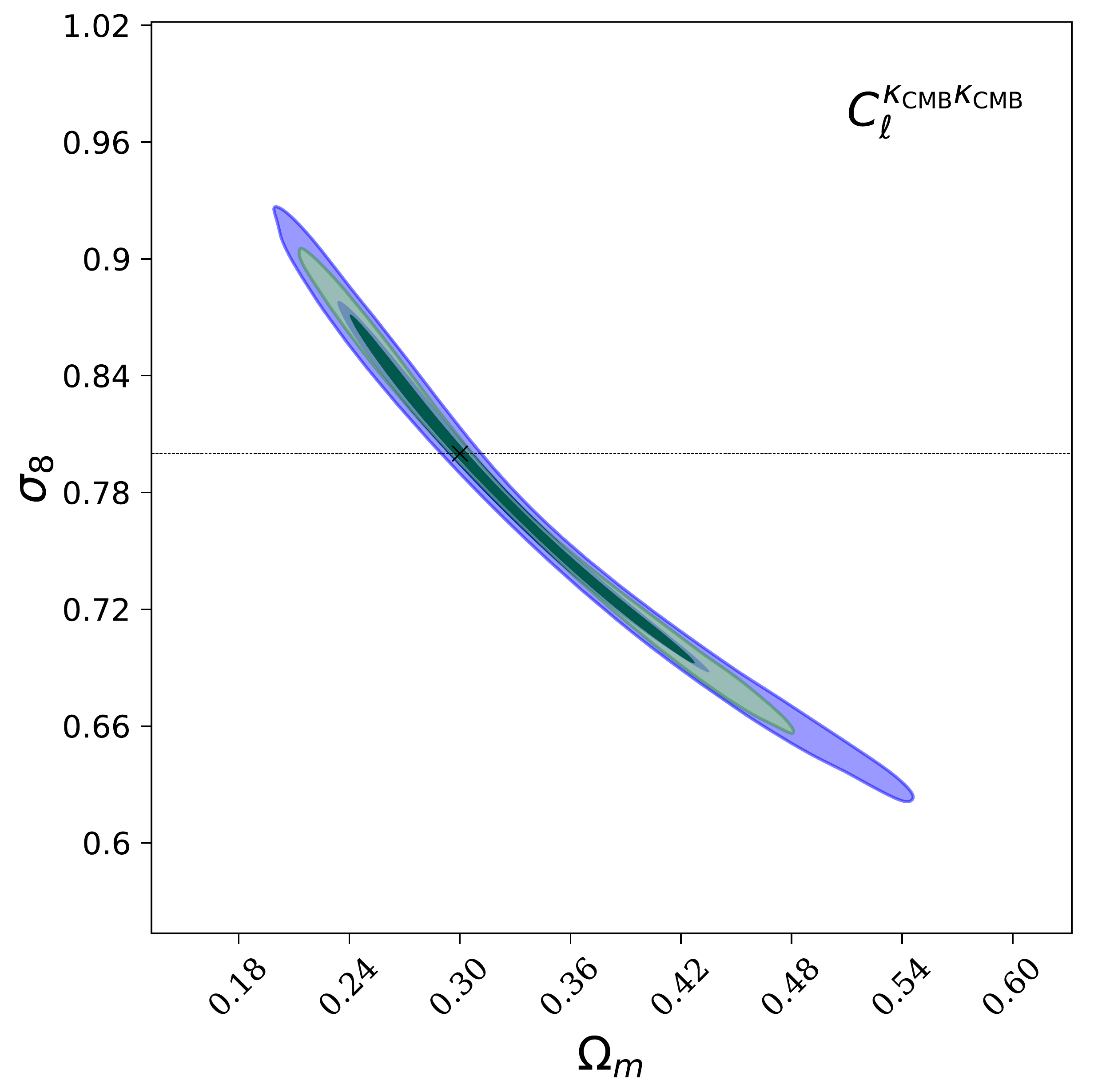} 
  \end{minipage} 
  \begin{minipage}[b]{0.37\paperwidth}
    \centering
    \includegraphics[width=.3\paperwidth]{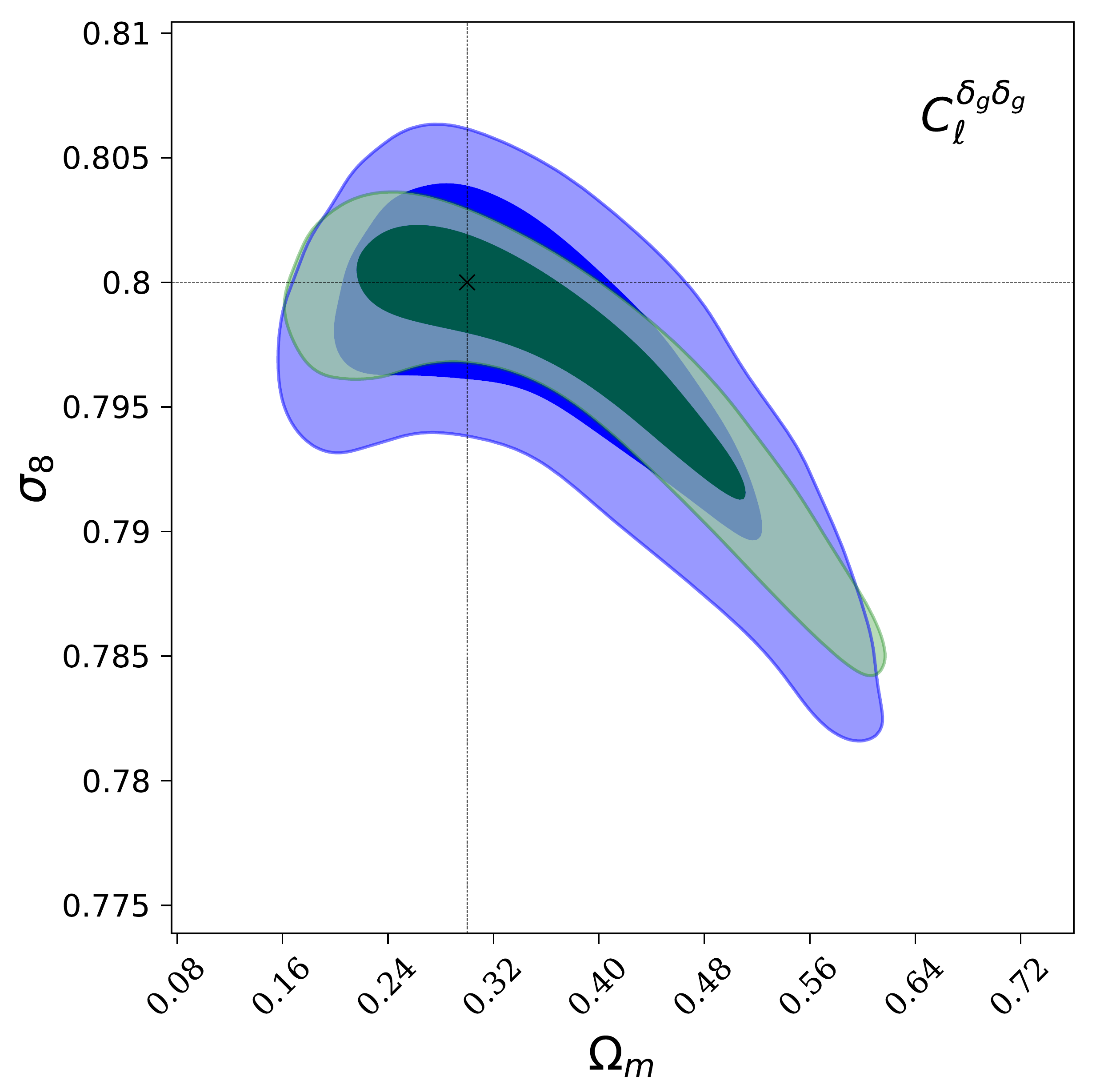} 
  \end{minipage}%% 
  \begin{minipage}[b]{0.28\paperwidth}
    \centering
    \includegraphics[width=.28\paperwidth]{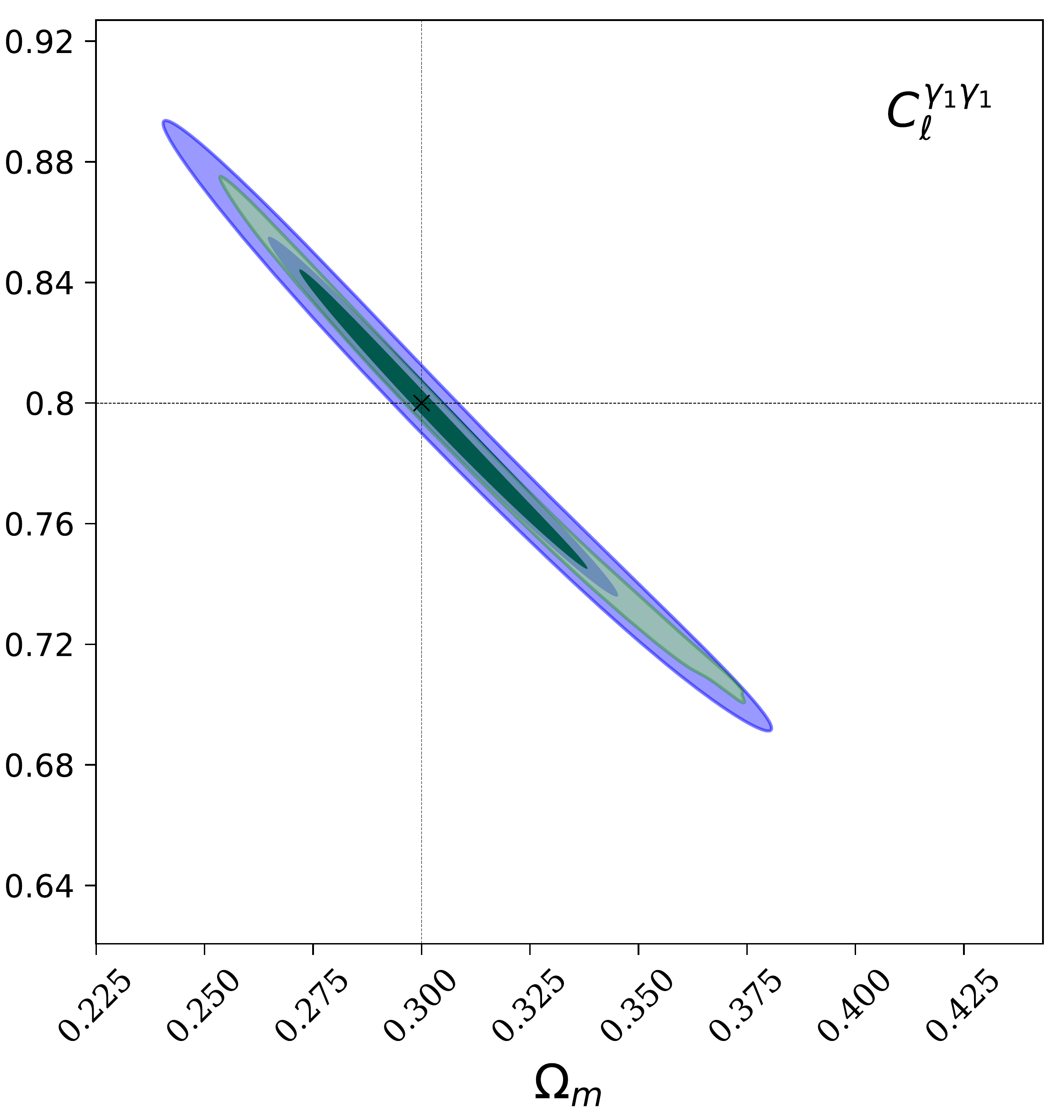} 
  \end{minipage} 
\caption{Cosmological parameter constraints for $\Omega_m$ and $\sigma_8$ derived from a single-probe analysis based on the auto-power spectra using a Gaussian (green contours) and simulation-based non-Gaussian (blue contours) covariance matrix. The inner (outer) contours depict the 68\% (95\%) confidence levels. Note that our forecast analysis is only limited by cosmic variance, i.e. no survey-specific noise has been added to our simulated maps.}
\label{single_autos}
\end{figure}

\begin{figure}[htbp!]
\centering
\includegraphics[width=14cm]{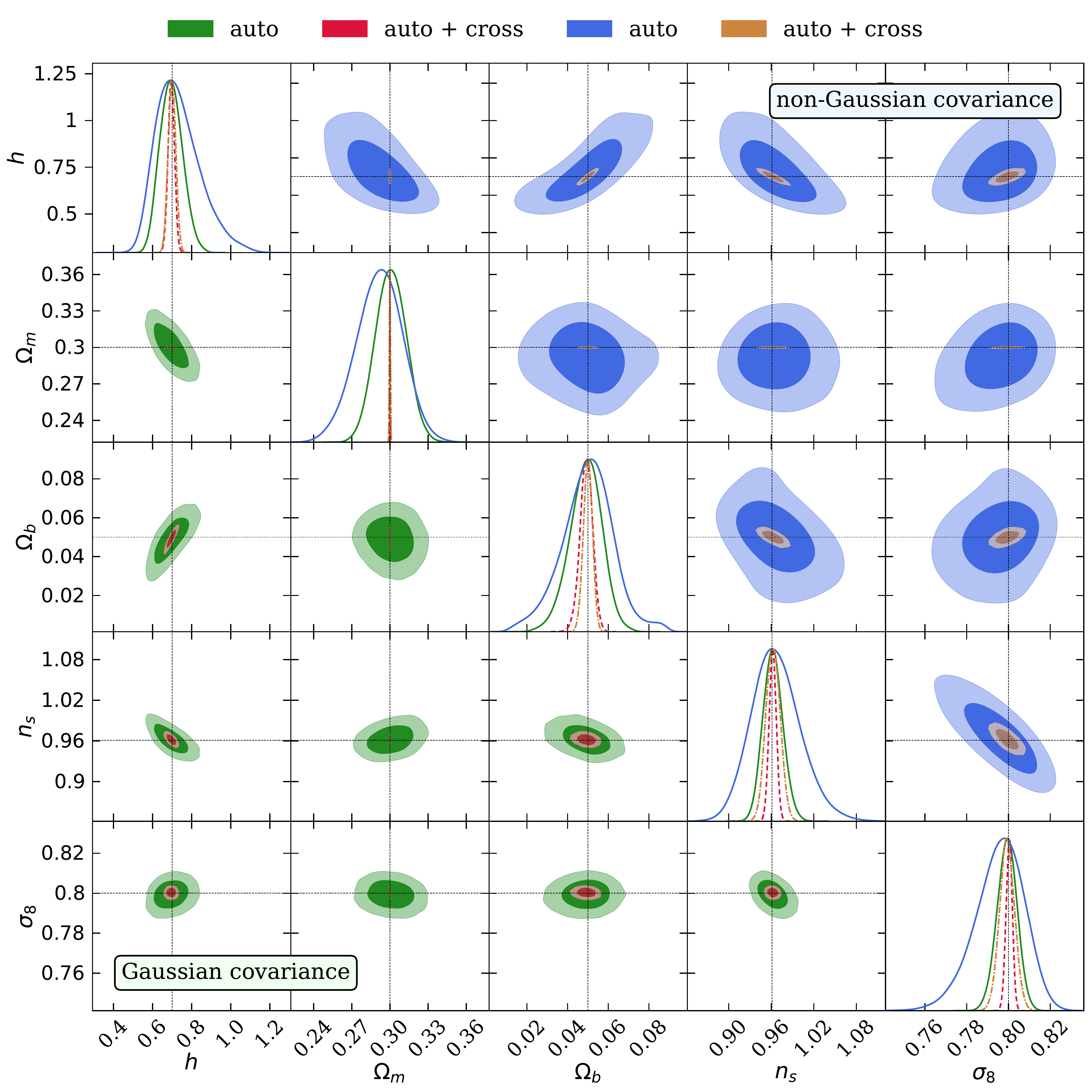}
\caption{Parameter constraints for the parameters described in section \ref{forecast} marginalized over the nuisance parameters. The lower (upper) triangle-plot show the contours obtained when using a Gaussian (non-Gaussian) covariance matrix. The green (blue) and red (brown) contours in the lower (upper) triangle-plot are obtained using only the auto-power spectra and the auto- and cross-power spectra respectively. The inner (outer) contours depict the 68\% (95\%) confidence levels. Note that our forecast analysis is only limited by cosmic variance, i.e. no survey-specific noise has been added to our simulated maps.}
\label{4probes_nonuisanceplot}
\end{figure}

\begin{figure}[htbp!]
  \begin{minipage}[b]{0.37\paperwidth}
    \centering
    \includegraphics[width=.37\paperwidth]{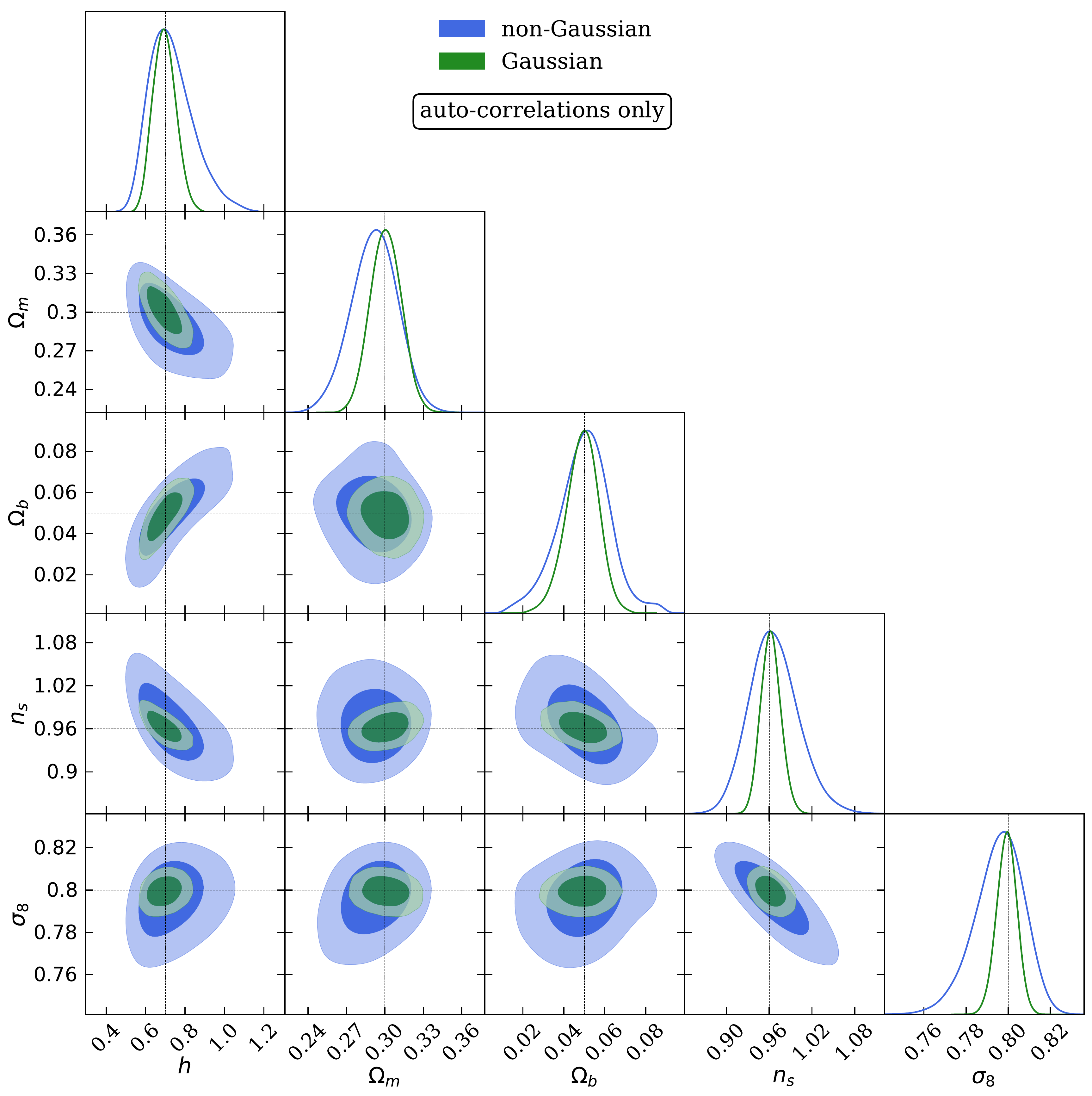} 
  \end{minipage}%%
  \begin{minipage}[b]{0.37\paperwidth}
    \centering
    \includegraphics[width=0.37\paperwidth]{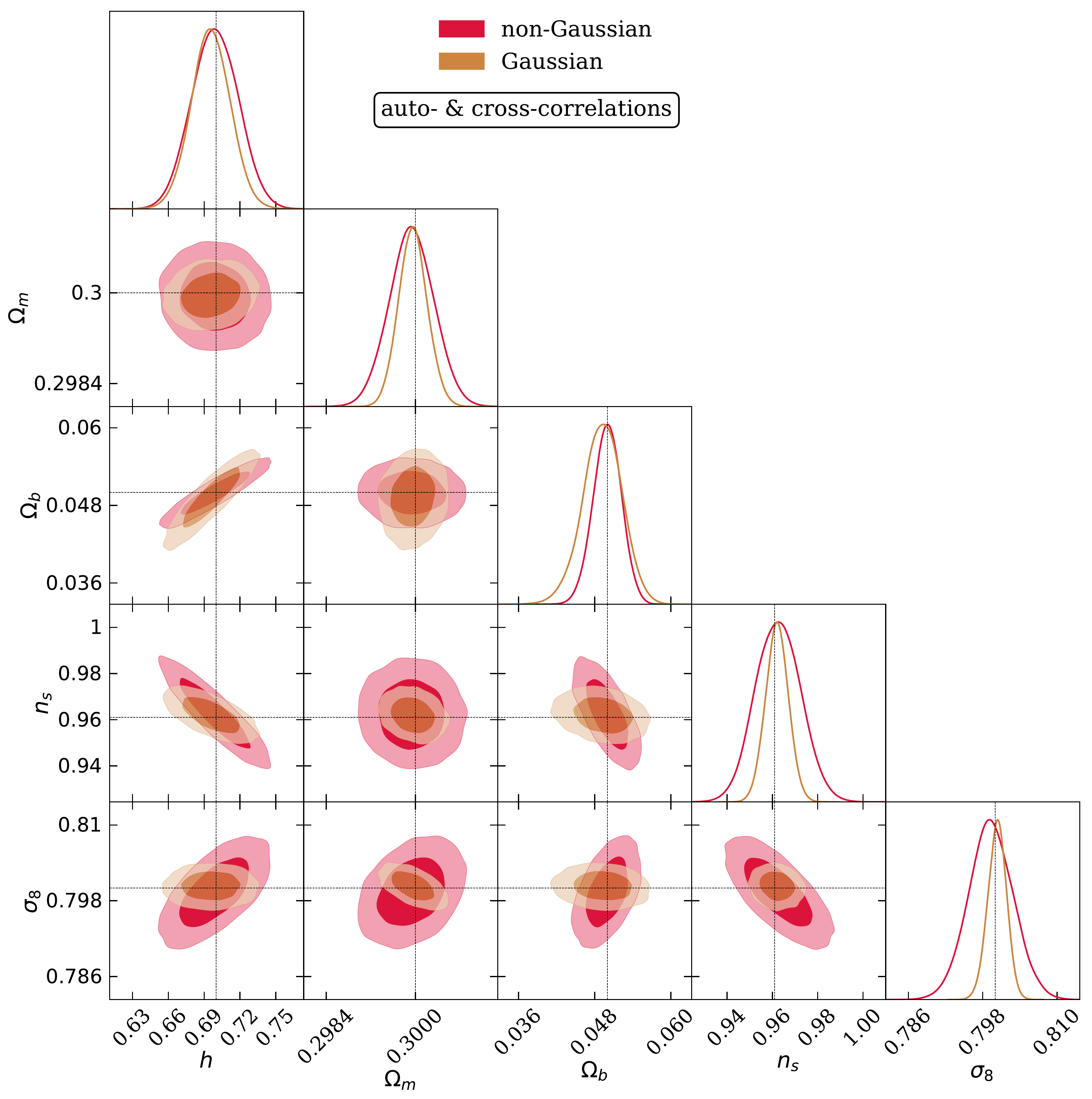} 
  \end{minipage} 
  \caption{Parameter constraints for the parameters described in section \ref{forecast} marginalized over the nuisance parameters (same constraints as in Figure \ref{4probes_nonuisanceplot}, but shown in a different combination). Left panel:  The blue (green) constraints are obtained by using only the auto-power spectra with the corresponding non-Gaussian (Gaussian) covariance matrix. Right panel: The red (brown) constraints are obtained by using the auto- and cross-power spectra with the corresponding non-Gaussian (Gaussian) covariance matrix. The inner (outer) contours depict the 68\% (95\%) confidence levels. Note that our forecast analysis is only limited by cosmic variance, i.e. no survey-specific noise has been added to our simulated maps.}
\label{4probes_dualplot}
\end{figure}

\begin{figure}[htbp!]
\centering
\includegraphics[width=15cm]{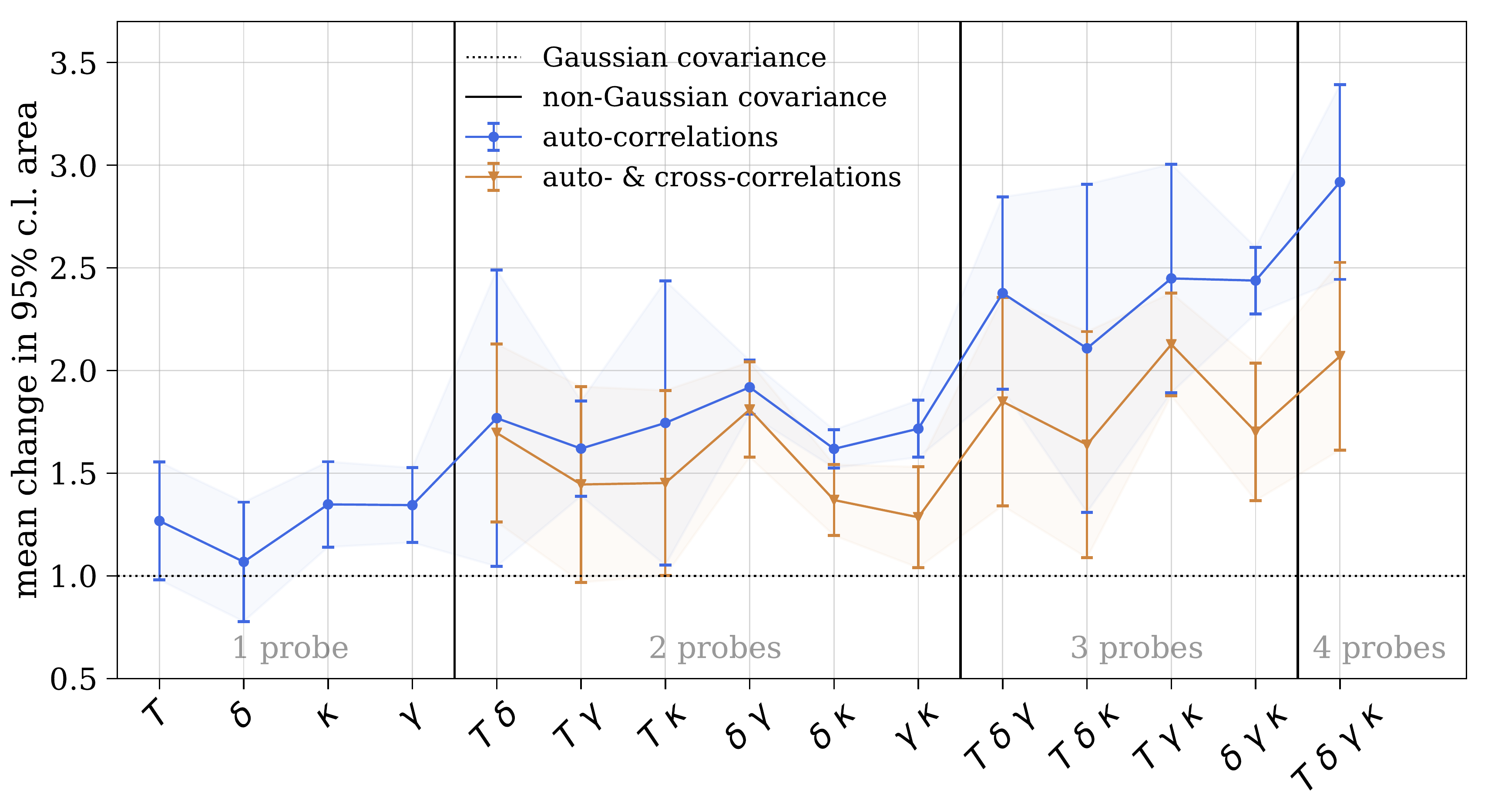}
\caption{The blue (brown) lines represent the mean change in 95\% confidence level area from using a Gaussian (dotted line) to using a non-Gaussian covariance matrix (solid lines). The error bars represent the $1\sigma$-error on the change in 95\% confidence level area for the 10 two-dimensional contours obtained for different combinations of cosmological parameters. Note that our forecast analysis is only limited by cosmic variance, i.e. no survey-specific noise has been added to our simulated maps.}
\label{area_plot_cross_auto}
\end{figure}

In this section we compute constraints on cosmological parameters in order to quantify the impact of the non-Gaussian contributions to the covariance matrix and of the combination of different auto- and cross-power spectra.\\
\\
\textbf{Survey specifications.} This analysis is performed for a stage IV-like survey area, i.e. we mask half of the sky (including masking a simplified galactic annulus) leading to a map geometry with an unmasked sky fraction of $f_\mathrm{sky} \sim 0.42$, as shown in the left panel of Figure \ref{survey_geometry}. 
The same survey mask is then applied to each map of the cosmological probes $\gamma$, $\delta_g$, $\kappa_\mathrm{CMB}$ and $\Delta T_\mathrm{ISW}$ for all the 630 sets of full-sky \textsc{UFalcon} maps. Each map covers a redshift range between $z=0.0$ and $1.75$. Furthermore, the cosmic shear $\gamma$ and galaxy clustering $\delta_g$ maps are generated using a Smail \textit{et al.} \cite{Smail1994} source galaxy distribution (given by equation (\ref{nz}) and shown in the right panel of Figure \ref{survey_geometry}).
We intentionally do not add survey-related noise to our simulated maps (such as shape noise for cosmic shear), which would increase the diagonal components of our covariance matrices and therefore potentially degrade the relative impact of the non-Gaussian error contribution to the covariance. Our forecast analysis is therefore only limited by cosmic variance.
\newpage
In order to parametrise the observational and sky systematics involved in the cosmological measurements, we introduce four different multiplicative bias parameters $m_i$ for the estimator $\hat{i}$ of each probe as
\begin{equation}
\hat{i} = (1 + m_i) i \, ,
\end{equation}
where  $i \in \{ T_\mathrm{ISW}, \kappa_\mathrm{CMB}, \delta_g, \gamma_1 \}$. Motivated by the cosmological analysis done in Nicola \textit{et al.} \cite{Nicola2016}, \cite{Nicola2017}, the introduction of scalar multiplicative bias parameters for the weak lensing shear can take into account calibration uncertainties \cite{Heymans2006} and uncertainties in the amplitude due to intrinsic alignment of unlensed galaxies (see Appendix A in \cite{Nicola2017}). Furthermore, a multiplicative bias parameter can be introduced to take into account the normalisation uncertainty of the CMB lensing convergence estimator \cite{Nicola2017}. We additionally introduce nuisance parameters for the CMB temperature anisotropies due to the ISW effect $\Delta T_\mathrm{ISW}$ and the galaxy overdensity $\delta_g$, which can parametrise possible foreground contamination. Concerning the latter, we set the linear galaxy bias parameter to $b=1$. The introduced nuisance parameters $m_i$ are simultaneously fit with the cosmological parameters $\theta$. Note that our analysis does not include other systematic effects, such as baryonic effects on the power spectrum or photometric redshifts uncertainties. Moreover, the present work represents a non-tomographic analysis.\\
\\
\textbf{Pseudo-$C_\ell$ estimation.} For our analysis we rely on the application of the pseudo-$C_\ell$ method, which was first proposed in \cite{Wandelt2001} and \cite{Hivon2002}, and is based on the original method in \cite{Peebles1973}. Extending this framework to spin-0 and spin-2 fields, we relate the underlying full-sky analytical predictions based on \textsc{PyCosmo} to the observed cut-sky power spectra $\tilde{\boldsymbol{C}}_\ell$ through
\begin{equation}
\tilde{\boldsymbol{C}}_\ell = \sum_{\ell'} \boldsymbol{M}_{\ell \ell'} \boldsymbol{C}_{\ell '}^\mathrm{PyCosmo} \, ,
\end{equation}
where $\boldsymbol{M}$ denotes the mode-coupling matrix describing the effect of the sky-cut applied to the data. A more detailed explanation of our implementation of this method is given in appendix \ref{pseudocl_appendix}.\\
\\
\textbf{Parameter inference.} We assume a joint likelihood given by \cite{Sellentin2016}, \cite{Sellentin2017}
\begin{equation}
\mathcal{L} (D | \theta, \hat{\Sigma}, N_s) \propto |\hat{\Sigma}|^{-1/2} \left[ 1 + \frac{\left( \boldsymbol{C}_{\ell}^\mathrm{obs} - \tilde{\boldsymbol{C}}_\ell \right)^{T} \hat{\Sigma}^{-1}\left( \boldsymbol{C}_{\ell}^\mathrm{obs} - \tilde{\boldsymbol{C}}_\ell \right)}{N_s -1} \right]^{- \frac{N_s}{2}} \, ,
\label{likelihood}
\end{equation}
which takes into account the uncertainty of the covariance matrix estimated from simulations. The covariance matrix is estimated using equation (\ref{cov2}) for 630 masked realizations based on simulations and based on synthetic Gaussian maps as described in section \ref{covariance}. We approximate the covariance matrices to be cosmology independent, i.e. they are based on simulation and Gaussian realizations for fixed fiducial parameters values $\boldsymbol{\theta}_\mathrm{fid}$. The calculation of a cosmology dependent covariance matrix would require running the same number of simulations for various cosmological parameters combinations and is therefore computationally very expensive.
\smallbreak
The 'mock observation' power spectrum vector is set to $\boldsymbol{C}_{\ell}^\mathrm{obs} = \boldsymbol{\tilde{C}}_{\ell} \left( \boldsymbol{\theta}_\mathrm{fid} \right)$, i.e. to the pseudo-$C_\ell$ vector based on the  analytical prediction computed with \textsc{PyCosmo} for our fiducial parameters $\boldsymbol{\theta}_\mathrm{fid}$. This choice of data vector allows us to isolate the effect of using different covariance matrices for the parameter inference. The resulting constraints are therefore expected to be centered around the fiducial cosmology and not experience any change in area / position due to added noise on the data vector. The analytical prediction is computed as $\boldsymbol{C}_{\ell}^\mathrm{th} = \boldsymbol{\tilde{C}}_{\ell}\left( \boldsymbol{\theta} \right)$ for varying parameters $\boldsymbol{\theta}$ and has the same dimensionality as $\boldsymbol{C}_{\ell}^\mathrm{obs}$. The total power spectrum vector with dimensionality $d$ has the form 
\begin{equation}
\boldsymbol{C}_{\ell} = \left( C_\ell^{TT}\,\, C_\ell^{T \delta}\,\,C_\ell^{\delta \delta}\,\,C_\ell^{\kappa T}\,\,C_\ell^{\kappa \delta}\,\,C_\ell^{T \gamma}\,\,C_\ell^{\gamma \delta}\,\,C_\ell^{\kappa \gamma}\,\,C_\ell^{\gamma \gamma}\,\,C_\ell^{\kappa \kappa}\,\, \right)\, ,
\label{datavec}
\end{equation}
where $T = \Delta T_\mathrm{ISW}$, $\delta = \delta_g$, $\kappa = \kappa_\mathrm{CMB}$ and $\gamma = \gamma_1$ for notational brevity. Furthermore, the power spectrum vectors and the covariance matrices are binned linearly with $\delta \ell = 100$ within a multipole range $\Delta \ell = [10^2, 10^3]$. Note that since we examine the effect of combining different auto- and cross-power spectra, the combination of power spectra and the dimensionality $d$ in the covariance matrix and the above vectors depend on the considered probes.\\
\\
We sample the joint likelihood given by equation (\ref{likelihood}) in a Monte Carlo Markov Chain (MCMC) to obtain cosmological parameters constraints using  \textsc{uhammer}\footnote{\href{http://cosmo-docs.phys.ethz.ch/uhammer/}{http://cosmo-docs.phys.ethz.ch/uhammer/}}, which is based on the \textsc{emcee} sampler \cite{emcee}. Therefore we vary 5 cosmological parameters $\{h, \Omega_m, \Omega_b, n_s, \sigma_8 \}$ and up to 4 nuisance parameters $\{m_{\Delta T_\mathrm{ISW}}, m_{\kappa_\mathrm{CMB}}, m_{\delta_g}, m_{\gamma_1}\}$, depending on the probes considered for the analysis. We assume flat priors for all the parameters with width $h \in \left[0.2, 1.2\right]$, $\Omega_m \in \left[0.1, 0.7\right]$, $\Omega_b \in \left[0.01, 0.09\right]$, $n_s \in \left[0.1, 1.8\right]$, $\sigma_8 \in \left[0.4, 1.5\right]$ and $m_i \in \left[-0.2, 0.2\right]$. Note that we do not observe any change in the resulting parameter constraints for any combination of probes when increasing the (flat) prior range for the nuisance parameters $m_i$ beyond $\left[-0.15, 0.15\right]$. Our choice of flat priors $m_i \in \left[-0.2, 0.2\right]$ is therefore not compromising our results and leading to a faster convergence of our MCMC runs compared to when using broader priors.\\
\\
\textbf{Cosmological constraints.} Figure \ref{single_probes_sim} shows parameter constraints for $\Omega_m$ and $\sigma_8$ (the associated 68\% and 95\% confidence limits) by using the auto-power spectra of the different probes separately and the corresponding simulation-based non-Gaussian covariance matrices. The combination of the different probes potentially allows us to break the degeneracy between $\Omega_m$ and $\sigma_8$. As can be seen from the figure, the contours computed from the $\Delta T_\mathrm{ISW}$ auto-power spectrum are broader than the ones obtained from the other auto-power spectra. 
%Note that the signal of the temperature anisotropies due to the ISW effect is weak compared to the full CMB temperature anisotropies and is %therefore only observable through the cross-correlation with other tracers of the LSS. 
The ISW effect leads to secondary CMB anisotropies on linear scales, which affects the CMB temperature power spectrum on large scales. On such scales, the primordial anisotropies and cosmic variance dominate the signal and therefore complicate a direct detection of the ISW signal in the full CMB temperature field. In order to detect the ISW signal, one can consider the cross-correlation of the CMB temperature fluctuations with other tracers of the gravitational potential \cite{Crittenden1996}. Thus, the inclusion of the $\Delta T_\mathrm{ISW}$ auto-power spectrum in our analysis remains somewhat theoretical.
\\
\\
In Figure \ref{single_autos}, we show the constraints in the $\Omega_m - \sigma_8$ plane when adopting a Gaussian covariance matrix (green contours) and a non-Gaussian covariance matrix measured from simulations (blue contours). As seen in all four panels for each individual probe, the effect of the non-Gaussian error contribution in the covariance matrix (in the absence of added noise, such as shape noise for weak lensing shear) is crucial for an accurate inference of cosmological parameters. The impact of non-Gaussian components in the covariance matrix in general increase the size of the obtained parameters constraints, as previously observed in Sato \textit{et al.} \cite{Sato2013} for the weak lensing power spectrum. More quantitatively and concerning our results, the 95\% confidence levels for the parameters $\Omega_m$ and $\sigma_8$ increase by $\sim 10 \%, \sim 20 \%, \sim 20 \%$ and $\sim 40 \%$ for the probes $\Delta T_\mathrm{ISW}, \kappa_\mathrm{CMB}, \delta_g, \gamma_1$ respectively when changing from using a Gaussian to using a non-Gaussian covariance matrix.
\smallbreak
We further study the impact of different probe-combinations on the cosmological parameter constraints in the context of adopting a Gaussian or non-Gaussian covariance matrix. In Figure \ref{4probes_nonuisanceplot}, we show the integrated parameter constraints obtained when performing a joint fit of the auto-power spectra $\boldsymbol{C}_\ell = \left(C_\ell^{TT}, C_\ell^{\delta \delta}, C_\ell^{\gamma \gamma}, C_\ell^{\kappa \kappa} \right)$ with total dimension $d=36$ (each power spectrum has 9 multipole bins, i.e. dimension $d = 9$) and jointly fitting all auto- and cross-power spectra of total dimension $d = 90$ (given by equation (\ref{datavec})), for a Gaussian and non-Gaussian covariance matrix respectively. The parameter contours displaying the 4 varied nuisance parameters are shown in Appendix \ref{appendix_constraints}.
\smallbreak
For both choices of covariance matrices the inclusion of the cross-correlations increases the information gain in the parameter constraints significantly. This effect is enhanced through our use of additional nuisance parameters for each probe. As one can clearly see from Figure \ref{4probes_nonuisanceplot}, adopting a non-Gaussian covariance matrix increases the size of the inferred constraints compared to when using a Gaussian approximation. We observe this effect when fitting only the auto-power spectra (shown in the left panel of Figure \ref{4probes_dualplot}) and also when including the cross-power spectra in the analysis (shown in the right panel of Figure \ref{4probes_dualplot}). This behavior is expected, as the non-Gaussian contributions introduce non-diagonal elements in the covariance matrix, which in turn lead to broader constraints in general. 
\smallbreak
In order to investigate our results for the different setup configurations, we compute the change in area of the 95\% confidence levels between using a Gaussian and using a simulation-based non-Gaussian covariance matrix. Figure \ref{area_plot_cross_auto} shows the change in area averaged over all cosmological parameter combinations for different probe-combinations. In general, we observe that the impact of the non-Gaussian error contribution to the covariance matrix becomes increasingly important as further cosmological probes are incorporated in the analysis. Invoking only the auto-power spectra, we observe that the mean area of our obtained contours increase by a factor of $\sim 3$. When we further include the cross-power spectra, the impact of the non-Gaussian covariance matrix on the size of the contours is smaller: The additional constraining power coming from the cross-correlations between the probes counteracts the effect of adding non-Gaussian corrections to the covariance matrix to some degree. We expect this effect to be enhanced by our use of nuisance parameters. Considering the $\Omega_m\, - \, \sigma_8$ plane, we observe a increase by a factor of $\sim 3.4$ and $\sim 2.5$ when using only the auto-power spectra and the auto- and cross-power spectra respectively.
\section{Conclusion}
\label{conclusion}
In the present work, we have added to our \textsc{UFalcon} package the functionality to generate a \textit{set} of full-sky maps for different cosmological probes in a self-consistent way based on the same underlying simulated matter density field with a minimal runtime of $\sim 1$ walltime-hour. This set contains full-sky maps of different cosmological probes for a redshift range between $z=0.0$ and $1.75$ such as weak lensing shear, galaxy clustering including RSD, CMB lensing and CMB temperature anisotropies from the ISW effect.
\smallbreak
The past-lightcone is thereby constructed by first replicating the simulation volume 6 times along each axis and then performing a weighted projection of the DM particles onto the sphere for cosmic shear, galaxy clustering and CMB convergence. In order to obtain continuous full-sky maps of the ISW induced temperature fluctuations and CMB lensing potential maps, the 3D-matter density field is interpolated and evaluated along each line of sight. The replication procedure allows us to cover a large enough survey volume with a sufficiently high resolution. We increase the number of quasi-independent realizations obtained from one $N$-Body simulation by applying random transformations to the periodically arranged density fields. A version of the \textsc{UFalcon} package containing the weak gravitational lensing part of the pipeline is publicly available and a short description is given in section \ref{codebase}.
\smallbreak
We then computed various statistical quantities, such as the auto- and cross- spherical harmonic power spectra between the different simulated full-sky maps. We find better than $2\%$ agreement between the analytical predictions and our simulation-based auto- and cross-power spectrum results in the multipole range $\ell = 10^2$ to $10^3$ for the probes $\gamma$, $\delta_g$ and $\kappa_\mathrm{CMB}$. Our results agree within $\sim 5\%$ to the analytical predictions in the same multipole range when considering auto- and cross-correlations including $\Delta T_\mathrm{ISW}$, with the exception of the cross-power spectrum $C_{\ell}^{\Delta T_\mathrm{ISW} \gamma}$, which shows a lower agreement. From an ensemble of 630 simulated realizations, we have estimated a multi-probe covariance matrix for the multipole range $\ell \in [10^2, 10^3]$. 

We further have analysed the impact of the non-Gaussian covariance matrix based on simulations on cosmological parameter constraints by performing a joint likelihood analysis for a stage-IV-like survey geometry. Note that we have neglected any additional survey-specific noise-contribution to the covariance matrix in order to isolate the effect of the non-Gaussian term. We first studied the impact of using a non-Gaussian covariance matrix on the parameter constraints when performing separate inferences for each auto-power spectrum of the different probes. Here we observe an increase of $\sim 10 \%, \sim 20 \%, \sim 20 \%$ and $\sim 40 \%$ in the area of the 95\% confidence level in the $\Omega_m - \sigma_8$ plane for the probes $\Delta T_\mathrm{ISW}, \kappa_\mathrm{CMB}, \delta_g$ and $\gamma$ respectively when using a non-Gaussian covariance matrix instead of a Gaussian approximation.
\smallbreak
Furthermore, we studied the effect of different probe-combinations in a joint likelihood analysis using both a Gaussian and a simulation-based non-Gaussian multi-probe covariance matrix. We have therefore introduced a multiplicative bias parameter for each probe considered. From this analysis, we observe that the inclusion of the cross-correlation significantly increases the information gain. The importance of using a covariance matrix which includes non-Gaussian contributions compared to using a Gaussian approximation becomes even more apparent in the context of a multi-probe analysis: We observe a mean increase of the 95\% confidence level area over all cosmological parameters by a factor of $\sim 3$ and $\sim 2$ when using only the auto-power spectra and when including auto- and cross-power spectra in the analysis respectively.
\newpage
The combination of different probes of the large-scale structure of the Universe can lead to significant information gain in cosmological parameters. Further taking into account the cross-correlations between multiple probes based on different physical fields offer more stringent tests of the systematics and can potentially reveal new physics. This is crucial as future surveys become increasingly systematics limited. As combining multiple probes increases the size of the data vector and the retrievable information content, a accurate estimate of the covariance matrix is of uttermost importance. Although using a Gaussian approximation might be sufficient for some ongoing surveys, it will ultimately be crucial to take into account non-Gaussianity for future large surveys.
%The main intent of this paper is to introduce our extended \textsc{UFalcon} pipeline and to perform a first demonstration of the importance of including non-Gaussian contributions in a multi-probe covariance matrix when performing a joint parameter inference. This framework lays the foundation for further theoretical studies of the full non-Gaussian statistics recovered from simulations, such as:
%\begin{itemize}
%\item The impact on parameter constraints for varying $\ell$-cuts and systematics used in the multi-probe parameter inference.
%\item A more detailed analysis of the impact on the lensing of the CMB temperature and polarization fields.
%\end{itemize}
\acknowledgments

We thank Douglas Potter and Joachim Stadel from the University of Z\"urich for the distribution of the code \textsc{PkdGrav3} and for their continuous support with the code. We would also like to thank Bj\"{o}rn Malte Sch\"{a}fer and Aurel Schneider for very helpful discussions concerning the pipeline and Christiane Lorenz for careful reading of the manuscript. Furthermore, we thank Uwe Schmitt for his help with the computing implementation. AR is grateful for the hospitality of KIPAC at Stanford University/SLAC where part of his contribution was made. We acknowledge support by the Swiss National Science Foundation grant 200021\_169130. AA is supported by a Royal Society Wolfson Fellowship. This research made use of \texttt{IPython}, \texttt{NumPy}, \texttt{SciPy}, \texttt{Matplotlib}, \texttt{Healpy}, \texttt{PyCosmo}, \texttt{UHammer} and \texttt{GetDist}.


\bibliography{bibliography_probes_paper}
\bibliographystyle{ieeetr}

%%%%%%%%%%%%%%%%%%%%%%%%%%%%%%%%%%%%%%%%%%%%%%%%%%%%%%%%%%%

\begin{appendix}

\section{Analytical Prediction for the ISW Auto-Correlation}
\label{appendixA}
This detailed derivation of the auto-correlation of the CMB temperature anisotropies due to the ISW effect follow closely Appendix F in Nicola \textit{et al.} \cite{Nicola2016}, where the analytical prediction for the cross-correlation between the CMB temperature anisotropies and weak lensing shear is derived. The ISW effect leads to the generation of CMB temperature anisotropies given by equation (\ref{iswdef}), which can be decomposed into spherical harmonics with multipole coefficients
\begin{equation}
\Delta T_{\mathrm{ISW}, \ell m} = 4 \pi i ^\ell \, 2\, T_\mathrm{CMB} \int^{\eta_0}_{\eta_r} \mathrm{d} \eta \int \frac{\mathrm{d} \vec{k}}{(2 \pi)^3} \frac{\mathrm{d}}{\mathrm{d} \eta} \Phi (\vec{k}, z) j_\ell (k \chi(z)) Y^\ast (\hat{n}_k) \, .
\label{iswcoeff}
\end{equation}
The spherical harmonic power spectrum for the CMB temperature anisotropies due to the ISW effect is defined as
\begin{equation}
\left< \Delta T_{\mathrm{ISW}, \ell m} \Delta T_{\mathrm{ISW}, \ell' m'} \right> = C_\ell^\mathrm{ISW} \delta_{\ell \ell'} \delta_{m m'} \, ,
\end{equation}
and can be written as
\begin{equation}
\begin{split}
C_\ell^\mathrm{ISW} = (4 \pi)^2 \, 4\, T_\mathrm{CMB}^2 &\left< \int^{z_\ast}_0 \mathrm{d} z \int \frac{k^2 \mathrm{d}k}{(2 \pi)^3} \frac{\mathrm{d}}{\mathrm{d} z} \left[ D(z) (1+z) \right] \Phi (\vec{k}, z = 0) j_\ell (k \chi(z))\right.\\
&\left. \times \int^{z_\ast}_0 \mathrm{d} z' \int \frac{k'^2 \mathrm{d}k'}{(2 \pi)^3} \frac{\mathrm{d}}{\mathrm{d} z'} \left[ D(z') (1+z') \right] \Phi (\vec{k}', z' = 0) j_\ell (k' \chi(z')) \right> \, ,
\end{split}
\label{cellisw}
\end{equation}
where we used equation (\ref{iswcoeff}) and the fact that in linear perturbation theory one can separate the time- and scale-dependence of the gravitational potential $ \Phi (k, z) = \Phi(k, z=0) D(z) (1 + z)$. The linear power spectrum of the gravitational potential at present time is defined as
\begin{equation}
\left< \Phi(\vec{k}, z=0) \Phi(\vec{k}', z'=0) \right> = (2 \pi)^3 P^\mathrm{lin}_{\Phi \Phi} (k, z = 0) \delta^\mathrm{D} (\vec{k} - \vec{k}') \, ,
\label{phidef}
\end{equation}
and is related to the matter power spectrum through Poisson's equation as
\begin{equation}
P^\mathrm{lin}_{\Phi \Phi} (k, z = 0) = \left( \frac{3}{2} \right)^2 \frac{\Omega_m^2 H_0^4}{c^4} \frac{P^\mathrm{lin}_{\delta \delta} (k, z = 0)}{k^4} \, ,
\label{phimat}
\end{equation}
where $\delta^\mathrm{D} (\vec{k} - \vec{k}')$ denotes the Dirac delta function. The expression for the spherical harmonic power spectrum given by equation (\ref{cellisw}) can be further simplified using equations (\ref{phidef}) and (\ref{phimat}) to

\begin{equation}
\begin{split}
C_\ell^\mathrm{ISW} &= (4 \pi)^2 \,  T_\mathrm{CMB}^2 \left( \frac{3 \Omega_m H_0^2}{c^2} \right)^2 \int^{z_\ast}_0 \mathrm{d} z \int \frac{k^2 \mathrm{d}k}{(2 \pi)^3} \frac{\mathrm{d}}{\mathrm{d} z} \left[ D(z) (1+z) \right] \\
& \times \int^{z_\ast}_0 \mathrm{d} z' \frac{\mathrm{d}}{\mathrm{d} z'} \left[ D(z') (1+z') \right] \frac{P^\mathrm{lin}_{\delta \delta} (k, z = 0)}{k^4} j_\ell (k \chi(z)) j_\ell (k \chi(z')) \, .
\end{split}
\label{ciswexact}
\end{equation}
We further use the Limber approximation (\cite{Limber1953}, \cite{Kaiser1992}, \cite{Kaiser1998}) to simplify the calculation of equation (\ref{ciswexact}) giving
\begin{equation}
\begin{split}
C_{\ell}^\mathrm{ISW} &= T_\mathrm{CMB}^2 \left( \frac{3 \Omega_m H_0^2}{c^2} \right)^2 \frac{1}{(\ell + 1/2)^4} \int \mathrm{d}z \frac{\mathrm{d}}{\mathrm{d}z} \left[ D(z) (1+z) \right]^2 \chi(z)^2\\
& \times P^{\mathrm{lin}}_{\delta \delta} \left(k = \frac{\ell + 1/2}{\chi(z)}, 0 \right) \, .
\end{split}
\end{equation}

\section{Shot Noise Estimation}
\label{snappendix}

\begin{figure}[htbp!]
\centering
\includegraphics[width=10cm]{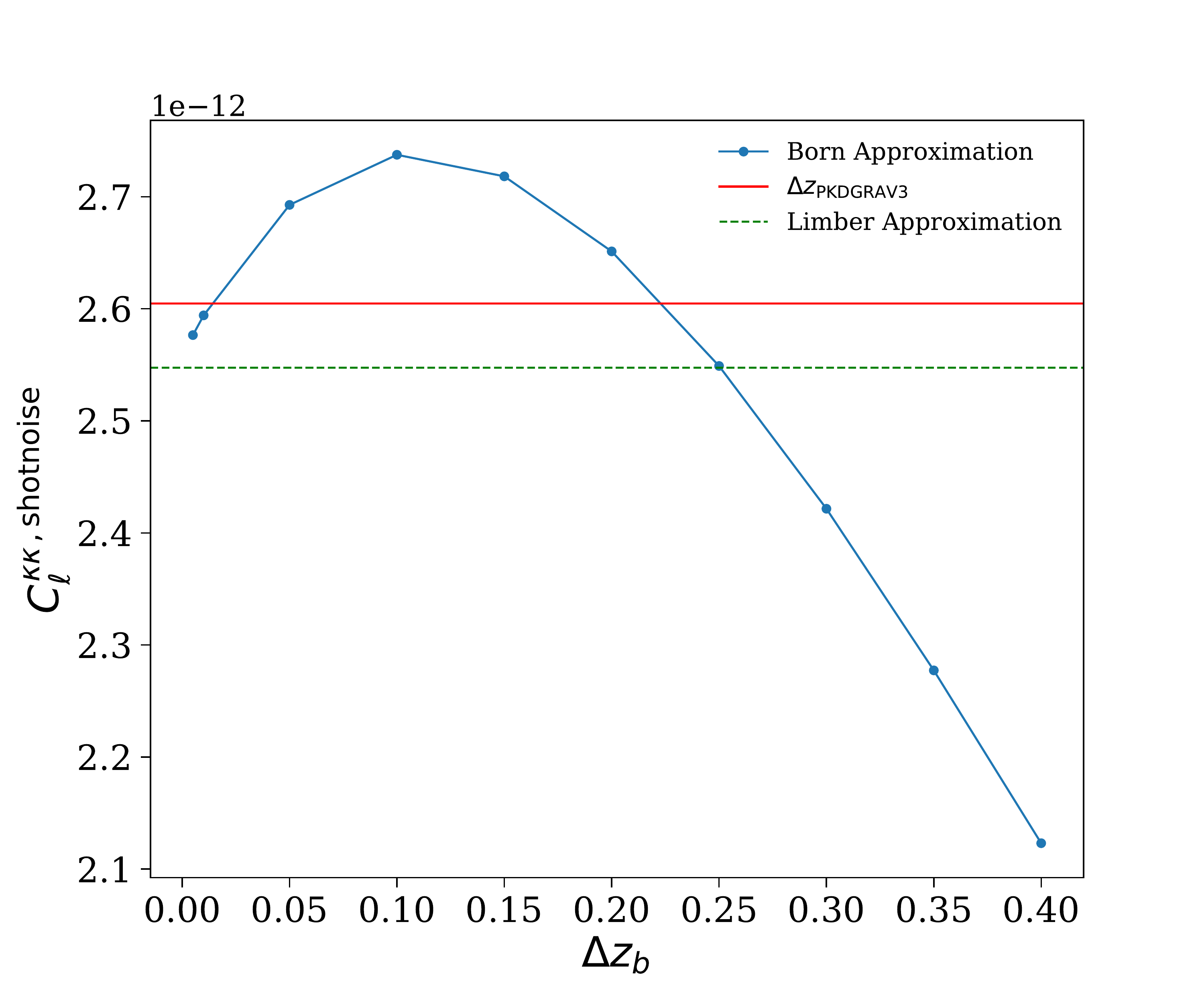}
\caption{The shot noise based on the Born approximation is given by the blue line and depends on the choice of shell thickness $\Delta z_b$, which generally increases when choosing thinner shells. The drop in power for shells smaller than $\sim 0.1$ is explained by the $1/ \mathcal{D}^2 (z_b)$ factor in equation (\ref{snborn}). The horizontal red solid line represents the shot noise obtained when using the non-constant redshift-spacing from our \textsc{PkdGrav3} simulation run. The shot noise based on the Limber approximation is given by the horizontal green dashed line and is independent of shell-thickness.}
\label{shotnoise_label}
\end{figure}

In this section we will derive a estimation for the shot noise contribution to the convergence spherical harmonic power spectrum $C_\ell^{\mathrm{sn}, \, \mathrm{Born}}$ based on the Born approximation. The shot noise of each shell is inverse proportional to the mean particle density of the sphere, which can be calculated analytically. This calculation assumes that the shot noise of different shells are uncorrelated and that the signal and the shot noise are also uncorrelated. With this assumption, we can write the overdensity on a spherical shell as
\begin{equation}
\delta = \frac{\rho - \bar{\rho}}{\bar{\rho}} \, ,
\end{equation}
where the density $\rho$ describes a Poisson point process on that shell and is proportional to the number of particles in the simulation $N_p$. The Poisson shot noise contribution to the spherical harmonic power spectrum is then given by
\begin{equation}
C_{\ell}^\mathrm{shotnoise} = \frac{4 \pi}{N_p} \, .
\end{equation}
Our pipeline \textsc{UFalcon} computes the convergence from $N$-Body simulation outputs using the Born approximation as
\begin{equation}
\kappa(\mathrm{pix}) \approx \frac{3}{2} \Omega_m \sum_b W_b \frac{H_0}{c} \left[ \frac{N_\mathrm{pix}}{4 \pi} \frac{V_\mathrm{sim}}{N_p^\mathrm{sim}} \left( \frac{H_0}{c} \right)^2 \frac{n_p}{\mathcal{D}^2(z_b)} \right] \, ,
\end{equation}
where $n_p = n_p(\mathrm{pix}, \Delta z_b)$ is the number of particles per pixel in the shell $\Delta z_b = z_b^\mathrm{max} - z_b^\mathrm{min}$. The comoving volume of shell $b$ (in Mpc) can be written as
\begin{equation}
V_b = \frac{4}{3} \pi \left( \frac{c}{H_0} \right)^3 \left[ \mathcal{D}^3(z_b^\mathrm{max}) - \mathcal{D}^3(z_b^\mathrm{min}) \right] \, .
\end{equation}
The mean density of a shell (per pixel) is then given by
\begin{equation}
\bar{n}_p = \rho \cdot \frac{V_b}{N_\mathrm{pix}} = \left( \frac{N_p^\mathrm{sim}}{V_\mathrm{sim}} \right) \frac{V_b}{N_\mathrm{pix}} \, .
\end{equation}
The terms in the expression for the convergence can then be rearranged in order to obtain the shot noise contribution of each individual shell
\begin{equation}
\begin{split}
\kappa(\mathrm{pix}) & \approx \frac{3}{2} \Omega_m \sum_b W_b \frac{H_0}{c} \left[ \frac{N_\mathrm{pix}}{4 \pi} \frac{V_\mathrm{sim}}{N_p^\mathrm{sim}} \left( \frac{H_0}{c} \right)^2 \frac{\bar{n}_p}{\mathcal{D}^2(z_b)} \frac{n_p}{\bar{n}_p} \right] \\
& = \frac{3}{2} \Omega_m \sum_b W_b \left( \frac{H_0}{c} \right)^3 \frac{1}{4 \pi} \frac{V_b}{\mathcal{D}^2(z_b)} \left[\frac{n_p}{\bar{n}_p} \right] \, .
\end{split}
\end{equation}
The shot noise contribution to the spherical harmonic power spectrum of the convergence using the Born approximation is therefore given by
\begin{equation}
C_\ell^{\mathrm{sn}, \, \mathrm{Born}} = \left[\frac{3}{2} \Omega_m \left( \frac{H_0}{c} \right)^3 \frac{1}{4 \pi} \right]^2 \sum_b \left[ W_b \frac{V_b}{\mathcal{D}^2(z_b)} \right]^2 \frac{4 \pi}{N_b} \, ,
\label{shotnoisecl}
\end{equation}
where $N_b$ is the number of particles in the shell $b$ and can be expressed as
\begin{equation}
N_b = \frac{N_p^\mathrm{sim}}{V_\mathrm{sim}} V_b \, .
\end{equation}
In Figure \ref{shotnoise_label} we show the comparison of the shot noise contribution to the weak lensing convergence power spectrum based on the Limber (equation (\ref{snlimber}) in section \ref{shotnoise}) and Born (equation \ref{shotnoisecl}) approximations.

%The shot noise contribution given by equation (\ref{shotnoisecl}) is $\Delta z_b$
\section{Correlated Gaussian Maps}
\label{GiannantonioRoutine}

Here we describe our routine used in section \ref{forecast} to generate sets of correlated Gaussian spin-0 maps from prescribed auto- and cross-power spectra, which can be used to estimate the power spectrum covariance matrix.

Our implementation is based on the general algorithm presented in Giannantonio \textit{et al.} \cite{Giannantonio2008} and implemented and extended to spin-2 field in Nicola \textit{et al.} \cite{Nicola2016, Nicola2017}. This algorithm is based on the \textsc{Healpix} subroutine \texttt{synfast}, which is used to generate maps from an input spherical harmonic power spectrum $C_\ell^{ii}$ and probe $i$. Thereby a random phase $\xi$ with mean zero $\left< \xi \right> = 0$ and unit variance $\left< \xi \xi^* \right> = 1$ is assigned to each spherical harmonic mode $\ell$. The correct power spectrum is then ensured by setting
\begin{equation}
a_{\ell m}^i = \sqrt{C_\ell^{ii}} \xi \, ,
\end{equation}
which is effectively implemented in \texttt{synfast}. In the general case, we consider $n$ maps and therefore require $n$ different phases, which are assumed uncorrelated  $\left< \xi_i \xi_{j}^* \right> = \delta_{ij}$. The spherical harmonics are then given by
\begin{eqnarray} 
a_{\ell m}^1 &=& \xi_1 T_{11} \\
a_{\ell m}^2 &=& \xi_1 T_{21} + \xi_2 T_{22} \\
a_{\ell m}^3 &=& \xi_1 T_{31} + \xi_2 T_{32} + \xi_3 T_{33} \\
\cdots \, ,
\end{eqnarray} 
which satisfy the constraints $\left< a_{\ell m}^i a_{\ell' m'}^{j*} \right> = C_\ell^{ij} \delta_{\ell \ell'} \delta_{m m'}$. This gives $n(n+1)/2$ equations for the unknown amplitudes $T_{ij}$ given by
\begin{eqnarray} 
C_\ell^{11} &=& T^2_{11} \\
C_\ell^{12} &=& T_{11} T_{21} \\
C_\ell^{22} &=& T^2_{21} + T^2_{22} \\
C_\ell^{13} &=& T_{11} T_{31} \\
\cdots \, .
\end{eqnarray} 
The general recursive relation for the amplitudes is given by \cite{Giannantonio2008}
\begin{eqnarray} 
T_{ij} &=& \sqrt{C_{\ell m}^{ji} - \sum_{k=1}^{j-1} T^2_{ik}}\quad , \quad \mathrm{if} \quad i=j \\
T_{ij} &=& \frac{C_{\ell m}^{ji} - \sum_{k=1}^{j-1} T_{ik}T_{jk}}{T_{jj}} \quad , \quad \mathrm{if} \quad i>j\, .
\end{eqnarray}
In practice, we have implemented this relation using \texttt{synfast} with $n = 4$ different \textit{random seeds}, corresponding to the different phases $\xi$, to generate correlated Gaussian maps for the probes $\Delta T_\mathrm{ISW}$, $\kappa_\mathrm{CMB}$, $\delta_g$ and $\kappa_\mathrm{smail}$.

\section{Pseudo-$C_{\ell}$ Estimation}
\label{pseudocl_appendix}
It is well known that limited coverage of the celestial sphere poses a real challenge for the measurement of the auto- and cross-power spectra. Since the discovery of the CMB radiation anisotropy by the \textit{COBE} satellite (Smoot \textit{et al.} \cite{Smoot1992}), numerous power spectrum estimators for the analysis of large CMB temperature data sets have been developed (an overview is given in \cite{Efstathiou2004}).

\begin{figure}[htbp!]
\centering
\includegraphics[width=1\textwidth]{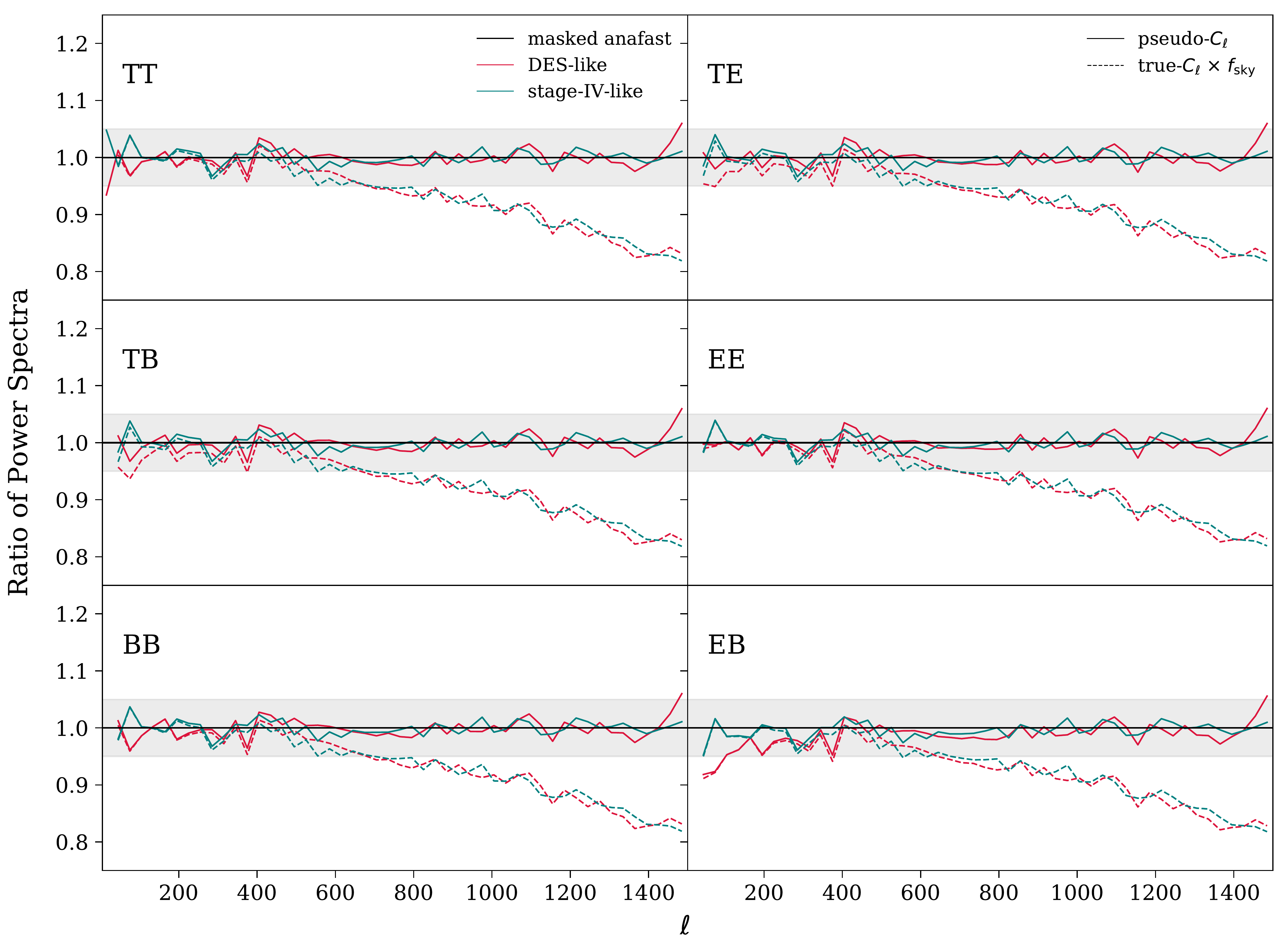}
\caption{Ratio between one realization of the masked power spectra (obtained by applying the mask on a map level) and the pseudo-$C_\ell$ (using equation (\ref{pseudoclmethod}) on the true-$C_\ell$; solid lines) and by multiplying the true-$C_\ell$ with $f_\mathrm{sky}$ (dashed lines). The different panels show the ratios of the weak lensing convergence ($T$) and shear ($E$ and $B$) auto- and cross-spherical harmonic power spectra for a stage IV-like (dark cyan) and a DES Y3-like (red) mask. The grey band denotes 5\% deviation from the masked power spectra.}
\label{pseudocl_ratios}
\end{figure}

Our implementation of the pseudo-$C_\ell$ method is based on \cite{Brown2005} and \cite{Kogut2003}, which applies to temperature- and polarisation-type fields denoted by $T$ and $P$. The effect of finite window functions is then given by \cite{Brown2005}
\begin{equation}
\tilde{T} (\hat{n}) = W_T (\hat{n}) T (\hat{n}) \quad , \quad \tilde{P} (\hat{n}) = W_P (\hat{n}) P (\hat{n}) \, ,
\end{equation}
where $W_T (\hat{n})$ and $W_P (\hat{n})$ denote the window functions for the two field types pointing in direction $\hat{n}$ on the sky and are in general not equal. Note that the window functions vanish outside of the survey area. Within the survey, the window function can also take into account weighting of the field, which depends on the specific survey considered. The observed cut-sky power spectra are related to the underlying full-sky analytical predictions based on \textsc{PyCosmo} by the matrix relation
\begin{equation}
\tilde{\boldsymbol{C}}_\ell = \sum_{\ell'} \boldsymbol{M}_{\ell \ell'} \boldsymbol{C}_{\ell '}^\mathrm{PyCosmo} \, ,
\label{pseudoclmethod}
\end{equation}
where $\boldsymbol{M}$ is the mode-coupling matrix describing the effect of the mask applied to the data. This relation can be expanded as
\begin{equation}
\begin{pmatrix}
\tilde{C}_{\ell}^{TT} \\
\tilde{C}_{\ell}^{TE} \\
\tilde{C}_{\ell}^{TB} \\
\tilde{C}_{\ell}^{EE} \\
\tilde{C}_{\ell}^{EB} \\
\tilde{C}_{\ell}^{BB}
\end{pmatrix} 
= \sum_{\ell '}
\begin{pmatrix}
M_{\ell \ell '}^{TT, TT} & 0 & 0 & 0 & 0 & 0 \\
0 & M_{\ell \ell '}^{TE, TE} & M_{\ell \ell '}^{TE, TB} & 0 & 0 & 0 \\
0 & M_{\ell \ell '}^{TB, TE} & M_{\ell \ell '}^{TB, TB} & 0 & 0 & 0 \\
0 & 0 & 0 & M_{\ell \ell '}^{EE, EE} & M_{\ell \ell '}^{EE, EB} & M_{\ell \ell '}^{EE, BB} \\
0 & 0 & 0 & M_{\ell \ell '}^{EB, EE} & M_{\ell \ell '}^{EB, EB} & M_{\ell \ell '}^{EB, BB} \\
0 & 0 & 0 & M_{\ell \ell '}^{BB, EE} & M_{\ell \ell '}^{BB, EB} & M_{\ell \ell '}^{BB, BB}
\end{pmatrix} 
\begin{pmatrix}
C_{\ell '}^{TT} \\
C_{\ell '}^{TE} \\
C_{\ell '}^{TB} \\
C_{\ell '}^{EE} \\
C_{\ell '}^{EB} \\
C_{\ell '}^{BB}
\end{pmatrix}\, ,
\end{equation}
where the temperature $T$ and the $E$ and $B$ modes represent any spin-0 and spin-2 fields respectively. The coupling terms can be simplified using the symmetry and orthogonality properties of the Wigner-3j symbols \cite{Rotenberg1959}. Our implementation of the non-zero mode-coupling terms is based on equations (A12) - (A17) in \cite{Brown2005}, and is given by the following set of expressions:
\begin{equation}
\begin{split}
M_{\ell \ell '}^{TT, TT} &= \frac{(2 \ell ' + 1)}{4 \pi} \sum_{\ell ''} \left(2 \ell '' + 1 \right) \mathcal{W}_{\ell ''}^{TT}
\begin{pmatrix}
\ell & \ell ' & \ell '' \\
0 & 0 & 0
\end{pmatrix} ^2 \\
M_{\ell \ell '}^{TE, TE} &= M_{\ell \ell '}^{TB, TB} = \frac{(2 \ell ' + 1)}{4 \pi} \sum_{\ell ''} \left(2 \ell '' + 1 \right) \mathcal{W}_{\ell ''}^{TP}
\begin{pmatrix}
\ell & \ell ' & \ell '' \\
0 & 0 & 0
\end{pmatrix}
\begin{pmatrix}
\ell & \ell ' & \ell '' \\
2 & -2 & 0
\end{pmatrix}\\
M_{\ell \ell '}^{EE, EE} &= M_{\ell \ell '}^{BB, BB} = \frac{(2 \ell ' + 1)}{8 \pi} \sum_{\ell ''} \left(2 \ell '' + 1 \right) \mathcal{W}_{\ell ''}^{PP} \left[1 + (-1)^{\ell + \ell ' + \ell ''} \right]
\begin{pmatrix}
\ell & \ell ' & \ell '' \\
2 & -2 & 0
\end{pmatrix} ^2 \\
M_{\ell \ell '}^{EE, BB} &= M_{\ell \ell '}^{BB, EE} = \frac{(2 \ell ' + 1)}{8 \pi} \sum_{\ell ''} \left(2 \ell '' + 1 \right) \mathcal{W}_{\ell ''}^{PP} \left[1 - (-1)^{\ell + \ell ' + \ell ''}\right]
\begin{pmatrix}
\ell & \ell ' & \ell '' \\
2 & -2 & 0
\end{pmatrix} ^2 \\
M_{\ell \ell '}^{EB, EB} &= \frac{(2 \ell ' + 1)}{4 \pi} \sum_{\ell ''} \left(2 \ell '' + 1 \right) \mathcal{W}_{\ell ''}^{PP}
\begin{pmatrix}
\ell & \ell ' & \ell '' \\
2 & -2 & 0
\end{pmatrix} ^2 \, .
\end{split}
\label{couplingterms}
\end{equation}
The kernels in equation (\ref{couplingterms}) depend on the geometry of the cut sky and are expressed in terms of the auto- and cross-power spectra of the window functions $W_T (\hat{n})$ and $W_P (\hat{n})$, which are given by
\begin{equation}
\begin{split}
\mathcal{W}_{\ell}^{TT} &= \frac{1}{2 \ell + 1} \sum_m w_{\ell m}^T  \left(w_{\ell m}^T \right)^*\, , \\
\mathcal{W}_{\ell}^{PP} &= \frac{1}{2 \ell + 1} \sum_m w_{\ell m}^P  \left(w_{\ell m}^P \right)^*\, , \\
\mathcal{W}_{\ell}^{TP} &= \frac{1}{2 \ell + 1} \sum_m w_{\ell m}^T  \left(w_{\ell m}^P \right)^*\, ,
\end{split}
\end{equation}
where $w_{\ell m}^T$ and $w_{\ell m}^P$ are the spherical harmonic coefficients of the window functions
\begin{equation}
w_{\ell m}^T = \int \mathrm{d} \hat{n} W_T(\hat{n}) Y_{\ell m} ^* \quad , \quad w_{\ell m}^P = \int \mathrm{d} \hat{n} W_P(\hat{n}) Y_{\ell m} ^* \, .
\end{equation}
Note that for the term $M_{\ell \ell '}^{EE, BB}$ we have adopted the correct sign given by (A17) in \cite{Kogut2003}, which differs from the sign used in equation (A15) in \cite{Brown2005}. Furthermore, we have explicitly written out the prefactors $\left(2 \ell '' + 1 \right)$ in the coupling terms (analogous to equation (A31) in \cite{Hivon2002}). We numerically compute the Wigner-3j symbols using the \textsc{pyshtools}\footnote{\href{https://pypi.org/project/pyshtools/}{https://pypi.org/project/pyshtools/}} package.
\begin{figure}[htbp!]
    \centering
    \subfloat[]{{\includegraphics[width=6.5cm]{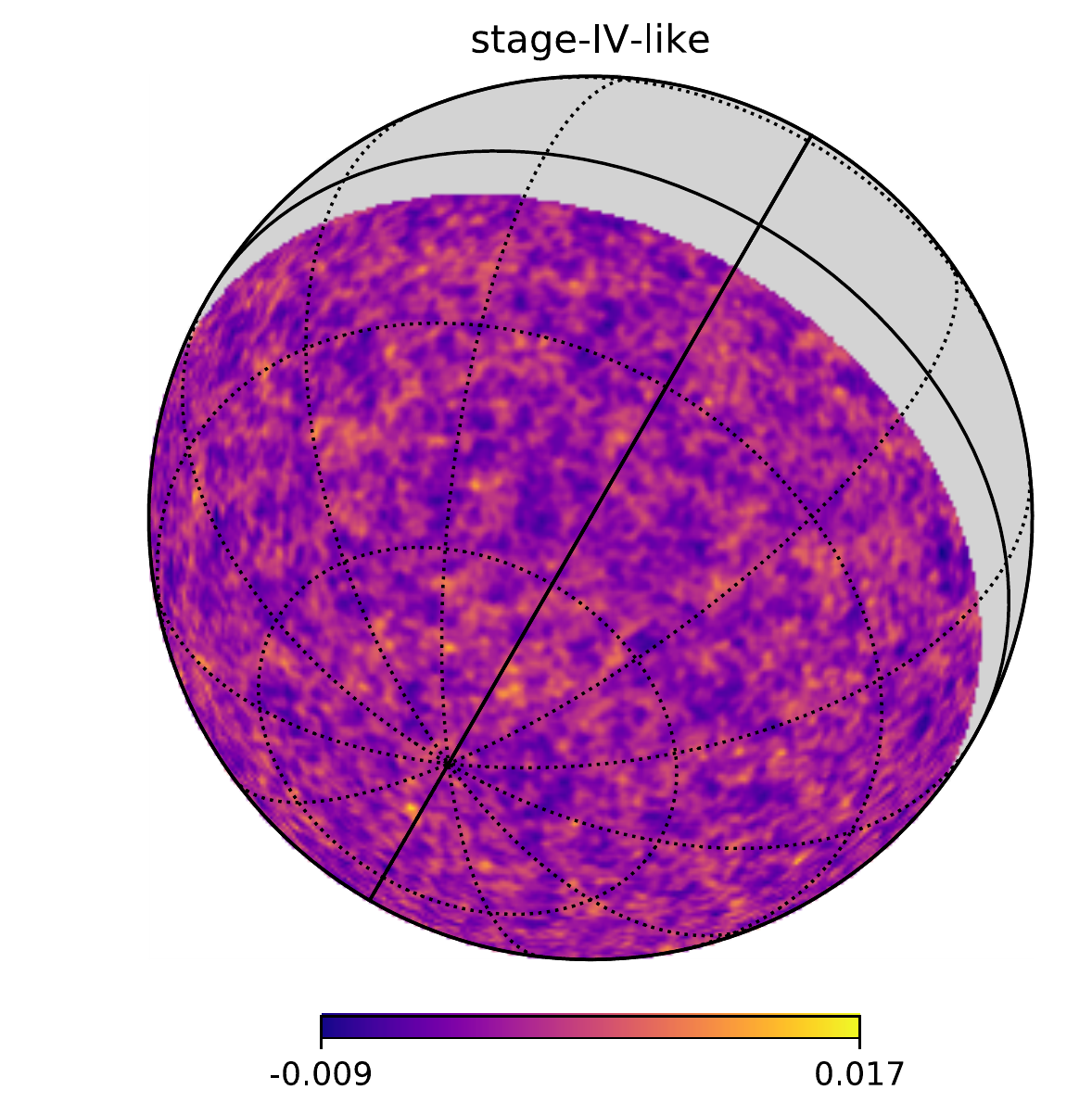} }}%
    \,
    \subfloat[]{{\includegraphics[width=6.5cm]{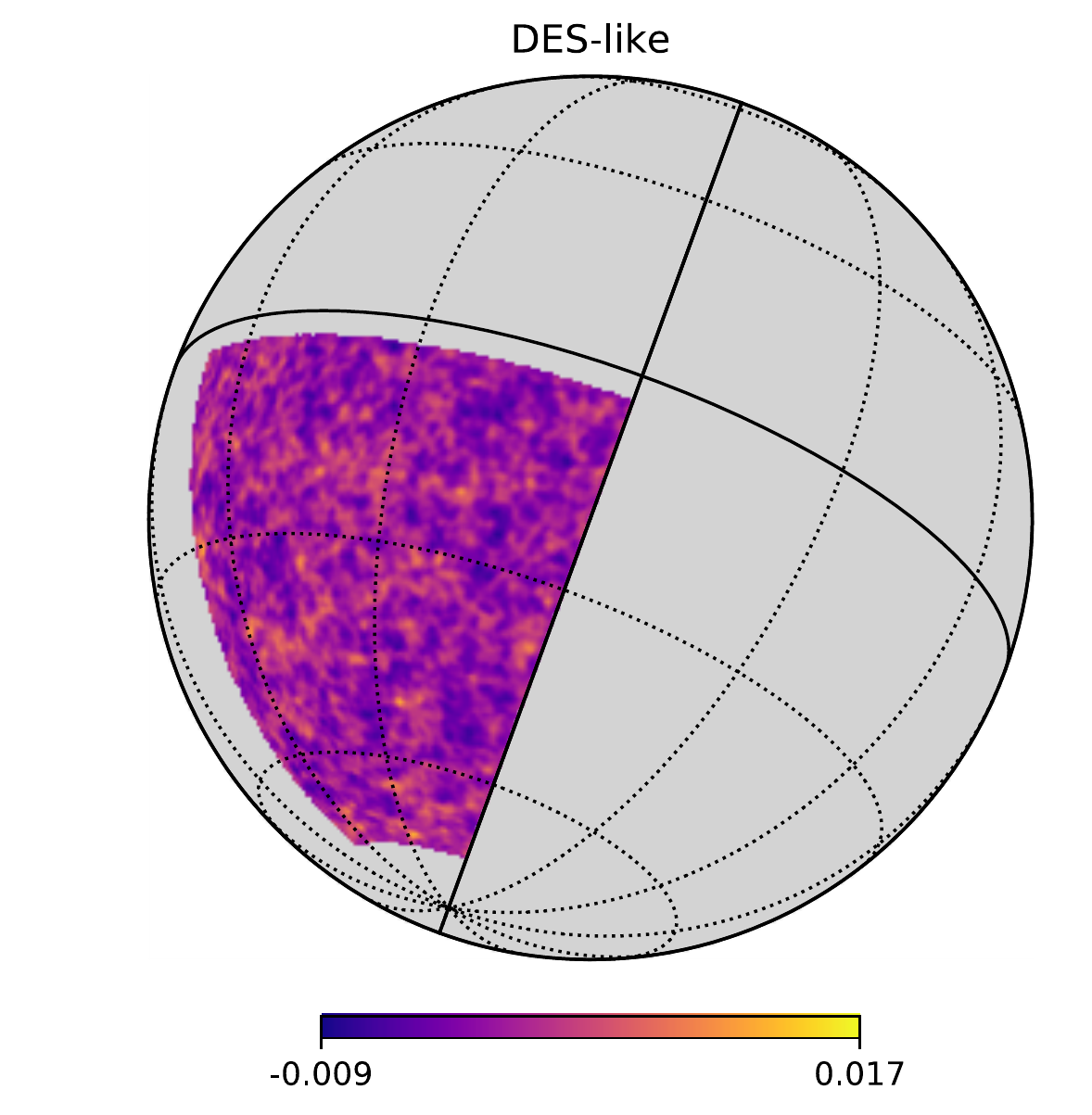} }}%
    \caption{Weak lensing mass maps for different sky cuts. Left: stage IV-like geometry as used in this work. Right: DES Y3-like survey footprint.}
    \label{masks}
\end{figure}
\\
For testing purposes, we applied our pseudo-$C_\ell$ method on the weak lensing convergence (denoted by $T$) and shear (denoted by $E$ and $B$) spherical harmonic power spectra. Hereby we used a stage-IV-like and DES Y3-like survey geometry, as shown in Figure \ref{masks}. We first computed analytical predictions for the power spectra using \textsc{PyCosmo}, denoted by true-$C_\ell$. To obtain a reference for the effect of the sky cut, we generated Gaussian maps from the true-$C_\ell$'s using \texttt{synfast}, applied the masks on a map level and computed the masked power spectra using \texttt{anafast}. In Figure \ref{pseudocl_ratios}, we show the ratios between the masked power spectra described above and the pseudo-$C_\ell$ based on our method and when simply multiplying the true-$C_\ell$'s with $f_\mathrm{sky}$. Note that we only consider one synthetic Gaussian realization for the masked power spectrum. We observe that our results for the pseudo-$C_\ell$'s closely follow the masked power spectra. The results using the true-$C_\ell \, \times \, f_\mathrm{sky}$ are significantly lacking power at higher multipoles, starting at $\ell \sim 500$. A simple multiplication with $f_\mathrm{sky}$ only corrects the true-$C_\ell$'s for the covered fraction of the sky, but does not take into account the coupling between the different modes, which is especially important when applying complicated sky cuts.

\pagebreak

\section{CMB Lensing}
\label{cmb_lensing_appendix}
The weak gravitational lensing of the CMB represents one of the most important mechanisms which can generate secondary anisotropies and induce non-Gaussian features in the observed CMB sky \cite{Lewis2006}. In this section we show the capability of our CMB lensing potential maps, generated using \textsc{UFalcon}, to perform weak gravitational lensing of the different CMB fields without emphasis on quantitative results. 

\begin{figure}[htbp!]
\centering
\includegraphics[width=1\textwidth]{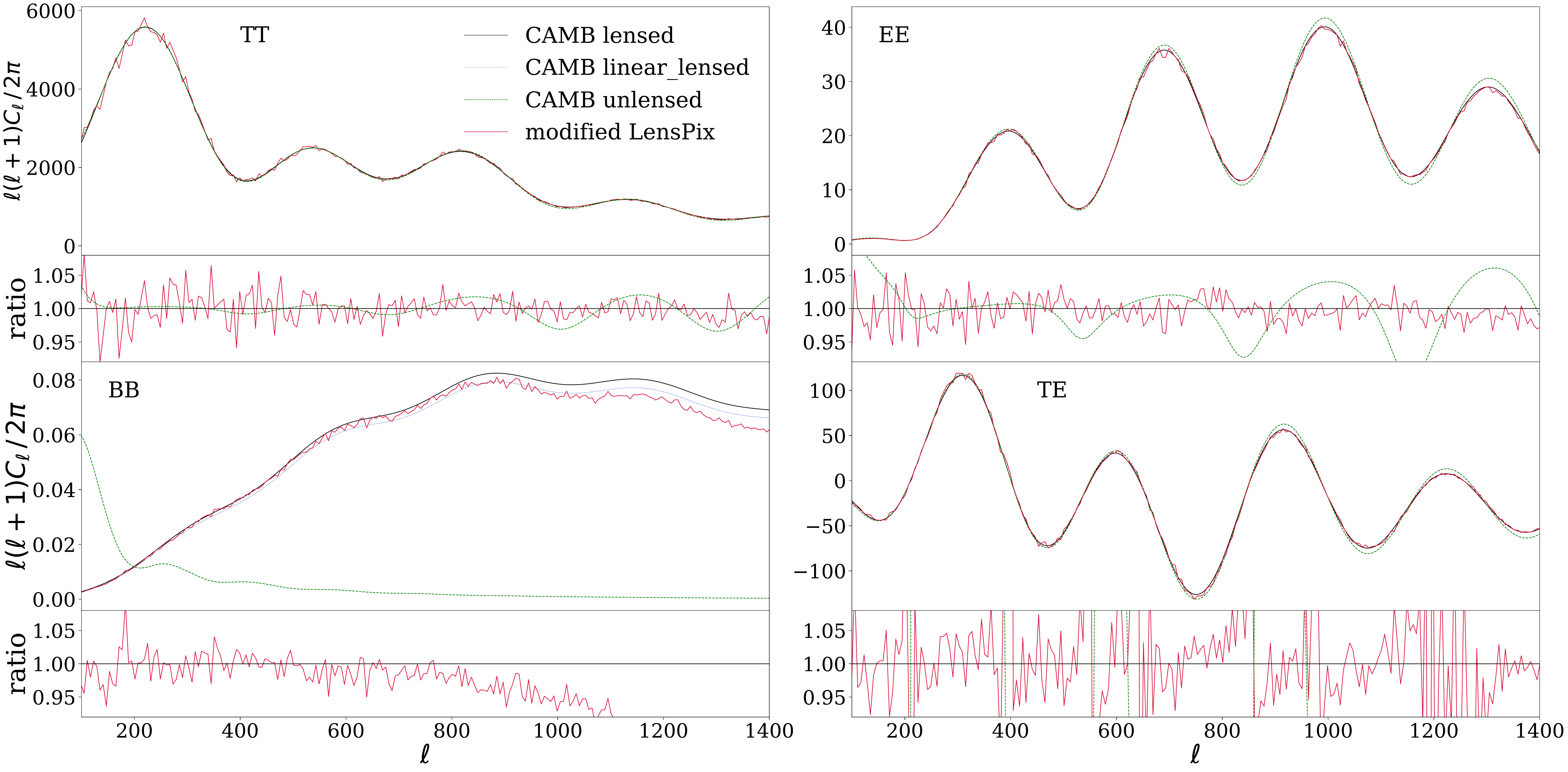}
\caption{Simulated lensed $TT$-, $EE$-, $TE$- and $BB$-spherical harmonic power spectra (in $\mu K^2$) (solid red lines) generated using the modified-\textsc{LensPix} code with our \textsc{UFalcon}-based deflection angle. Our results are compared to the nonlinearly and linearly lensed and unlensed CAMB predictions (solid black, dotted blue and dashed green lines respectively). Our simulation results correspond to one realization and is binned with $\delta \ell = 10$.}
\label{all_lensing} 
\end{figure}

The temperature and polarisation fields are moved from a initial angular position $\vec{\theta}$ to a new position $\vec{\theta} + \vec{\alpha}$, where $\vec{\alpha} = \nabla_{\hat{n}} \psi$ represents the deflection angle and $\nabla_{\hat{n}}$ is the two-dimensional transverse derivative w.r.t. the line-of-sight pointing in direction $\hat{n} \equiv (\theta, \phi)$ on the sky. This requires accurate interpolation on the pixels (lensed rays are not at centers of pixels), which is why we used the \textsc{LensPix}\footnote{\href{https://cosmologist.info/lenspix/}{https://cosmologist.info/lenspix/}} package \cite{Lewis2005}, \cite{Challinor2005}. The original code uses a Gaussian realization of the lensing potential power spectrum obtained using the publicly available Code for Anisotropies in the Microwave Background (CAMB\footnote{\href{http://camb.info/}{http://camb.info/}}). We modified the \texttt{HealpixInterpLensedMapGradPhi} subroutine in the code in order to read the deflection angle we obtained from our simulation, capturing the non-linear and non-Gaussian features. This approach has previously been applied in \cite{Carbone2009} and \cite{Takahashi2017}. In Figure \ref{intmaps} (in the main text) we show the deflection angle modulus $|\vec{\alpha}| = \sqrt{(\Delta \theta)^2 + (\Delta \phi)^2}$ we obtained by directly performing the angular derivative on our lensing potential map with nside=1024. In Figure \ref{all_lensing} we show all the spherical harmonic power spectra binned with $\delta \ell = 10$ obtained when using \textsc{LensPix} with our simulation-based deflection angle as input. Our results are compared to the lensed and unlensed CMB power spectra based on the full-sky correlation function technique implemented in CAMB \cite{Challinor2005}. Note that we have used $\ell_\mathrm{max} = 1500$ in \textsc{LensPix} and chosen our fiducial cosmological parameters given by equation (\ref{fid_params}).
\\
Our modified-\textsc{LensPix} results for the $TT$-, $EE$- and $TE$- power spectra agree within $5\%$ with the nonlinearly lensed CAMB predictions up to $\ell \sim 1500$, where CMB lensing has led to a transfer of power from larger to smaller scales, smearing out the acoustic peaks. For these cases we do not show the linearly lensed CAMB power spectra, since we expect the nonlinear effects to become important at $\ell < 2500$ \cite{Carbone2009}. In the $BB$-case, the nonlinearities present in the lensing potential are affecting all scales of the lens-induced $B$-mode power spectrum. This effect is expected to be already of the order of $\sim 7.5\%$ for multipoles $\ell \leq 10^3$ \cite{Carbone2009}. Our modified-\textsc{LensPix} $B$-mode results agree to the nonlinearly lensed CAMB prediction up to $\ell \sim 500$ and to the linearly lensed CAMB prediction up to $\ell \sim 10^3$. This result clearly indicates the missing lensing signal in our simulation-based lensing potential, which covers the range from $z = 0$ to $1.75$. Even though the most relevant redshift for CMB lensing is at about $z \sim 1$, we therefore expect to neglect a significant amount of the lensing power. We leave a more detailed treatment of CMB lensing by extending the integration range in our pipeline to future work.

\section{Parameter Constraints with Nuisance Parameters}
\label{appendix_constraints}

In this section we show the parameter contours discussed in \ref{forecast} including the 4 nuisance parameters $\{m_T, m_\delta, m_\gamma, m_\kappa\}$, with $T = \Delta T_\mathrm{ISW}$, $\delta = \delta_g$, $\kappa = \kappa_\mathrm{CMB}$ and $\gamma = \gamma_1$ for notational brevity. We assume flat priors for all the nuisance parameters $m_i \in \left[-0.2, 0.2\right]$.

\begin{figure}[htbp!]
\centering
\includegraphics[width=15cm]{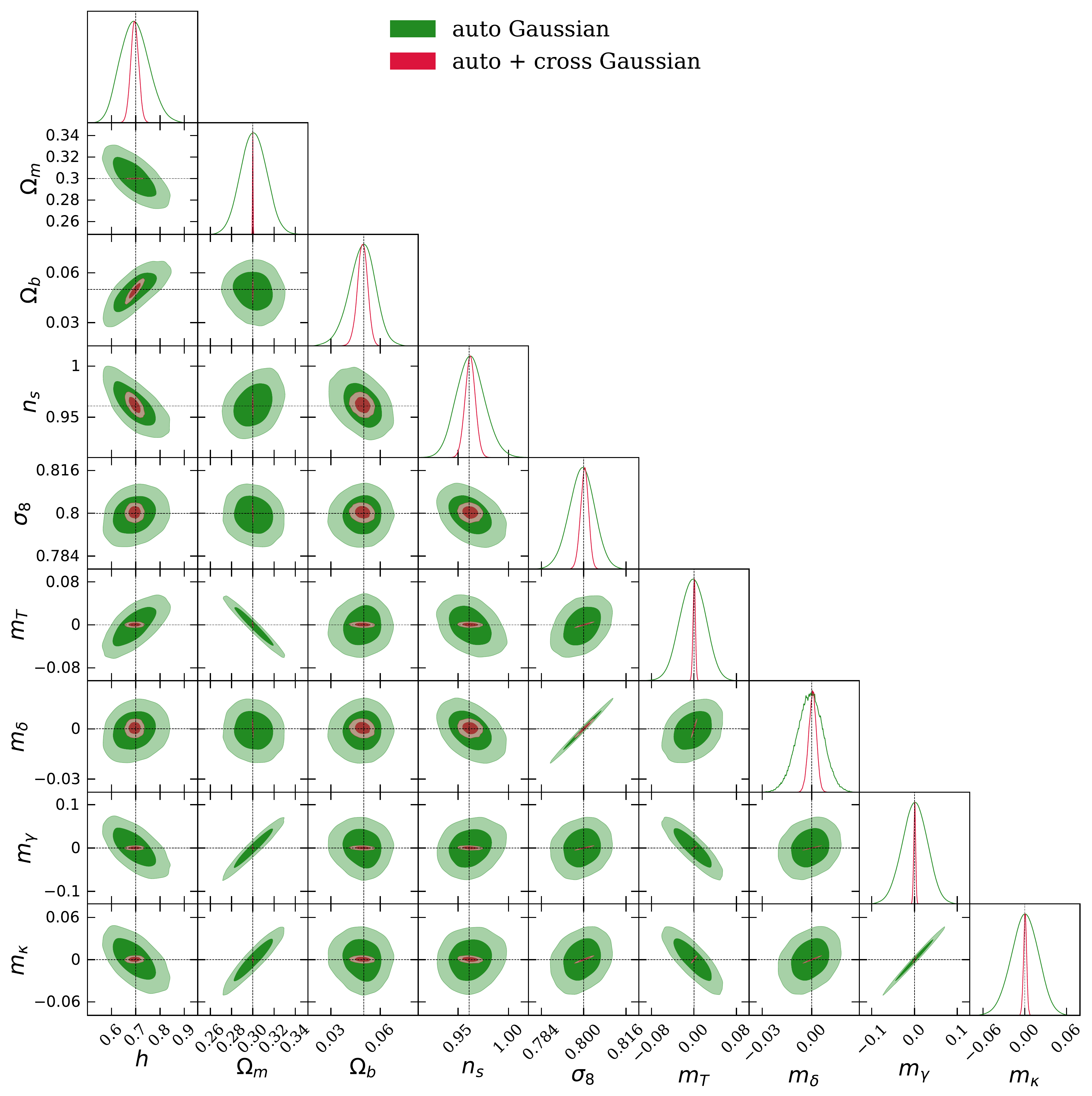}
\caption{Parameter constraints for the parameters described in section \ref{forecast} including the 4 nuisance parameters $\{m_T, m_\delta, m_\gamma, m_\kappa\}$. The triangle-plot show the contours obtained when using a Gaussian covariance matrix. The green and red contours are obtained using only the auto-power spectra and the auto- and cross-power spectra respectively. The inner (outer) contours depict the 68\% (95\%) confidence levels.}
\label{4probes_gaussian}
\end{figure}

\begin{figure}[htbp!]
\centering
\includegraphics[width=15cm]{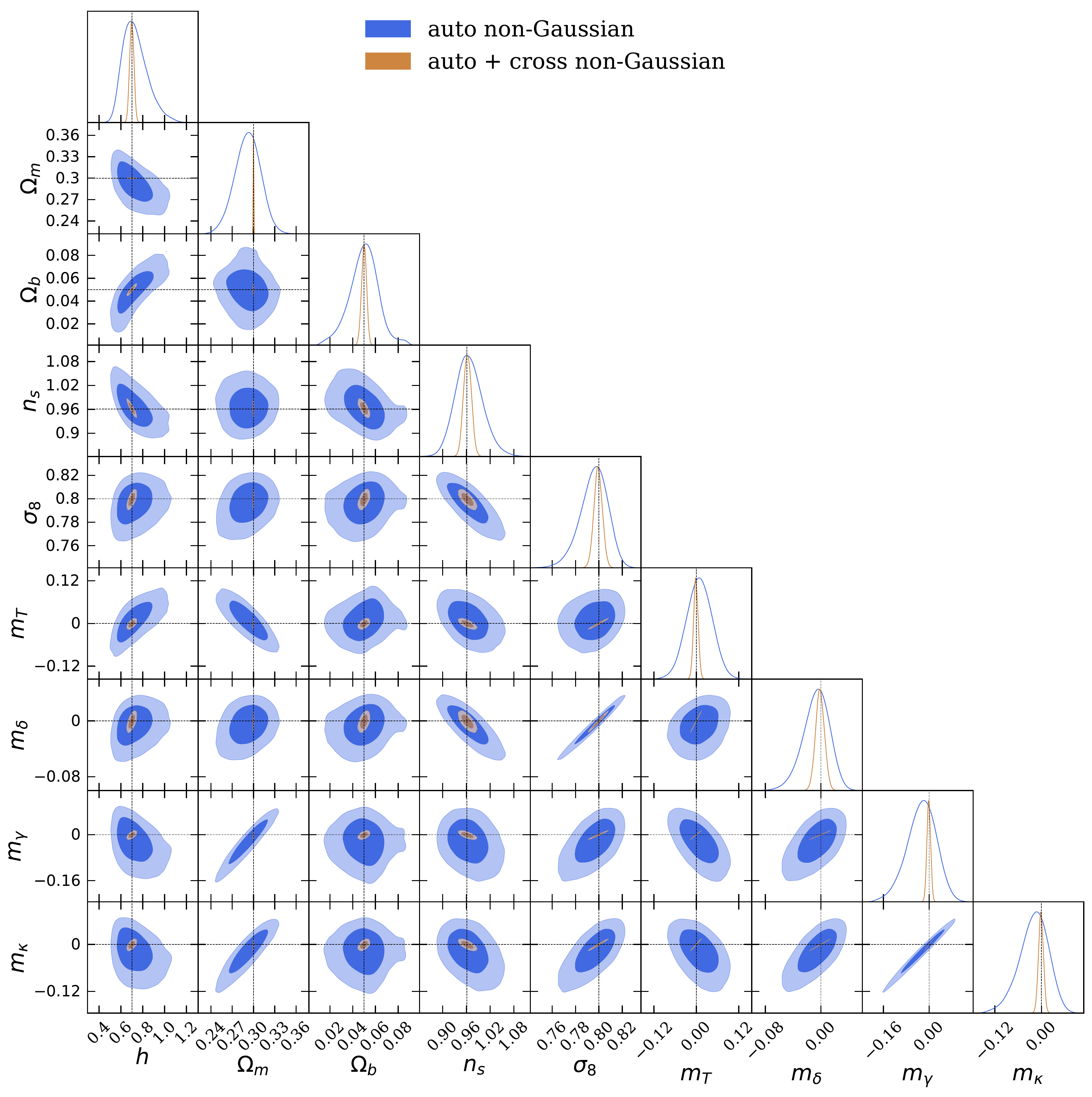}
\caption{Parameter constraints for the parameters described in section \ref{forecast} including the 4 nuisance parameters $\{m_T, m_\delta, m_\gamma, m_\kappa\}$. The triangle-plot show the contours obtained when using a simulation-based non-Gaussian covariance matrix. The blue and brown contours are obtained using only the auto-power spectra and the auto- and cross-power spectra respectively. The inner (outer) contours depict the 68\% (95\%) confidence levels.}
\label{4probes_simulation}
\end{figure}

\begin{figure}[htbp!]
\centering
\includegraphics[width=15cm]{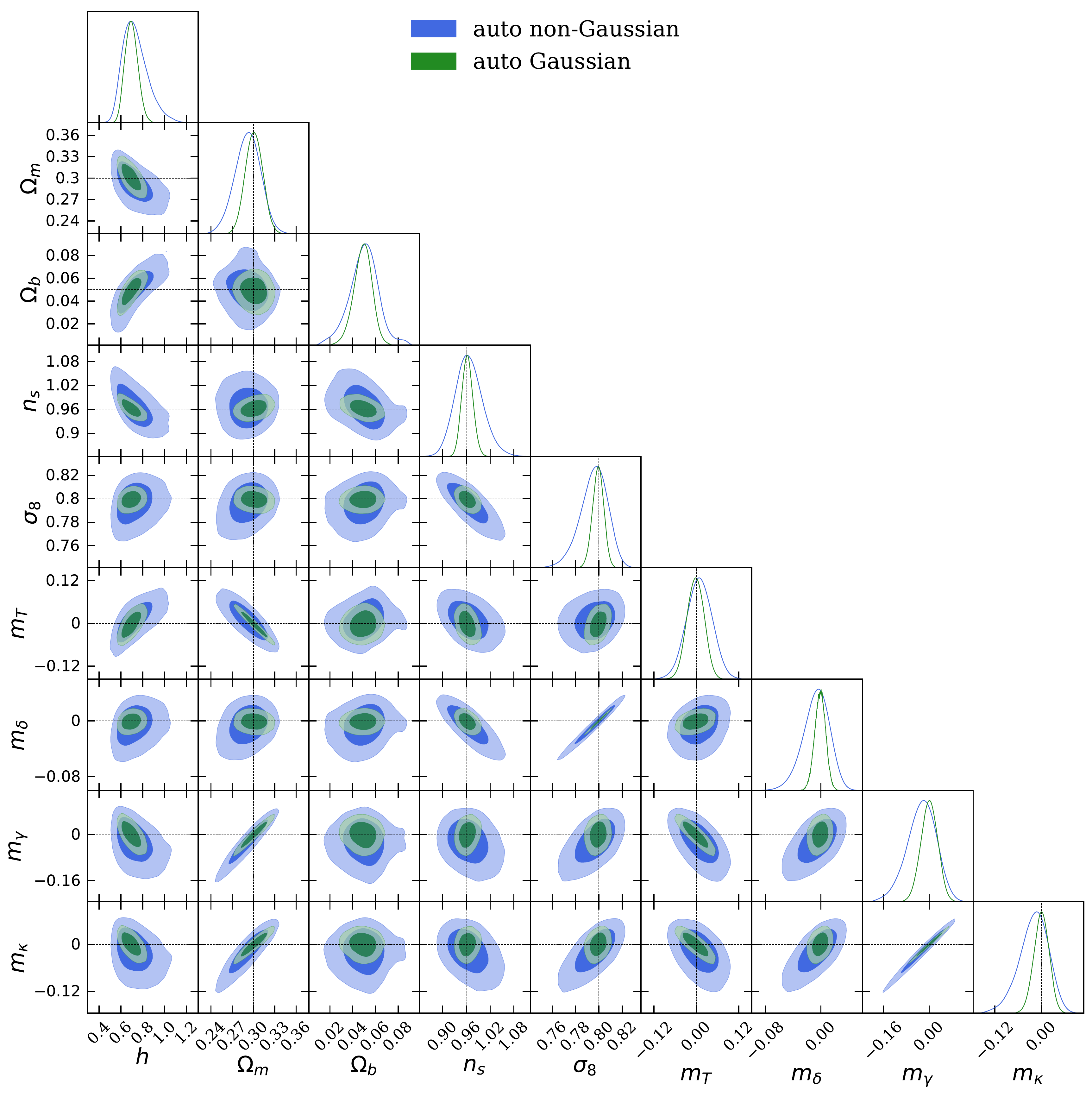}
\caption{Parameter constraints for the parameters described in section \ref{forecast} including the 4 nuisance parameters $\{m_T, m_\delta, m_\gamma, m_\kappa\}$. The triangle-plot show the contours obtained when using the auto-power spectra only. The blue (green) contours are obtained by using a non-Gaussian (Gaussian) covariance matrix. The inner (outer) contours depict the 68\% (95\%) confidence levels.}
\label{4probes_auto}
\end{figure}

\begin{figure}[htbp!]
\centering
\includegraphics[width=15cm]{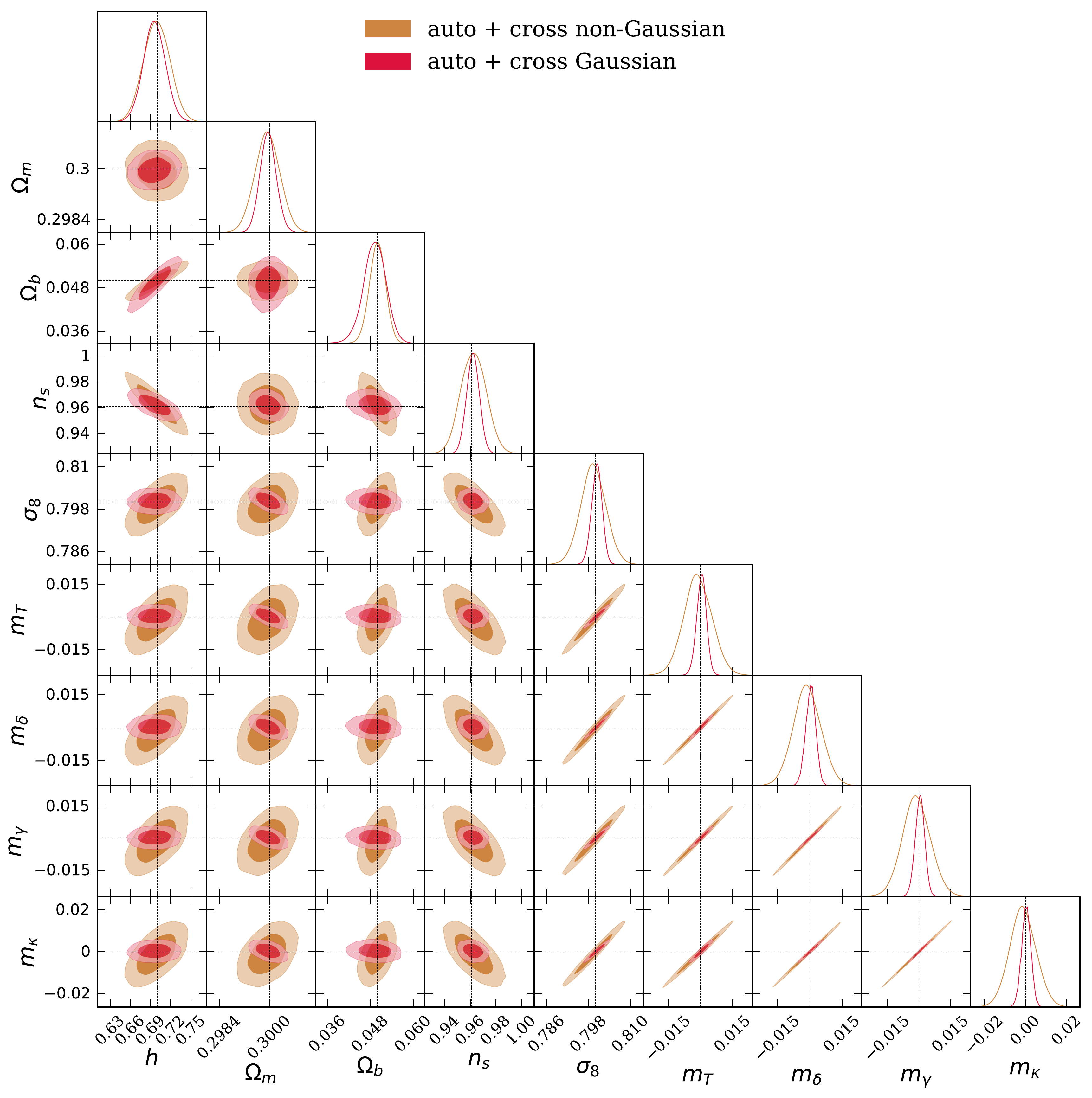}
\caption{Parameter constraints for the parameters described in section \ref{forecast} including the 4 nuisance parameters $\{m_T, m_\delta, m_\gamma, m_\kappa\}$. The triangle-plot show the contours obtained when using the auto- and cross-power spectra. The brown (red) contours are obtained by using a non-Gaussian (Gaussian) covariance matrix. The inner (outer) contours depict the 68\% (95\%) confidence levels.}
\label{4probes_cross}
\end{figure}

\end{appendix}

%\begin{thebibliography}{99}
%\end{thebibliography}
\end{document}